\documentclass[prd,amsfonts,twocolumn,nofootinbib,showpacs]{revtex4-1}
\usepackage{hyperref}
\usepackage{amsmath}
\usepackage{amssymb}
\usepackage{graphicx}
\usepackage{comment} 
\usepackage{array}
\usepackage{natbib}

\newcommand{\beq}{\begin{equation}}
\newcommand{\eeq}{\end{equation}}
\renewcommand{\d}{\delta}
\newcommand{\dv}{\delta v}
\newcommand{\du}{\delta u}
\newcommand{\dx}{\delta x}
\newcommand{\dth}{\delta\theta}
\newcommand{\dvf}{\delta\varphi}
\newcommand{\vf}{\varphi}
\newcommand{\eps}{\epsilon}
\newcommand{\rdot}{\dot{r}}

\renewcommand{\th}{\theta}

\begin{document}

\title{Gravitational self-force from radiation-gauge metric perturbations}
\author{Adam Pound, Cesar Merlin, and Leor Barack}
\affiliation{Mathematical Sciences, University of Southampton, Southampton, SO17 1BJ, United Kingdom}

\begin{abstract}

Calculations of the gravitational self-force (GSF) on a point mass in
curved spacetime require as input the metric perturbation in a
sufficiently regular gauge. A basic challenge in the program to compute
the GSF for orbits around a Kerr black hole is that the standard
procedure for reconstructing the metric perturbation is formulated in a class of
``radiation'' gauges, in which the particle singularity is non-isotropic
and extends away from the particle's location. Here
we present two practical schemes for calculating the GSF using a
radiation-gauge reconstructed metric as input. The schemes are based on
a detailed analysis of the local structure of the particle singularity
in the radiation gauges. We show that three types of radiation gauge
exist: two containing a radial string-like singularity emanating from
the particle, either in one direction (``half-string'' gauges) or both
directions (``full-string'' gauges); and a third type containing no
strings but with a jump discontinuity (and possibly a delta function)
across a surface intersecting the particle. Based on a flat-space
example, we argue that the standard mode-by-mode reconstruction
procedure yields the ``regular half'' of a half-string solution, or
(equivalently) either of the regular halves of a no-string solution. For
the half-string case, we formulate the GSF in a locally deformed
radiation gauge that removes the string singularity near the particle.
We derive a mode-sum formula for the GSF in this gauge, which is
analogous to the standard Lorenz-gauge formula but requires a correction
to the values of the regularization parameters. For the no-string case,
we formulate the GSF directly, without a local deformation, and we
derive a mode-sum formula that requires no correction to the
regularization parameters but involves a certain averaging procedure. We
explain the consistency of our results with Gralla's invariance theorem
for the regularization parameters, and we discuss the correspondence
between our method and a related approach by Friedman {\it et al}.

\end{abstract}

\date{\today}

\maketitle


\section{Introduction}
 
The gravitational self-force (GSF) acting on a point particle in curved
spacetime has been formulated rigorously only in a class of gauges
satisfying certain regularity conditions. The original formulations by
Mino, Sasaki, and Tanaka \cite{Mino-Sasaki-Tanaka:97} and Quinn and Wald
\cite{Quinn-Wald:97} relied explicitly on a specific choice of
gauge---the Lorenz gauge---and their main outcome was an equation of
motion with a GSF constructed from the metric perturbation (MP) in
that particular gauge. The Lorenz gauge is singled out in GSF
formulations for two main reasons: the linearized Einstein Field
Equations (EFE) take a manifestly hyperbolic form in that gauge, which
guarantees the well-posedness of an initial-value formulation; and
the point-particle singularity takes a locally isotropic form
in the particle's local frame. Reformulations of the GSF aimed at practical calculations have also generally been given in the Lorenz gauge~\cite{Barack:09,Poisson-Pound-Vega:11}. In particular, the standard mode-sum formulation~\cite{Barack-Ori:00}, which has been the dominant method of numerically calculating the GSF, originally required as input {\it specifically} the multipole modes of the Lorenz-gauge MP.
 
Unfortunately, while the Lorenz gauge is ideal for describing the local particle singularity, it is less well suited to numerical calculations of MPs in black-hole spacetimes. In a Schwarzschild background, the linearized EFE in the Lorenz gauge constitute a complicated set of coupled equations, even though they admit a full separation into tensorial-harmonic and frequency modes. The situation is worse in a Kerr background, where (to our knowledge) the equations cannot be separated in terms of any known set of harmonics. Significant progress has been made over the past decade in tackling these Lorenz-gauge perturbation equations, leading to successful GSF calculations in both Schwarzschild~\cite{Barack-Lousto:05,Barack-Sago:07,Berndtson:07,Barack-Sago:10,Dolan-Barack:13,Barack-Ori-Sago:08,Akcay:11,Akcay-Warburton-Barack:13} and Kerr~\cite{Dolan-Wardell-Barack:14}, but such direct numerical attacks involve complicated algorithms and are computationally intensive. There is therefore a strong motivation to pursue an alternative route, starting with an extension of the GSF formulation (and of practical schemes derived from it, like mode-sum regularization) to a broader class of gauges, including the standard gauges of black hole perturbation theory.
 
Indeed, there has already been steady progress in broadening the class
of admissible gauges for GSF calculations. Barack and Ori
\cite{Barack-Ori:01} showed how the GSF can be
obtained in any gauge related to the Lorenz gauge by a continuous
transformation. They also showed that the mode-sum formula (and the
values of the ``regularization parameters'' involved in it) is
invariant under such transformations. Gralla and Wald
\cite{Gralla-Wald:08,Gralla-Wald:11} went on to show how the GSF can be obtained in
gauges related to Lorenz by a transformation whose generator may have
a direction dependence at the particle (but is bounded there, and smooth
elsewhere). Their method, however, still began in the
Lorenz gauge, and their final expression for the GSF in alternative gauges had the
form of a transformation away from Lorenz. More
recently, for a subset of the Gralla-Wald class satisfying a certain parity condition near the particle, Gralla~\cite{Gralla:11} eliminated the preferred role of the Lorenz gauge, showing that the GSF in this ``parity-regular'' class can be obtained by averaging a certain ``full''
force exerted by the MP over a small sphere around the particle. Gralla also showed that the mode-sum formula in its
standard form (with the standard Lorenz-gauge parameter values) is invariant within this class. 
 
In the present work we seek a practical formulation---specifically, a mode-sum formulation---of the GSF starting from the MP in a gauge that lies outside any of the above classes: a \emph{radiation gauge}. This goal is motivated by the fact that the MP in a radiation gauge can be obtained in Kerr from the solution to a fully separable (spin-weighted) scalar field equation, in contrast to the nonseparable tensorial field equation that must be solved in the Lorenz gauge. The means of finding the MP in this way is the reconstruction procedure developed by Chrzanowski \cite{Chrzanowski:75} and Cohen and Kegeles \cite{Cohen-Kegeles:75,Cohen-Kegeles:79} (henceforth CCK), and in later work by Wald~\cite{Wald:78} and Stewart~\cite{Stewart:79}, for vacuum perturbations of algebraically special spacetimes. In the CCK procedure, the MP in a (traceless) radiation gauge is given as a second-order differential operator acting on a certain scalar field $\Psi$, called a Hertz potential, that in the Kerr case satisfies one or the other of the spin-$\pm2$ Teukolsky equations \cite{Teukolsky:73}. The Hertz potential is found not by solving the Teukolsky equation, but by solving an equation of the form ${\cal D}\Psi^*=\psi$, where $\psi$ is one of the Weyl scalars $\psi_0$ or $\psi_4$ corresponding to the physical MP, ${\cal D}$ is a certain fourth-order linear partial differential operator, and a $*$ denotes complex conjugation. This equation can be solved mode by mode in spin-weighted spheroidal harmonics~\cite{Lousto-Whiting:02,Ori:03}. The reconstruction procedure begins, therefore, with the Weyl scalars, which in Kerr can be found mode by mode by solving the separated Teukolsky equation with physical boundary conditions. Because the Weyl scalars determine the MP only up to certain stationary and axisymmetric pieces corresponding to mass and angular-momentum perturbations \cite{Wald:73}, the final step in the CCK procedure is to ``complete" the MP by finding and adding those pieces. The completion piece of the MP can be added in any gauge, and we shall refer to the gauge of the full MP as a \emph{completed radiation gauge}. 
 
The benefits of this procedure are obvious. Given a rigorous formulation of the GSF in terms of a completed-radiation-gauge MP, it reduces the most laborious numerical component of a GSF calculation to solving ordinary differential equations (ODEs) for scalar quantities. Unfortunately, utilizing the CCK procedure for the GSF is a delicate endeavor. The procedure was designed for vacuum, while we are interested in an MP sourced by a point particle. An extension of the procedure to nonvacuum scenarios was given by Ori \cite{Ori:03}. He showed that the extension introduces a nontrivial complication: a reconstructed MP invariably develops irregularities
in the presence of matter. These irregularities extend into the vacuum
region {\em outside} the matter; in the point particle case, they
extend off the particle's worldline, placing the MP outside the classes of gauges in which the GSF has been formulated. The problem was first
identified by Barack and Ori~\cite{Barack-Ori:01}, who considered
the form of a radiation-gauge MP in the elementary example of a static particle in flat space, and it was
reaffirmed by Ori, who showed it to be a generic feature of the CCK procedure in the presence of matter. 
 
Despite this issue, substantial progress has been made in the use of the CCK procedure for GSF calculations in a research program pursued by Friedman and collaborators \cite{Keidl-etal:07,Keidl-etal:10,Shah-etal:11,Shah-etal:12}. That effort has led to successful numerical reconstructions of point particle MPs, recently culminating in a landmark calculation of a gauge-invariant effect of the GSF in Kerr spacetime~\cite{Shah-etal:12}. However, several essential points remain to be addressed: precisely what MP is obtained from numerical reconstruction, precisely how does it relate to the standard classes of gauges used in formulations of GSF, and most importantly, how can we use it to calculate the GSF in a rigorous yet practical way?
 
Our work aims to address these issues and thereby erect a clear framework for combining CCK reconstruction with GSF calculations. To begin, we perform a thorough analysis of the form of a reconstructed MP near the particle. Our analysis confirms that the irregularity in a completed radiation gauge MP is never confined to the particle's worldline. We identify three categories
of radiation gauges: ``half string'' gauges, in which the MP is singular (at
any given time) along a radial ray starting at the particle and
extending either inward or outward; ``full string'' gauges, in which the MP is
singular along the entire radial axis through the particle; and ``no
string'' gauges, in which the MP has no string singularities but instead has a
discontinuity (and possibly a delta function) across a surface
containing the particle. Each radiation gauge belongs to one or the
other of these categories, so none has a singularity that is confined
to the particle's location. 

The string singularity in any of the
half-string or full-string solutions is quite strong: some of the
MP components blow up like the inverse of the squared distance to the
string (in a suitable local frame). As a result, a multipole expansion
is not obviously defined on any sphere intersecting a string, and a mode-by-mode CCK reconstruction procedure is likely to fail everywhere in spacetime for a full-string solution and in the entire ``irregular half" of a half-string solution. Based on this fact, we argue that a mode-by-mode numerical implementation of the CCK procedure can only produce the ``regular half'' of a half-string solution (or, equivalently, either of the regular halves of a no-string solution). We substantiate our argument with an analytical mode-by-mode MP reconstruction in the case of a static particle in flat space.
 
In formulating our GSF schemes we therefore assume that the input MP is
given in either a half-string or a no-string completed radiation gauge. We deal with each of these two cases separately.
 
For the half-string case we choose to remain within the Barack-Ori
class of gauges; that is, we formulate the GSF in a gauge related to
Lorenz via a continuous transformation. Starting from the MP in a
half-string gauge, we perform a gauge transformation that
``deforms'' the MP locally towards the Lorenz gauge. The resulting
\emph{locally Lorenz} (LL) MP is within the Barack-Ori class, meaning the GSF associated with it can be
obtained using the standard mode-sum formula. That formula involves two ingredients: the modes of a full force (distinct, in general, from the full force mentioned in conjunction with Gralla's results) and a set of analytically known ``regularization parameters" denoted $A_{\alpha}$, $B_{\alpha}$, $C_{\alpha}$, and $D_\alpha$~\cite{Barack-Ori:00}. When we apply the formula in the LL gauge, the input ``full'' modes are in the LL gauge. To make the
scheme practical, we re-express each of these LL modes as
a completed-radiation-gauge mode plus a gauge correction. This results in a mode-sum formula in which the input
modes are given (as desired) in a half-string completed radiation gauge, and the
gauge-transformation terms are expressed as corrections to the
standard regularization parameters. The input modes are to be obtained numerically as a directional limit from the ``regular'' side of the particle, while the gauge correction to the regularization parameters can be derived
analytically by considering the multipole expansion of the local gauge
transformation. We describe the calculation of the corrections for generic orbits in Kerr, and as concrete examples we give explicit results for generic orbits in
Schwarzschild and for circular, equatorial orbits in Kerr.
We find that for all orbits, there is no correction to the parameters $A_{\alpha}$, $B_{\alpha}$, and $C_{\alpha}$, but
generically, there occurs a nonzero correction to $D_{\alpha}$.
 
The above method uses as input ``one-sided'' MP information, namely modes of the
(half-string) completed-radiation-gauge MP and their derivatives at the
particle, calculated via a directional (radial) limit from either
``outside'' or ``inside'' the particle's orbit---whichever is the ``regular side" of the spacetime. Our second GSF formulation uses as input
the {\em no-string} completed-radiation-gauge MP, and it seeks to take advantage of
{\em both} one-sided pieces of MP information available in this case. Applying the LL formalism for this purpose proves to be difficult. Consequently, we
approach the problem without resorting to a local gauge deformation by appealing
to general transformation laws for the GSF, derived by Gralla and Wald. More precisely, we utilize a slightly generalized version of
Gralla and Wald's result, by extending their class of gauges to
include ones related to Lorenz by a gauge vector with certain irregularities away from the particle (and ones that are unbounded at the particle). We accomplish this by taking a step back to inspect the fundamental definition of
motion and GSF in the framework of matched asymptotic expansions. Once
we have a suitably generalized notion of GSF, we use it to derive a
mode-sum formula that takes as input the {\em average} of the two
one-sided full-force modes computed on either side of the particle in a
no-string gauge. In this formulation, the mode-sum formula involves only the standard Lorenz-gauge regularization parameters, with no corrections required. As a corollary, we also show that the GSF in a no-string gauge is given by the average of a full force over a sphere around the particle, as in Gralla's class of gauges.
 
The basic idea for our first (half-string) formulation was suggested
over a decade ago in Ref.\ \cite{Barack-Ori:01}; here
we develop this idea in full for the first time. Our second (no-string)
formulation is original, to the best of our knowledge.
 
Using our generalized notion of the GSF, we also briefly
discuss direct formulas for the GSF,  {\em without} a local
gauge deformation, in half-string or full-string solutions. In a certain class of full-string solutions we show, like in the no-string case, that the GSF is equal to an average over a sphere around the particle. In the
half-string case we derive a practical mode-sum formula with corrected
regularization parameters (which differ from those computed
using the gauge-deformation method). However, we consider these formulations less appealing than the alternatives---the full-string case is not amenable to a numerical mode-sum scheme, and in the half-string case the underlying definition of motion
is not intuitive---so we relegate this part of our discussion to an
appendix.
 
Much of our analysis is concerned with general results in any background spacetime admitting a radiation gauge. But from a practical point of view, the main outcomes of our work are two
alternative mode-sum formulas applicable in Kerr, given in Eqs.~\eqref{MS4} and \eqref{mode-sum-no-string}. The former
can be used with reconstructed (and completed) radiation-gauge modes
evaluated at the particle from either ``outside'' or ``inside'' the
orbit; the latter requires both of these one-sided values. The formula~\eqref{MS4}
contains a correction to the Lorenz-gauge value of the regularization
parameter $D_{\alpha}$, while \eqref{mode-sum-no-string} requires no such correction. In either
case, our analysis supplements the mode-sum formula for the GSF with two
important pieces of information: (1) an equation of motion that makes
clear the physical meaning of the prescribed GSF; and (2) a way
of calculating the MP associated with the prescribed GSF, in the same
gauge, given a completed-radiation-gauge solution. 
This should provide the complete information from which one can deduce any
physical effect of the GSF at first order in the mass ratio, for generic
orbits in Kerr.

     \subsection{Structure of this paper}
 
Section\ \ref{gauges} sets the stage for our analyses with a survey of the classes of gauges in which the GSF has been formulated, along with a review of work that has been done to relate the completed radiation gauges to these classes. Using a specially constructed Fermi-like coordinate system near the particle, in Sec.\ \ref{Fermi-analysis} we perform a thorough analysis of the local singularity structure in the completed radiation gauges, classifying them into full-string,
half-string, and no-string subclasses and clearly identifying the relationship of each to the classes of gauges used to formulate the GSF.

Starting from these results, in Sec.~\ref{LL-method} we devise our scheme for calculating the GSF in a locally deformed half-string gauge, and as examples we calculate the explicit corrections to the regularization parameters for generic orbits in Schwarzschild and for circular, equatorial orbits in Kerr. 

In Sec.\ \ref{force-no-string} we turn to our second method: a direct formulation
of the GSF in an (undeformed) no-string gauge, with a corresponding mode-sum formula. 
 
Section \ref{reconstruction-completion} validates the compatibility of these two methods with a mode-by-mode CCK
reconstruction and completion. We perform an analytical reconstruction and completion in the
elementary example of a static particle in flat space, illustrating explicitly how half-string and no-string solutions arise from this
procedure. We highlight the role of the completion in
ensuring that the EFE are satisfied globally in the no-string solution.
 
To keep the presentation relatively streamlined, we relegate a substantial amount of material to appendices. Appendix~\ref{Rad-force} forms the backbone for the body of the paper. It reviews the foundations of GSF in matched asymptotic expansions, and it describes the procedure we use to derive expressions for the GSF in undeformed radiation gauges, based on a method outlined by Gralla and Wald~\cite{Gralla-Wald:08}. It sketches, using ideas from Colombeau theory, how to rigorously interpret the GSF in the undeformed no-string gauge despite the gauge's irregularities away from the particle. Contrarily, it shows the drawbacks of formulating the motion in an undeformed half- or full-string gauge; for completeness, despite those drawbacks, Appendix~\ref{force-with-string} derives expressions for the GSF in the gauges with strings. 

Appendices~\ref{trans_to_global}, \ref{Schw-example}, and Appendix~\ref{dF-arbitrary} are concerned with expansions in the limit of small coordinate distances. Appendix~\ref{trans_to_global} presents the transformation of the local gauge vector from Fermi-like coordinates to any arbitrarily chosen ones. As a complement, Appendix~\ref{Schw-example} derives the local gauge vector directly in a global coordinate system, without reference to local Fermi-like coordinates, in the particular case of a Schwarzschild background. Appendix~\ref{dF-arbitrary} establishes general properties of (and necessary formulas for) the gauge transformation of the full force.

     \subsection{Notation and conventions}
 
Throughout this work we use geometrized units (with $G=c=1$) and the metric signature ${-}{+}{+}{+}$. For gauge transformations generated by a vector $\xi^\alpha$, we use the sign convention $x^\alpha\to x^\alpha-\xi^\alpha$. Greek indices $\alpha,\ \beta,\ \gamma$ run from 0 to 4. In Sec.~\ref{reconstruction-completion}, lowercase Latin indices $a,\ b,\ c$ refer to the $(t,r)$ plane in a spherical polar coordinate system $(t,r,\theta,\varphi)$, and uppercase Latin indices $A,\ B,\ C$ refer to the coordinates $\theta^A=(\theta,\varphi)$ on the spheres of constant $(t,r)$. In all other sections, lowercase Latin indices refer to spatial coordinates and run from 1 to 3, uppercase Latin indices refer to the first two of those coordinates and run from 1 to 2, and these indices are raised and lowered with a Kronecker delta $\delta_{ab}$ or $\delta_{AB}$. Unless explicitly stated otherwise, $\ell^\alpha$ can be taken to refer to either an ingoing principal null vector or an outgoing one.


\section{Gauge and motion}\label{gauges}
Before setting about deriving expressions for the GSF, we must establish a framework in which to perform and interpret those derivations. Doing so requires defining the particle's position, the GSF that governs its evolution, and how each is affected by the choice of gauge. For a review of the formalism necessary for this, developed by Gralla and Wald in Ref.~\cite{Gralla-Wald:08} (and expanded in Refs.~\cite{Pound:10a,Pound:10b}), we refer the reader to Appendix~\ref{Rad-force}. In this section our emphasis is on one particular aspect of the formalism: the classes of gauges compatible with it, and how motion is defined in each.  

Imagine that the ``point mass'' is in fact a very small, compact extended object. It moves through an external background spacetime $g_{\mu\nu}$, creating as it does an MP $\varepsilon h_{\mu\nu}+O(\varepsilon^2)$, where $\varepsilon\equiv1$ counts powers of its very small mass $\mu$. We write the object's worldline as the perturbative expansion\footnote{An alternative ``self-consistent" treatment of the motion, used often in the literature and put on a systematic basis in Refs.~\cite{Pound:10a,Pound:10b,Pound:12a,Pound:12b}, instead describes the trajectory in its unexpanded form $z^\mu(\tau,\varepsilon)$. Although our work could be made compatible with the self-consistent description, throughout this paper we opt to use a perturbative expansion of the worldline, as presented by Gralla and Wald~\cite{Gralla-Wald:08}.} 
\beq
z^\mu(\tau,\varepsilon) = z_0^\mu(\tau) + \varepsilon z_1^\mu (\tau)+O(\varepsilon^2).
\eeq
The leading term, $z^\mu_0(\tau)$, is the coordinate description of a geodesic $\Gamma$ of the background spacetime $g_{\alpha\beta}$, and $\tau$ is proper time on $\Gamma$. At linear order in $\mu$, the object's MP $h_{\mu\nu}$ is that of a point mass on $\Gamma$~\cite{Gralla-Wald:08}. Inspired by that fact, in later sections we will refer to $z^\mu_0(\tau)$, rather than $z_0^\mu(\tau) + \varepsilon z_1^\mu (\tau)$, as the ``particle's position". The next term, $z^\mu_1$, is a vector field defined on $\Gamma$. It describes the first-order deviation of the object's center of mass from $\Gamma$, where the center of mass is defined by the object's mass dipole moment in a locally inertial frame centered on $\Gamma$. In the Lorenz gauge, this first-order correction to geodesic motion is governed by 
\beq\label{motion-in-Lorenz}
\mu\frac{D^2z_{1{\rm Lor}}^\alpha}{d\tau^2} = -\mu R^\alpha{}_{\mu\beta\nu}u^\mu z_{\rm 1Lor}^\beta u^\nu + F^\alpha_{\rm Lor},
\eeq
where $u^\mu=\frac{dz_0^\mu}{d\tau}$, and $F^\alpha_{\rm Lor}\propto\mu^2$ is the Lorenz-gauge GSF produced by the MP of a point mass moving on $\Gamma$. In addition to the GSF, the equation of motion contains the term $-R^\alpha{}_{\mu\beta\nu}u^\mu z_{\rm 1Lor}^\beta u^\nu$, which is purely a background effect, familiar from the geodesic deviation equation; it expresses the fact that if $F^\alpha_{\rm Lor}$ forces the small object slightly off $\Gamma$, the object continues to move relative to $\Gamma$ due to the background curvature. 

The GSF in the Lorenz gauge can be written in numerous (equivalent) forms, of which we will require two in this paper: the mode-sum form
\beq\label{mode-sum_form}
F^\alpha_{\rm Lor} = \sum_\ell\left[(\tilde F^\alpha_{\rm Lor})^\ell-A^\alpha L -B^\alpha -C^\alpha/L\right]-D^\alpha,
\eeq
and what we will call the Quinn-Wald-Gralla form
\beq\label{Gralla-average}
F^\alpha_{\rm Lor} = \lim_{s\to0}\frac{1}{4\pi s^2}\int\tilde F^\alpha_{\rm Lor}dS.
\eeq
In the first expression $L=\ell+\frac{1}{2}$, and $(\cdot)^\ell$, roughly speaking, denotes a spherical-harmonic mode (defined precisely in Sec.~\ref{LL-method}). In the second expression the integral is over a small two-sphere centered on $\Gamma$ with a geodesic radius $s$ perpendicular to $\Gamma$, $dS=s^2d\Omega+O(s^4)$ is the surface element on that sphere (with $d\Omega$ being the surface element on a unit sphere), and the integration is performed componentwise in a local coordinate frame centered on $\Gamma$. In both expressions $\tilde F^\alpha$ is a ``full gravitational force" exerted by the MP, given in its simplest form by
\beq\label{full_force1}
\tilde F^\alpha \equiv -\frac{1}{2}\mu(g^{\alpha\beta}+\tilde u^{\alpha}\tilde u^{\beta})(2\nabla_{\!\delta}h_{\beta\gamma}-\nabla_{\!\beta}h_{\gamma\delta})\tilde u^\gamma \tilde u^\delta.
\eeq
The tildes will be explained momentarily. If $h_{\alpha\beta}$ were some smooth external perturbation (e.g., an incoming gravitational wave), then the full force evaluated at the particle would reduce to the gravitational force on $\mu$ due to $h_{\alpha\beta}$. In our case, where $h_{\alpha\beta}$ is the field of the particle itself, the full force is defined only off $\Gamma$; on $\Gamma$, it diverges. To define it as a field off the worldline, we have introduced $\tilde u^\alpha$ as some smooth extension of $u^\alpha$ off $\Gamma$. The extension can be chosen freely in the mode-sum formula, so long as the regularization parameters $A^\alpha$, $B^\alpha$, $C^\alpha$, and $D^\alpha$ are calculated accordingly. In the Quinn-Wald-Gralla formula it is defined by parallel propagation along geodesics perpendicular to $\Gamma$. (A more general form of the full force will be described in Sec.~\ref{LL-method}, along with a more complete review of the mode-sum formula.) 

Now suppose we begin from Eq.~\eqref{motion-in-Lorenz} and wish to find the GSF in a gauge other than Lorenz. Finding the GSF in this alternative gauge can be reduced to determining how $z_1^\mu$ transforms under a gauge transformation, since, after all, the GSF is merely a term in the evolution equation for $z^\mu_1$. To frame the discussion, let us foliate the spacetime near $\Gamma$ with three-dimensional spatial hypersurfaces $\Sigma$ that intersect $\Gamma$ orthogonally, and let $x^a$ be local Cartesian coordinates on each $\Sigma$, with $x^a=0$ at $\Gamma$; this will be a recurrent construction throughout later sections. We can arrange for $z^\mu_1$ to be orthogonal to $\Gamma$ and then focus on the spatial component $z^a_1$. Under a gauge transformation generated by a vector $\xi_\alpha$, $z_1^a$ transforms as $z^a_1\to z^a_1+\Delta z^a_1$, with 
\beq
\Delta z_1^a = -\lim_{s\to 0}\frac{3}{4\pi}\int n^a n^b\xi_b d\Omega. \label{Delta z}
\eeq
The notation follows Eq.~\eqref{Gralla-average}, and $n^a$ is the unit normal to the two-sphere. 

We again refer the reader to Appendix~\ref{Rad-force} for a derivation of this result. Our concern here lies only with the class of gauges for which the derivation is valid (discussed in more detail in the appendix). Principally, the spatial components $\xi_a$, those tangent to $\Sigma$, must be bounded in the limit to $\Gamma$, behaving locally as $\xi_a(x^b)=Z_a(0)+K_a(n^b)+o(1)$, while the component perpendicular to $\Sigma$ can diverge in the limit but no more strongly than $\ln s$ (and in a spherically symmetric way). Among other things, these conditions imply that the divergence of the first-order MP in the new gauge, $h_{\alpha\beta}=h^{\rm Lor}_{\alpha\beta}+2\xi_{(\alpha;\beta)}$, is no stronger than in the Lorenz gauge, behaving as $\sim 1/s$ near the particle. If $\xi_\alpha$ satisfies these conditions, \emph{and} the right-hand side of Eq.~\eqref{Delta z} evaluates to a finite, $C^2$ function along $\Gamma$, we say the gauge is \emph{sufficiently regular} to define the GSF.

Let us now examine the behavior of $z_1^a$ in the three classes of gauges we referred to in the introduction: in order of increasing generality, the Barack-Ori class, the Gralla class, and the Gralla-Wald class.

A gauge in the Barack-Ori class is related to the Lorenz gauge by a continuous gauge vector $\xi_\alpha$, allowing us to write $\xi_\alpha(x^a)=\xi_\alpha(0)+o(1)$.\footnote{In fact, the Barack-Ori results are valid for any transformation generated by a gauge vector with a well-defined limit to the particle, meaning cosmetic singularities such as $s\ln s$ at the particle are allowed. Singularities of this sort must be allowed in our analysis. We make all such functions continuous at $x^a=0$ by defining their values there to be equal to their limits.} Using the easily established identity $\int n^a n^bd\Omega=\frac{4\pi}{3} \delta^{ab}$, we can evaluate Eq.~\eqref{Delta z} to find 
\beq
\Delta z_1^a=-\xi^a\rvert_\Gamma.
\eeq
The right-hand side is simply the transformation of the coordinates $x^a\to x^a-\xi^a$ evaluated on $\Gamma$. In other words, transformations within the Barack-Ori class translate the center of mass just as they translate any other point. This corresponds to the simplest and most intuitive notion of the object's position. 

Near the particle, the first-order MP in the Barack-Ori class has the isotropic, Lorenz-gauge form~\cite{Poisson-Pound-Vega:11} 
\beq
h_{\alpha\beta} = \frac{2\mu}{s}(g_{\alpha\beta}+\tilde u_\alpha\tilde u_\beta)+o(s^{-1}),
\eeq
where $\tilde u_\alpha$ is any smooth extension of $u_\alpha$ off $\Gamma$.

Next, a gauge in the Gralla class is related to the Lorenz gauge by a gauge vector $\xi_\alpha$ that is smooth off $\Gamma$ but is allowed a certain type of ill-defined limit to $\Gamma$. Specifically, the vector must be bounded at $\Gamma$ and its spatial components must have the local form $\xi_a(x^b)=Z_a(0)+K_a(n^b)+O(s)$ with $K_a$ having odd parity, $K_a(-n^b)=-K_a(n^b)$, under the parity transformation $n^a\to-n^a$. We say any $\xi_\alpha$ is parity-regular if its spatial components have that leading-order form $Z_a(0)+K_a(n^b)$ with odd $K_a$. Noting that the integral of $n^an^bK_b(n^c)$ vanishes because $K_b$ is odd and $n^a n^b$ is even, for such a gauge vector we can reduce Eq.~\eqref{Delta z} to the simple average 
\beq
\Delta z_1^a = -\frac{1}{4\pi}\lim_{s\to 0}\int \xi^a d\Omega,\label{position-av}
\eeq
which evaluates to $\Delta z_1^a=-Z^a(0)$. Depending upon one's predilections, this type of transformation of the object's position might be deemed just as sensible as the result $\Delta z_1^a=-\xi^a|_\Gamma$: if the shift in position of a point depends on the direction one approaches it from, then the average over all directions plausibly yields the net shift. Gralla also showed that for any MP in his class, the GSF is given by the same simple spherical average~\eqref{Gralla-average} as in the Lorenz gauge. This form was originally taken as an axiom by Quinn and Wald in their derivation of the GSF in the Lorenz gauge~\cite{Quinn-Wald:97}. Gralla's work shows, without assuming it as an axiom, that it holds true in a large class of gauges; hence the name Quinn-Wald-Gralla we have given it. Additionally, Gralla showed, based on this result, that in his class of gauges the GSF can be written in the standard mode-sum form~\eqref{mode-sum_form}, with the standard Lorenz-gauge parameter values, lending great utility to these gauges. 

It follows from the conditions on $\xi_\alpha$, together with the fact that a partial derivative reverses the parity of a function it acts on, that any MP in the Gralla class must be smooth off $\Gamma$ and must be parity-regular. By the latter we mean it must have the local form 
\beq
h_{\alpha\beta} = \frac{\mu F_{\alpha\beta}(n^a)}{s}+O(1),
\eeq
and the leading-order spatial components $\frac{\mu F_{ab}(n^c)}{s}$ must have even parity under $n^a\to-n^a$.

Last, a gauge in the Gralla-Wald class is related to the Lorenz gauge by a gauge vector $\xi_\alpha$ that is smooth off $\Gamma$ but is allowed an arbitrary (bounded) direction-dependent limit to $\Gamma$. This means we can write $\xi_a(x^b)=Z_a(0)+K_a(n^b)+O(s)$ with $K_a$ now allowed any smooth dependence on $n^a$. Under these conditions, the simple averaging result \eqref{position-av} does not generically hold true, because any piece of $K_a(n^b)$ with even parity will contribute a finite amount to the integral in \eqref{Delta z}. $K_a(n^b)$ is referred to as a supertranslation. For each angle from which one approaches $s=0$, $K_a$ yields a different translation at $\Gamma$. If $\xi_\alpha$ is parity-regular, the supertranslations do not alter the position of the particle. Conversely, a parity-irregular transformation [i.e., when $K_a(n^b)$ has even or indefinite parity] is one in which $\xi_\alpha$ contains supertranslations that do alter the position of the particle, and a parity-irregular MP is one related to a parity-regular MP by a parity-irregular transformation. Historically, parity-irregular supertranslations have created challenges~\cite{Regge-Teitelboim:74,Ashtekar-Romano:92,Compere-Dehouck:11,Virmani:12} in canonical descriptions of spacetimes, as well as in defining angular momentum, because the group of supertranslations is infinite, while we would wish for a canonical 3-momentum, for example, to be associated with the three-dimensional group of translations. Here we might have similar problems in a canonical description of the motion of the small object. But even if we put those difficulties aside, we note that unlike transformations within the Barack-Ori and Gralla classes, the effects of a parity-irregular transformation simply do not seem to correspond to any physically intuitive idea of the object's position relative to $\Gamma$.

It follows from the conditions on $\xi_\alpha$ that any MP in the Gralla-Wald class must be smooth off $\Gamma$ and have the local form 
\beq\label{Gralla-Wald-MP}
h_{\alpha\beta} = \frac{\mu F_{\alpha\beta}(n^a)}{s}+O(1),
\eeq
where now, unlike in the Gralla class, $F_{\alpha\beta}(n^a)$ is allowed an arbitrary (smooth) dependence on $n^a$. 

The three classes we have described do not cover all possible gauges sufficiently regular to define the GSF. In particular, gauge vectors that are unbounded in the limit to $\Gamma$ are possible, so long as the spatial components remain bounded in that limit; and gauge vectors that are not smooth off the particle are possible, so long as the integral in Eq.~\eqref{Delta z} remains well defined. For the reasoning behind allowing these irregularities, we again refer the reader to Appendix~\ref{Rad-force}.

Nevertheless, the three subclasses, and their attendant notions of position, will provide a touchstone throughout our analysis. They are also of great importance because  prior discussions of the radiation gauge in the context of GSF have hinged upon whether any radiation gauge (or completed radiation gauge) falls within one of the three classes. Barack and Ori~\cite{Barack-Ori:01} first noted that the radiation gauge includes string singularities emanating from the particle, and so they ruled it out of their class of gauges. Later, Friedman and collaborators~\cite{Keidl-etal:10,Shah-etal:11} argued that a reconstructed MP is parity-regular and therefore the GSF in the gauge of that MP is given by the Quinn-Wald-Gralla formula~\eqref{Gralla-average}. Based on that argument, they devised a method of calculating the GSF by subtracting from the full force $\tilde F^\alpha$ a numerically determined singular force field $F^\alpha_S$ satisfying two conditions: (a) $\lim_{s\to0}\int F^\alpha_Sd\Omega=0$; and (b) $\tilde F^\alpha-F^\alpha_S$ is continuous at $\Gamma$. Gralla~\cite{Gralla:11} called into question these arguments by noting that since the radiation gauge is irregular off the particle, it does not fall within his class of gauges. However, he conjectured that because the expressions remain well defined even in the case of off-particle (integrable) irregularities, the Quinn-Wald-Gralla form of the force and the mode-sum formula would remain valid in the reconstructed even-parity radiation gauge.

Our analysis in the bulk of this paper largely agrees with these results and conjectures, but it also reveals many subtle complications in them. In the next section, by analyzing the local behavior of the gauge vectors $\xi_\alpha$ that bring the MP from the Lorenz gauge to a completed radiation gauge, we confirm that there do exist parity-regular completed radiation gauges, and the MP in these gauges can be obtained from a CCK reconstruction procedure---these are the no-string gauges mentioned in the introduction. However, we show that the MP in these gauges generically contains nonzero jump discontinuities that occur in both the divergent and bounded pieces of the MP at the particle. Despite the discontinuities, in Sec.~\ref{force-no-string} we do find that in the no-string gauges, the GSF is given by the angular-average formula~\eqref{Gralla-average}, as conjectured by Gralla. Yet the discontinuities pose a seemingly intractable problem in the numerical scheme of Friedman and collaborators, because they imply that the conditions (a) and (b) mentioned above cannot be simultaneously satisfied in general: terms finite at the particle must be included in the singular force $F^\alpha_S$ in order to satisfy condition (b), but those terms generically possess a finite spherical average that violates condition (a). We also find that even though the Quinn-Wald-Gralla formula is valid, Gralla's result concerning the invariance of the mode-sum formula for parity-regular gauges is {\em not} applicable in a no-string gauge: the correct mode-sum  formulation requires a two-sided average of the full modes and parameters. We explain how the failure of the mode-sum formula to be invariant arises directly as a result of the discontinuity away from the particle, which violates the smoothness condition in Gralla's class of gauges. 

Our analysis also shows that outside the no-string gauges, no completed radiation gauges are parity-regular (with the exception of a certain subclass of full-string gauges, which we deem too singular to be considered in a numerical implementation). In particular, all half-string gauges are found to be parity-irregular, with MPs of the general form of Eq.~\eqref{Gralla-Wald-MP}, but with the functions $F_{\alpha\beta}(n^a)$ diverging on the string. Because of the counterintuitive interpretation of motion in parity-irregular gauges, we handle the half-string gauges by performing local deformations to transform them into parity-regular gauges within the Barack-Ori class, where the motion is most intuitive. To complement that calculation, we also calculate expressions for the GSF directly in the undeformed parity-irregular gauges, but we confine those results to Appendix~\ref{force-with-string}.

Regardless of parity, our analysis will show that all classes of radiation gauges fall outside the Barack-Ori, Gralla, and Gralla-Wald classes, because all contain irregularities not just at the particle, but \emph{away} from the particle (in addition, they are related to the Lorenz gauge by vectors that are logarithmically unbounded at the particle). Nevertheless, they do fall within the class of gauges sufficiently regular to define the GSF, and expressions for the GSF can be readily calculated in each of them. Due to their irregularities away from the particle, they do introduce certain technical pitfalls into the formalism upon which Eq.~\eqref{Delta z} is based. In Appendix~\ref{Colombeau} we describe those pitfalls and a means of bypassing them in the no-string gauge (and the parity-regular subclass of the full-string gauges) using the Colombeau theory of nonlinear distributions; we also show that the same circumvention appears to fail in the half-string gauges (and most full-string gauges), for reasons related to the gauges' parity-irregularity. However, those concepts are not necessary for our concrete calculations, and we continue apace without them.


\section{Local singularity structure in radiation gauges}\label{Fermi-analysis}

We start by analyzing the local structure of the radiation-gauge MP near a point particle. We do this by working out explicitly a gauge transformation from the Lorenz-gauge MP (in which the local singularity structure is known) to the radiation-gauge MP, accurate to leading order in the singularity. We will show that, even fully exhausting the freedom to perform gauge transformations within the class of radiation gauges, it is not possible to construct a radiation gauge in which the singularity is supported solely on the particle. We will introduce a classification of radiation gauges according to the form of the irregularity away from the particle, identifying three classes. A preliminary analysis has already been carried out, in Ref.\ \cite{Barack-Ori:01}, considering the elementary problem of a static particle in flat space. Here we present a much more complete analysis, and we extend it to a particle in an arbitrary geodesic motion in an arbitrary algebraically special vacuum background. The restriction to algebraically special backgrounds is necessary because radiation gauges have been shown to exist only when there is a repeated principal null direction~\cite{Price-Shankar-Whiting:07}; we restrict to vacuum because the local form of the Lorenz-gauge MP is known only for a vacuum backgrounds. Beyond those specifications, we leave the analysis general. 

Our analysis in this section is applicable without modification to either the ``ingoing'' or ``outgoing'' radiation gauge \cite{Keidl-etal:07}. We write the radiation gauge condition as $h_{\alpha\beta}\ell^\beta=0$ in either case, where, counter to common usage, we denote by $\ell^\beta$ either an ingoing or outgoing principal null vector.  Elsewhere in the paper we will specialize to the ingoing gauge, corresponding to $\ell^\beta$ being an outgoing null vector; we will state clearly when we do so. 

We base our analysis on a judicious choice of a Fermi-like coordinate system based at the particle's worldline, and we will start below by describing the construction of that system. At the end of our calculation, we transform our results to an arbitrary coordinate system. For the benefit of readers less at ease with Fermi-type coordinates, in Appendix~\ref{Schw-example} we repeat our analysis using Eddington-Finkelstein coordinates in the particular case of a Schwarzschild background.   

\subsection{Fermi-like coordinates}\label{local_coords}

Let $x^{\alpha}=x^{\alpha}_{p}(\tau)$ be some (for now arbitrary) coordinates on the particle's zeroth-order, geodesic orbit $\Gamma$. We introduce the notation $x^\alpha_p$ in place of $z_0^\alpha$ for future convenience. 

To perform our local analysis, we adopt Fermi-like coordinates $(\tau,x^a)$ centered on $\Gamma$. Fermi normal coordinates are adapted for convenient calculations near a worldline; here, we slightly modify them to accommodate the preferred direction that the principal null vector introduces into our calculations. 

Ordinary Fermi normal coordinates are constructed by first erecting an orthonormal basis $(u^\alpha, e^\alpha_a)$, $a=1,2,3$, that is parallelly propagated along $\Gamma$. In a neighborhood of $\Gamma$, the spacetime is then foliated with spatial hypersurfaces $\Sigma_\tau$, each of which is generated by spatial geodesics that intersect $\Gamma$ orthogonally at a point $x_p(\tau)$. On $\Sigma_\tau$, a Cartesian coordinate system is established, with coordinates defined as $x^a\equiv-e^a_{\bar\alpha}\nabla^{\bar\alpha}\sigma(\bar x, x)$. Barred indices correspond to the point $\bar x\equiv x_p(\tau)$, and $\sigma(\bar x,x)$ is Synge's world function, equal to one-half the squared geodesic distance from $\bar x$ to $x$. With this definition, $x^a$ has a magnitude
\beq
s\equiv\sqrt{\delta_{ab}x^a x^b}
\eeq
equal to the geodesic distance to $x$, and it has a direction along a triad leg $e^\alpha_a$. On the worldline, we have $x^a=0$. By labelling each point on $\Sigma_\tau$ with the time $\tau$, one arrives at a 4D coordinate system $(\tau,x^a)$. In these coordinates $g_{\mu\nu}$ takes the locally flat form $\eta_{\mu\nu}+O(s^2)$, where $\eta_{\mu\nu}={\rm diag}(-1,1,1,1)$, with Christoffel symbols $\Gamma^\alpha_{\beta\gamma}=O(s)$. 

For our purposes, on each $\Sigma_\tau$ we wish to single out the direction along the principal null vector $\ell^\alpha$. We let $x^a=(x^A,z)$, $A=1,2$, and we keep the spatial piece of $\ell^\alpha$ fixed in the positive $z$ direction at $s=0$, such that
\beq
\ell^a = \hat\ell\delta^a_z+O(s), \qquad \hat\ell>0.
\eeq
Since $\ell^\alpha$ is null, we also have $\ell^\tau=\hat\ell+O(s)$. By keeping the orientation of our coordinates fixed relative to $\ell^\alpha$ in this way, we cease to parallel propagate the spatial triad $e^\alpha_a$ along $\Gamma$. Instead, we allow it to rotate according to
\beq
\frac{D e^\alpha_a}{d\tau} = \omega_a{}^b e^\alpha_b,
\eeq
where $\omega_a{}^b$ is a time-dependent rotation matrix. More specifically, we have chosen one of our triad legs to be
\begin{equation}\label{e3}
e^\alpha_3 = \frac{P^\alpha{}_\beta\ell^\beta}{\sqrt{P_{\mu\nu}\ell^\mu\ell^\nu}},
\end{equation} 
where 
\beq\label{Projection}
P_{\alpha\beta}\equiv g_{\alpha\beta}(x_p)+u_{\alpha}u_{\beta}
\eeq
is the operator (defined along $\Gamma$) that projects a vector onto $\Sigma_{\tau}$. We have thereby forced a pursuant rotation of the triad. Despite this rotation, the rest of the coordinate construction is identical to the Fermi construction, with the exception that due to the non-inertial rotation, we now have $g_{\mu\nu}=\eta_{\mu\nu}+O(s)$ and $\Gamma^\alpha_{\beta\gamma}=O(1)$. 

Besides computational convenience, any Fermi-like coordinates have the distinct advantage of allowing us to directly examine the parity of our solutions under the parity transformation defined by Gralla. In terms of our coordinates, we will be interested in the parity of {\em spatial} components, and the parity transformation simply reads $x^a\to-x^a$.

Before proceeding with the calculation of the gauge vector, we introduce a few more pieces of notation. We define the quantity 
\beq
\varrho=\sqrt{\delta_{AB}x^Ax^B},
\eeq
which is the geodesic distance in the direction orthogonal to both $u^\alpha$ and $\ell^a$. We also define the unit vector 
\beq
n^a = x^a/s
\eeq 
and its analogue
\beq
N^A = x^A/\varrho,
\eeq
which satisfy $\delta_{ab}n^an^b=1$ and $\delta_{AB}N^AN^B=1$. These vectors obey the useful rules $\partial_a s = n_a$ and $\partial_A\varrho = N_A$. Further details of Fermi normal coordinates can be found in \cite{Poisson-Pound-Vega:11}, and further details of our Fermi-like coordinates in Appendix~\ref{covariant_expansion}.

\subsection{Local gauge transformation}\label{Fermi-gauge-transformation}

The Lorenz-gauge MP, denoted $h^{\rm Lor}_{\alpha\beta}$, satisfies the gauge conditions
\beq
g^{\beta\gamma}h^{\rm Lor}_{\alpha\beta;\gamma}=\frac{1}{2}g^{\beta\gamma}h^{\rm Lor}_{ \beta\gamma;\alpha},
\eeq
where $g_{\alpha\beta}$ is the background metric, and, as usual, indices are raised and lowered using $g_{\alpha\beta}$. Expressed in our Fermi-like coordinates, $h^{\rm Lor}_{\alpha\beta}$ has the leading-order singular form \cite{Poisson-Pound-Vega:11}
\beq\label{LL_sing}
h^{\rm Lor}_{\alpha\beta} = \frac{2\mu}{s}\delta_{\alpha\beta} + o(s^{-1}).
\eeq

Starting from $h^{\rm Lor}_{\alpha\beta}$, we wish to make a local gauge transformation to a (completed) radiation-gauge MP, assumed to have been computed via a CCK reconstruction and completion. We write the complete perturbation as
\beq
h^{{\rm Rad}'}_{\alpha\beta}=h_{\alpha\beta}^{\rm Rad}+h_{\alpha\beta}^{\rm Cmpl},
\eeq
where $h_{\alpha\beta}^{\rm Rad}$ is the CCK-reconstructed piece of the MP, given in a radiation gauge, and  $h_{\alpha\beta}^{\rm Cmpl}$ is the completion term, most likely to be given in practice in some non-radiation gauge that we will leave unspecified. For the purpose of our local analysis we will assume that the completion piece is given in a gauge regular enough that $h_{\alpha\beta}^{\rm Cmpl}$ has no contribution to the leading-order singular structure of the completed MP. That such a gauge can be chosen for $h_{\alpha\beta}^{\rm Cmpl}$ will be demonstrated via an explicit construction in Sec.\ \ref{reconstruction-completion} (for flat space) and in Ref.\ \cite{Barack-etal:14} (for Kerr).

The reconstructed field $h_{\alpha\beta}^{\rm Rad}$ satisfies the radiation gauge condition
\beq\label{Rad_gauge_cond}
h_{\alpha\beta}^{\rm Rad}\ell^{\beta}=0.
\eeq
The CCK reconstruction returns a field $h_{\alpha\beta}^{\rm Rad}$ that also satisfies the supplementary trace-free condition
\beq\label{trace_free_cond}
h_{\alpha\beta}^{\rm Rad}g^{\alpha\beta}=0,
\eeq
which is known to be consistent with (\ref{Rad_gauge_cond}) in vacuum \cite{Price-Shankar-Whiting:07}.

Let us now consider the $O(\mu)$ gauge transformation
$\xi_{\alpha}=\xi_{\alpha}^{{\rm Rad}'\to {\rm Lor}}$ taking $h_{\alpha\beta}^{{\rm Rad}'}$ to $h^{\rm Lor}_{\alpha\beta}$:\footnote{Logically, we should be considering here the opposite transformation, $\xi_{\alpha}^{{\rm Lor}\to {\rm Rad}'}=-\xi_{\alpha}$. We instead choose to work with  $\xi_{\alpha}^{{\rm Rad}'\to {\rm Lor}}$ for later convenience.}
\beq\label{transformation}
h_{\alpha\beta}^{\rm Lor}=h_{\alpha\beta}^{\rm Rad}+\xi_{\alpha;\beta}+\xi_{\beta;\alpha}+o(s^{-1}).
\eeq
Here the $o(s^{-1})$ terms account for the contribution to $h_{\alpha\beta}^{\rm Rad'}$ from $h_{\alpha\beta}^{\rm Cmpl}$. Contracting both sides with $\ell^\beta$ and using Eqs.\ (\ref{LL_sing}) and (\ref{Rad_gauge_cond}) leads to 
\begin{align}\label{Fermi_eq}
(\xi_{\alpha,\beta}+\xi_{\beta,\alpha})\ell^\beta =\frac{2\mu}{s}\ell_\alpha + o(s^{-1}).
\end{align}
Here we have replaced covariant derivatives with partial ones, making the assumption that the singularity in $\xi_{\alpha,\beta}$ is stronger than that in $\xi_{\alpha}$, so that connection terms are sub-dominant in Eq.\ (\ref{Fermi_eq}). We seek a solution for $\xi_{\alpha}$ that is well behaved as a function of time $\tau$, i.e., whose $\tau$ derivatives do not change the degree of singularity; more precisely, we assume $\partial_\tau\xi_\alpha\sim o(s^{-1})$, such that $\tau$ derivatives can be neglected in Eq.~\eqref{Fermi_eq}. This assumption could be done without, but it is sufficiently mild for us to maintain. 

With our choice of coordinates we have $\ell^\alpha=\hat\ell (\delta^\alpha_\tau+\delta^\alpha_z)+O(s)$, reducing the four components of Eq.~\eqref{Fermi_eq} to
\begin{align}
\partial_z\xi_{\tau} &= \frac{2\mu}{\sqrt{\varrho^2+z^2}}+o(s^{-1}),\label{xi_tau_eq}\\
2\partial_z\xi_z+\partial_z\xi_{\tau} &= \frac{2\mu}{\sqrt{\varrho^2+z^2}}+o(s^{-1}),\label{xi_z_eq}\\
\partial_z\xi_A+\partial_A\xi_{\tau}+\partial_A\xi_z &= o(s^{-1}),\label{xi_A_eq}
\end{align}
where we have divided out the common factor of $\hat\ell$. We have also replaced $s$ with $\sqrt{\varrho^2+z^2}$ to make the dependence on $x^A$ and $z$ more transparent. 
The supplementary condition (\ref{trace_free_cond})
further constrains $\xi_\alpha$ to satisfy 
\beq \label{trace-free-condition0}
2\xi^\alpha_{\ ,\alpha}=g^{\alpha\beta}h^{\rm Lor}_{\alpha\beta}+o(s^{-1}),
\eeq
which, in our Fermi-like coordinates, becomes
\beq
\partial_a\xi^a=\frac{2\mu}{s}+o(s^{-1}).\label{trace-free-condition}
\eeq

Our goal is now to solve Eqs.~\eqref{xi_tau_eq}--\eqref{xi_A_eq} together with Eq.~\eqref{trace-free-condition}.

\subsubsection{General solutions}

One can see by inspection that $\xi_\tau^{\pm}=\pm 2\mu\ln(s\pm z)$ are both solutions to Eq.~\eqref{xi_tau_eq}. To them we can add any function of $\tau$ and $x^A$, leading to the general solutions
\beq
\xi_\tau^{\pm} = \pm 2\mu\ln(s\pm z)+\zeta^{\pm}_\tau(\tau,x^A)+o(1).\label{xi_tau}
\eeq 
Inspection of Eqs.~\eqref{xi_z_eq} and \eqref{xi_A_eq} similarly yields the general solutions
\begin{align}
\xi_z^{\pm} &= \zeta^{\pm}_z(\tau,x^A)+o(1),\label{xi_z}\\
\xi^{\pm}_A &= \frac{2\mu x^A}{s\pm z}-z\partial_A\left[\zeta^{\pm}_\tau(\tau,x^A)+\zeta^\pm_z(\tau,x^A)\right]\nonumber \\
		&\quad+\zeta^{\pm}_A(\tau,x^A)+o(1),\label{xi_A}
\end{align}
where $\zeta^{\pm}_\alpha$ are all arbitrary functions of $\tau$ and $x^A$.

Our solutions must also satisfy the trace-free condition \eqref{trace-free-condition}. A straightforward calculation shows that the condition is satisfied when $\zeta^{\pm}_\alpha\equiv0$; therefore, the condition's only effect is to constrain the functions $\zeta^{\pm}_\alpha(\tau,x^A)$.  Substituting the general solutions \eqref{xi_tau}--\eqref{xi_A} into the trace-free condition reduces it to $\partial^A\xi^{\pm}_A=\frac{2\mu}{s}+o(s^{-1})$, which becomes $z\partial^A\partial_A\left(\zeta^{\pm}_\tau+\zeta^{\pm}_z\right)=\partial^A\zeta^{\pm}_A+o(s^{-1})$. Since the right-hand side is independent of $z$, each side must vanish independently at leading order, implying
\begin{align}
\partial^A\partial_A\left(\zeta_\tau^\pm+\zeta_z^\pm\right) &= o(s^{-2})\quad \text{for }z\neq0,\label{C_z-constraint}\\
\partial^A\zeta_A^\pm &= o(s^{-1}) .\label{C_A-constraint}
\end{align}
In words, at leading order the sum $\zeta^{\pm}_\tau+\zeta_z^\pm$ must be a harmonic function of $x^A$, and $\zeta^{\pm}_A$ must be divergenceless in the 2D flat space charted by $x^A$.

Note that the terms involving $\zeta_{\alpha}^{\pm}$ in the general solutions  \eqref{xi_tau}--\eqref{xi_A} represent homogeneous solutions to the gauge transformation equations \eqref{transformation} and \eqref{trace-free-condition}---i.e., solutions to $\xi_{\alpha;\beta}+\xi_{\beta;\alpha}=0$ and $\xi^{\alpha}_{,\alpha}=0$. They therefore arise from the freedom to perform gauge transformations within the family of radiation gauges. 

The solutions $\xi_{\alpha}^{\pm}$ in Eqs.\ \eqref{xi_tau}--\eqref{xi_A} are completely general. In the following analysis we will show that any particular solution falls into one of three classes, each with its own distinct type of irregularity away from the particle.

\subsubsection{Half-string solutions}
Let us, for the moment, set $\zeta^{\pm}_\alpha\equiv 0$, and consider the singular structure of the resulting particular solutions. These solutions obviously diverge on $\Gamma$ (where $s=0=z$), but they also diverge {\it away} from the particle. Recall $s\pm z=(\varrho^2+z^2)^{1/2}\pm z$, so $s+z$ vanishes on the (``radial'') half-ray $\varrho=0,\,z<0$, while $s-z$ vanishes on the half-ray $\varrho=0,\,z>0$. Hence, $\xi_{\alpha}^+$ is singular on the $z<0$ half-ray, and $\xi_{\alpha}^-$ is singular on the $z>0$ half-ray. More specifically, on the singular half-ray we have, taking the limit $\varrho\to 0$ at fixed $z\ne 0$,
\begin{equation}\label{divergence}
\xi_\tau^{\pm} \sim  \pm 4\mu \ln \varrho ,\quad\quad
\xi_A^{\pm}    \sim  \frac{4\mu |z|N_A}{\varrho} .
\end{equation}
In words, (i) the component of $\xi^{\pm}_{\alpha}$ tangent to $\Gamma$ diverges logarithmically on a half-ray emanating radially from the particle either inward (for $\xi^+_{\alpha}$) or outward (for $\xi^-_{\alpha}$), and (ii) the component of $\xi^{\pm}_{\alpha}$ orthogonal to both $\Gamma$ and $\ell^{\alpha}$ diverges like the inverse distance to the corresponding half-rays (with a directional dependence). 

The above particular solutions $\xi_{\alpha}^{\pm}$ (with $\zeta^{\pm}_\alpha\equiv 0$) are two examples of what we shall call {\em half-string} solutions.

We can use the freedom in our general solutions to switch between the above half-string solutions. Choosing $\zeta^{\pm}_\tau(\tau,x^A)=\mp2\mu\ln\varrho^2$ and $\zeta^{\pm}_z=0=\zeta^{\pm}_A$, we have
\begin{align}
\xi^{\pm}_\tau  &= \pm2\mu\ln\frac{s\pm z}{\varrho^2}+o(1)\nonumber \\
				&= \mp2\mu\ln(s\mp z)+o(1),
\end{align}
and 
\begin{align}
\xi^{\pm}_A &= \frac{2\mu x^A}{s\pm z}\pm 2\mu z\partial_A\ln\varrho^2+o(1)\nonumber\\
				&= \frac{2\mu x^A}{s\mp z}+o(1),
\end{align}
where we have used $\partial_A\varrho = x_A/\varrho$, and in each equation the second equality follows from $\varrho^2=(s+z)(s-z)$. One can easily verify that this choice of $\zeta^{\pm}_\alpha$ satisfies the constraints \eqref{C_z-constraint} and \eqref{C_A-constraint}. 

However, switching between half-string singularities in this way requires $\zeta_\alpha$ to diverge along $x^A=0$. If we restrict $\zeta^{\pm}_\alpha(\tau,x^A)$ to be continuous, then the string singularity is fixed on one side. Furthermore, restricting $\zeta^{\pm}_\alpha(\tau,x^A)$ to be continuous functions of $x^A$ implies $\zeta^{\pm}_\alpha(\tau,x^A)=\zeta^{\pm}_\alpha(\tau,0)+o(1)$, making the term $z\partial_A(\zeta^{\pm}_\tau+\zeta^{\pm}_z)$ of order $s$. We are then left with the half-string solutions of the form
\beq\label{xipm}
\xi^\pm_\alpha=\xi^{0\pm}_\alpha(x^a)+Z^{\pm}_\alpha(\tau)+o(1),
\eeq
where
\begin{align}
\xi^{0\pm}_{\tau} &= \pm 2\mu\ln(s\pm z),\label{xi0_t-half}\\
\xi^{0\pm}_z &= 0,\label{xi0_z-half}\\
\xi^{0\pm}_A &= \frac{2\mu x^A}{s\pm z},\label{xi0_A-half}
\end{align}
and where $Z^\pm_\alpha(\tau)\equiv \zeta^\pm_\alpha(\tau,0)$ . For simplicity, we assume $Z^\pm_\alpha(\tau)$ is smooth. 

Equation \eqref{xipm} defines a family of half-string solutions. A specific member $\xi^+_\alpha$ in this family is smooth for $z>0$ but it diverges on the string $\varrho=0=x^A$ for $z<0$ [in the manner of Eq.\ \eqref{divergence}]. Conversely, a specific member $\xi^-_\alpha$ is smooth for $z<0$ but it diverges on the string $\varrho=0=x^A$ for $z>0$ [again in the manner of Eq.\ \eqref{divergence}].

We note that the half-string solutions $\xi^\pm_\alpha$ of Eq.\ \eqref{xipm} have {\em no definite parity}. To see this, note that under a transformation $x^a\to-x^a$ we have $(z,x^A)\to  (-z,-x^A)$ and $s\to s$.

\subsubsection{Full-string solutions}

Since the half-string fields $\xi_\alpha^+$ and $\xi_\alpha^{-}$ of Eq.\ \eqref{xipm} are each solutions to Eqs.~\eqref{transformation} and \eqref{trace-free-condition0}, any linear combination $n\xi_\alpha^++(1-n)\xi_\alpha^-$, $n\in\mathbb{R}$, is also a solution. For $n\ne 0,1$, such \emph{full-string} solutions are singular on the ray $\varrho=0$, on both sides of the particle. We can write the gauge vector as
$\xi_\alpha^{(n)}=\xi^{0(n)}_\alpha+Z_\alpha(\tau)+o(1)$,
where $Z_\alpha(\tau)$ is arbitrary and $\xi^{0(n)}_\alpha=n\xi_\alpha^{0+}+(1-n)\xi_\alpha^{0-}$. Generically, the divergences on each side of the particle have differing magnitudes, respectively proportional to $n$ and $1-n$. 

As a special case, we can consider weighting the divergences identically by choosing $n=1/2$. This leads to 
the {\em equal-weight full-string} solutions
\beq \label{fullstring}
\xi_\alpha = \xi^0_\alpha(x^a)+Z_\alpha(\tau)+o(1),
\eeq
where
\begin{align}
\xi^0_{\tau} &= \mu\ln\frac{s+z}{s-z},\label{xi0_t-full}\\
\xi^0_z &= 0,\\
\xi^0_A &= \frac{2\mu s x^A}{\varrho^2},\label{xi0_A-full}
\end{align}
and we have defined $Z_\alpha(\tau)=\frac{1}{2}Z^+_\alpha(\tau)+\frac{1}{2}Z^-_\alpha(\tau)$. For simplicity, we again assume $Z_\alpha(\tau)$ is smooth. In these solutions, $\xi_\alpha$ diverges along the entire ray $\varrho=0$, for both $z>0$ and $z<0$: in the limit $\varrho\to 0$ at fixed $z\ne 0$ we have
\begin{equation}\label{divergence_full}
\xi_\tau \sim  - 2\mu\, {\rm sign}(z)\ln \varrho ,\quad\quad
\xi_A    \sim  \frac{2\mu |z|N_A}{\varrho} . 
\end{equation}

Like the half-string solutions, most full-string solutions have no definite parity. The exceptions are the equal-weight full-string solutions, which are parity-regular in the sense of Gralla: $\xi_a$ at leading order is comprised of an odd-parity piece $\xi^0_a(x^b)$ that is discontinuous at $x^b=0$, plus a piece $Z_\alpha$ that is independent of $x^a=0$.

\subsubsection{No-string solutions}\label{no-string-Fermi}

The full-string solutions were found by summing two half-string solutions. But we can also consider combining two half-string solutions in a different way: by gluing together the regular regions of each. 
The surface $\mathcal{S}$ along which we glue them can, in principle, be chosen almost arbitrarily, so long as the two half-strings lie on opposite sides of it. 
We take $\mathcal{S}$ to be smooth, such that at leading order, at each $\tau$ it can be approximated by a plane intersecting the particle. The equation for the plane can be written as $p_a(\tau)n^a=0$ for some $p_a$ perpendicular to it. Section~\ref{reconstruction-completion} will demonstrate that for MP reconstruction in Kerr, the most relevant choice of $\mathcal{S}$ is the sphere of constant Boyer-Lindquist $(t,r)$. Here we leave it arbitrary.

We define the {\em no-string} solution $\xi_\alpha=\xi^+\theta(p_ax^a) +\xi^-\theta(-p_ax^a)$, taking $p_a$ to point toward the regular ($z>0$) side of the ``+'' solution. Explicitly,
\beq\label{nostring}
\xi_\alpha = \xi^0_\alpha(x^a)+Z_\alpha(\tau,x^a)+o(1),
\eeq
where
\begin{align}
\xi^0_{\tau} &= 2\mu\ln(s+z)\theta(p_ax^a)-2\mu\ln(s-z)\theta(-p_ax^a),\label{xi0_t-no}\\
\xi^0_z &= 0,\label{xiz-no}\\
\xi^0_A &= \frac{2\mu x^A}{s+z}\theta(p_ax^a)+\frac{2\mu x^A}{s-z}\theta(-p_ax^a),\label{xi0_A-no}
\end{align}
and
\beq
Z_\alpha = Z^+_\alpha(\tau)\theta(p_ax^a)+Z^-_\alpha(\tau)\theta(-p_ax^a).\label{Z-no}
\eeq
We again assume each $Z^\pm_\alpha$ is a smooth function of $\tau$, but note that there is no requirement that $Z^+_\alpha=Z^-_\alpha$. The no-string solutions, considered as distributions, solve the transformation equations \eqref{xi_tau_eq}--\eqref{xi_A_eq} and the supplementary condition \eqref{trace-free-condition}, even on the surface $p_ax^a=0$, at the relevant order: delta-function terms arising from differentiating \eqref{nostring} are formally sub-leading, and are contained within the $o(s^{-1})$ terms in these equations. 

The no-string solutions thus constructed are smooth for both $p_ax^a>0$ and $p_ax^a<0$, but the divergences have been removed at the cost of introducing a jump discontinuity at $p_ax^a=0$. Although it is not immediately obvious, careful inspection reveals that $\xi_a^0$ has odd parity. Therefore, like the equal-weight full-string solutions and unlike the half-string ones, the no-string solutions are parity-regular. More accurately, they are very nearly, but not quite parity-regular. They come in the correct general form $\xi_a=\xi^0_a(n^i)+Z_a+o(1)$, where $\xi^0_a$ is odd under $n^i\to-n^i$, $\partial_b\xi^0_a\sim 1/s$, and $\partial_bZ_a\sim 1$. But here $Z_a$, rather than having the simple form $Z_a(\tau)$, depends in a discontinuous way on $x^a$. These facts will play an important role in later sections.

\subsection{Singular form of the metric perturbation}\label{metric-summary}

\renewcommand{\arraystretch}{2.5}
\begin{table*}[htb]
\caption{\label{metric} The leading-order singular form of the radiation-gauge metric perturbation near the particle. The half-string solutions in the left column, $h_{\alpha\beta}^{\pm}$, corresponds to $\xi_{\alpha}^{\pm}$ of Eq.\ \eqref{xipm}. The full-string and no-string solutions, middle and right columns, are constructed from the corresponding gauge transformations $\xi_{\alpha}$, given in Eqs.\ \eqref{fullstring} and \eqref{nostring}, respectively. The label Rad$'$ is omitted from the MP for brevity. $\theta^\pm$ denotes $\theta(\pm p_ax^a)$.}
\begin{ruledtabular} \begin{tabular}{ >{\centering\arraybackslash}p{.35\linewidth}  >{\centering\arraybackslash}p{.35\linewidth} >{\centering\arraybackslash}p{.3\linewidth}}
Half-string solutions & Full-string solution & No-string solution \\ \hline
$h^{\pm}_{\tau\tau} = \dfrac{2\mu}{s}$ & $h_{\tau\tau} = \dfrac{2\mu}{s}$ & $h_{\tau\tau} = \dfrac{2\mu}{s}$\\
$h^{\pm}_{\tau z} = -\dfrac{2\mu}{s}$ & $h_{\tau z} = -\dfrac{2\mu}{s}$ & $h_{\tau z} = -\dfrac{2\mu}{s}$\\
$h^{\pm}_{zz} = \dfrac{2\mu}{s}$ & $h_{zz} = \dfrac{2\mu}{s}$ & $h_{zz} = \dfrac{2\mu}{s}$\\
$h^{\pm}_{\tau A} = \mp\dfrac{2\mu x_A}{s(s\pm z)}$ & $h_{\tau A} = \dfrac{2\mu zx_A}{s\varrho^2}$ & $h_{\tau A} = h^+_{\tau A}\theta^++h^-_{\tau A}\theta^-$\\
$h^{\pm}_{zA} = \pm\dfrac{2\mu x_A}{s(s\pm z)}$ & $h_{zA} = -\dfrac{2\mu zx_A}{s\varrho^2}$ & $h_{zA} = h^+_{zA}\theta^++h^-_{zA}\theta^-$\\
$h^{\pm}_{AB} = \dfrac{2\mu}{s(s\pm z)^2}\left(2x_Ax_B-\varrho^2\delta_{AB}\right)$ & $h_{AB} = \dfrac{2\mu (s^2+z^2)}{s\varrho^4}(2x_Ax_B-\varrho^2\delta_{AB})$ & $h_{AB} = h^+_{AB}\theta^++h^-_{AB}\theta^-$\\
\end{tabular}\end{ruledtabular}\end{table*}

We can now determine the form of the local singularity in the completed radiation gauge. A distinct singularity is associated with each of the above classes of gauge transformations. Making $h_{\alpha\beta}^{\rm Rad'}$ the subject of Eq.\ \eqref{transformation} and substituting for $h_{\alpha\beta}^{\rm Lor}$ from Eq.\ \eqref{LL_sing}, we have 
\beq\label{hrad}
h_{\alpha\beta}^{\rm Rad'}=
\frac{2\mu}{s}\delta_{\alpha\beta}-\xi_{\alpha,\beta}-\xi_{\beta,\alpha}+o(s^{-1}).
\eeq
Substituting for $\xi^{\alpha}$ from Eqs.\ \eqref{xipm}, \eqref{fullstring} and \eqref{nostring} (in turn), we obtain expressions for the leading-order terms in, respectively, the half-string, equal-weight full-string, and no-string radiation-gauge MPs. The results are summarized in Table~\ref{metric}.

We see from the table that in the half-string and full-string solutions, the MP inherits the string singularities of the gauge transformation. As we would expect, the divergences here are stronger than they were in the gauge vector. Near the singular strings we have, as $\varrho\to 0$ with fixed $z\ne 0$, 
\begin{align}
h_{\tau A}^{\pm}&\sim \mp \frac{4\mu N_A}{\varrho}, \quad\quad
h_{z A}^{\pm}\sim \pm \frac{4\mu N_A}{\varrho}, \nonumber\\
h_{AB}^{\pm}&\sim \frac{8\mu |z| (2N_A N_B-\delta_{AB})}{\varrho^2}
\end{align}
for the half-string solutions, and 
\begin{align}
h_{\tau A}^{\pm}&\sim \frac{2\mu\,{\rm sign(z)}N_A}{\varrho}, \quad\quad
h_{z A}^{\pm}\sim - \frac{2\mu\,{\rm sign(z)}N_A}{\varrho}, \nonumber\\
h_{AB}^{\pm}&\sim \frac{4\mu |z| (2N_A N_B-\delta_{AB})}{\varrho^2}
\end{align}
for the full-string solutions. In words, $h_{\tau A}$ and $h_{z A}$ diverge towards the singular string as $1/\varrho$ (with a directional dependence), and the  components $h_{AB}$ diverge even faster: as $1/\varrho^2$ (also with a directional dependence). Note that the MP fails to be absolutely integrable over a two-dimensional surface intersecting the string, meaning a harmonic expansion of the MP is not straightforwardly defined. This calls into question whether the singular portion of the perturbed spacetime can actually be recovered from a mode-by-mode reconstruction procedure {\it a la} CCK. We discuss this issue further in the following subsection and explore it in detail in Sec.~\ref{reconstruction-completion}.

In the {\it no-string} solution, $h_{\tau A}$ and $h_{zA}$ inherit a discontinuity across $p_ax^a=0$ from the gauge transformation. Generically, these jump discontinuities will occur not just at leading order, but at all orders in $s$; this will be an important point when formulating our practical methods of calculating the GSF. Beyond leading order, we can also expect the no-string solution to inherit $\delta(p_ax^a)$ terms from differentiation of Heaviside step functions in the gauge vector. Such terms indeed manifest themselves in $h_{\alpha\beta}^{\rm Rad'}$ in the flat-space example we work out explicitly in Sec.~\ref{reconstruction-completion}, and it seems reasonable to conjecture that delta distributions are present on the surface of discontinuity also in more general cases. 

Our analysis of the general solutions \eqref{xi_tau}--\eqref{xi_A} for the gauge vector makes it clear that there exist no solutions more regular than those we have presented. Because the free functions $\zeta_\alpha(\tau,x^A)$ in the general solution cannot depend on $z$, while the strings and discontinuities do depend on $z$, the freedom in the solution cannot be used to remove the irregularities, but only to move between them. We thus conclude that the three classes of solutions displayed in Table I are the ``most regular'' MPs possible in the radiation gauge.

In each of the three classes of solutions, the MP inherits its parity from the corresponding gauge vector. Specifically, the half-string MP has no definite parity, while both the equal-weight full-string and no-string MPs are parity-regular: the leading-order pieces of the spatial components $h_{ab}$ are invariant under $x^a\to-x^a$. 

Before we proceed, let us comment on the validity of the MPs we have obtained. A wary reader might fear that the half- and full-string MPs are incomplete, even at leading order: due to the divergences in the gauge transformation, these MPs could fail to satisfy the point-particle linearized EFE on the strings. This concern can be done away with by noting that (a) $h^{\rm Lor}_{\alpha\beta}-2\xi_{(\alpha;\beta)}$ and $h^{\rm Rad'}_{\alpha\beta}$ are identical distributions, and (b) the linearized Einstein tensor constructed from any distribution of the form $\xi_{(\alpha;\beta)}$ vanishes as a distribution. Therefore, since $h^{\rm Lor}_{\alpha\beta}$ is a solution to the point-particle EFE, $h^{\rm Rad'}_{\alpha\beta}$ is as well. (The fact that $\xi_{(\alpha;\beta)}$ is well defined as a distribution follows from the fact that $\xi_\alpha$ is locally integrable.)

Similarly, the reader might be concerned that a no-string MP cannot be a solution to the EFE beyond leading order. After all, if one takes two regular metrics and glues them together at a surface in a discontinuous way, the resulting metric will contain a distributional source on the surface. However, this is \emph{not} what we imagine the (completed) no-string MP to be. Instead of joining the regular halves of two half-string MPs, we join the two half-string gauge vectors, $\xi^+_\alpha$ and $\xi^-_\alpha$. The resulting MP $h^{\rm Rad'}_{\alpha\beta}=h^{\rm Lor}_{\alpha\beta}-2\xi_{(\alpha;\beta)}$, with discontinuous $\xi_\alpha$, is guaranteed to be a solution to the point-particle EFE.

\subsection{Local gauge transformation in arbitrary coordinates}
Thus far our analysis has been restricted to our set of Fermi-like coordinates. For practical purposes, including devising mode-sum formulas, we shall require the gauge vector in some set of global coordinates useful for numerical implementation, such as Boyer-Lindquist coordinates in a Kerr background. Besides enabling explicit calculations in later sections, expressions in global coordinates provide a direct view of the way in which the parity of the Fermi-like components is preserved in an arbitrary coordinate system.

Here we present the gauge vector in some arbitrary coordinates $x^\alpha$. For each point $x$, we define a reference point $x^{\alpha'}$ on $\Gamma$, and we seek an expansion of the components of $\xi_\alpha$ in the limit of small coordinate distances $\d x^{\alpha'}=x^\alpha-x^{\alpha'}$. We transform our results from $(\tau, x^a)$ to $x^\alpha$ in two steps: first, we convert our expansion of $\{\xi_\tau,\xi_A,\xi_z\}$ in terms of $x^a$ into a single tensorial expansion of $\xi_\alpha$ in terms of the covariant directed distance $\sigma^{\bar\alpha}$; second, we convert that covariant expansion into a coordinate expansion of the components of $\xi_\alpha$.

To avoid delaying the development of our core methods in subsequent sections, in this section we merely present,  without proof, the end results of the coordinate expansion. The complete calculation can be found in Appendix~\ref{trans_to_global}. We write the final expression \eqref{xi_final} in the now familiar form 
\beq
\xi_\alpha=\xi^0_\alpha+Z_{\alpha'}+o(1).\label{xi-coords-final}
\eeq
As in Fermi-like coordinates, the quantity $\xi^0_\alpha$ represents the leading-order singular term in the expansion of $\xi_\alpha$, and it splits conveniently into two pieces:
\beq
\xi^0_\alpha = \xi_{\alpha\parallel} +\xi_{\alpha\perp}.\label{xi0-coords}
\eeq
The first piece, 
\beq
\xi_{\alpha\parallel}=-\xi^0_\tau u_{\alpha'},
\eeq
is parallel to $u^{\alpha'}$, and the second piece,
\beq
\xi_{\alpha\perp}=\xi Q_{\alpha'\beta'}\delta x^{\beta'}\label{xi_perp},
\eeq
is orthogonal to both $u^{\alpha'}$ and $\ell^{\alpha'}$. In these expressions, a primed index denotes a component evaluated at the reference point $x'$ on $\Gamma$. $Q_{\alpha\beta}$ is a projection operator given by
\beq
Q_{\alpha\beta}\equiv P_{\alpha\beta}-\frac{P_{\alpha\mu}P_{\beta\nu}\ell^{\mu}\ell^{\nu}}{(\ell^{\gamma}u_{\gamma})^2}.\label{Q-body}
\eeq
The scalars $\xi^{0}_\tau$ and $\xi$ are 
\beq
\xi^{0\pm}_\tau = \pm 2\mu\ln(s_0\pm z_0), \qquad \xi^\pm = \frac{2\mu}{s_0\pm z_0}\label{xi0xi-hs}
\eeq
in the half-string case,
\beq
\xi^0_\tau = \mu\ln\left(\frac{s_0+z_0}{s_0-z_0}\right), \qquad \xi = \frac{2\mu s_0}{s_0^2-z_0^2}
\eeq
in the full-string case, and 
\beq
\xi^0_\tau = \xi^{0+}\theta^++\xi^{0-}\theta^-, \qquad \xi = \xi^+\theta^++\xi^-\theta^-
\eeq
in the no-string case, where $s_0$ and $z_0$ are the leading-order terms in the coordinate expansion of $s$ and $z$,
\begin{align}
s_0^2 &\equiv P_{\alpha'\beta'}\delta x^{\alpha'}\delta x^{\beta'},\label{s0-body}\\
z_0 &\equiv-u_{\alpha'}\delta x^{\alpha'}-\frac{\ell_{\alpha'}\delta x^{\alpha'}}{\ell_{\beta'}u^{\beta'}},\label{z0-body}
\end{align}
and $\theta^\pm$ is a step function equal to unity on the side of $\mathcal{S}$ on which $\xi_\alpha^\pm$ is regular.

We observe that the coordinate transformation has preserved the parity of the vectors in a particular, useful sense: at leading order, the components of $\xi_{\alpha\perp}$ have that same parity under the transformation $\delta x^{\alpha'}\to-\delta x^{\alpha'}$ as did the components $\xi_a$ under $x^a\to-x^a$, regardless of the choice of coordinates $x^\alpha$. This fact will be essential for the calculations in later sections.

In all of the above expressions, the reference point $x'$ on $\Gamma$ is arbitrary. In Sec.~\ref{LL-method}, we choose it to be the point on $\Gamma$ with the same Boyer-Lindquist coordinate time $t$ as the point $x$, such that $\delta x^{t'}=t-t'=0$.

\subsection{Preliminary lessons for the GSF problem}

Our main interest here is in using radiation-gauge solutions to construct the GSF. A crucial question is whether the solutions constructed above fall within any of the general gauge categories discussed in Sec.\ \ref{gauges}, for which the formulation of the GSF is understood. Clearly, none of the radiation gauges belongs to the Barack-Ori category, because the general $\xi_{\alpha}$ solution contains a singular limit to the particle for any choice of the free functions. The Gralla-Wald category allows for discontinuities at the particle, but it does not allow for a logarithmic divergence, nor does it allow for irregularities away from the particle---hence, all radiation gauges fall outside the Gralla-Wald category too. For the same reason they also fall outside Gralla's class of gauges, with half-string solutions and most full-string solutions also failing to satisfy parity-regularity. 

We see that the notion of a ``GSF exerted by a radiation-gauge MP'' is not meaningful in any simple way within the GSF formulations in existing literature. To construct the GSF from a radiation-gauge MP in any of the permissible gauges---Barack-Ori, Gralla, or Gralla-Wald---requires a suitable local gauge deformation of the input MP. This, indeed, will be the strategy we will pursue in Sec.\ \ref{LL-definitions}, focusing on half-string gauges. 

On the other hand, all radiation gauges constructed above fall into the much wider class of ``sufficiently regular'' gauges defined in Sec.\ \ref{gauges}. In all cases (half-string, full-string, no-string) the gauge vector satisfies $\partial_\alpha\xi_{\beta}=O(s^{-1})$, and it yields a well defined, finite result for $\Delta z_1^{a}$ when used in Eq.\ \eqref{Delta z} (note that $\Delta z_1^{a}$ depends only on the spatial components of $\xi_{\alpha}$, and it is thus insensitive to the logarithmic singularity in $\xi_{\tau}$). Based on this fact, one should be able to formulate the GSF in any of the radiation gauges, without local deformation. We will pursue this option in Sec.\ \ref{force-no-string}, focusing on no-string gauges.

Before we move on to discuss the GSF, though, we must address a second crucial question. For a GSF scheme to be useful, it must use as input the actual output from a CCK-type mode-by-mode reconstruction procedure. But which of the radiation gauges discussed above does the CCK procedure actually pick out? 

In addressing this question, let us first remind the reader that up until now we have been considering the MP behavior in the immediate neighbourhood of the point particle. In particular, the string singularities and surface discontinuity we have identified are {\em local} features of the MP. It seems, however, reasonable to expect these features to extend beyond the local neighbourhood of the particle. We expect the string singularity in half-string and full-string solutions to extend to infinity (or down through the event horizon where relevant), and we expect the discontinuous surface of the no-string solution to either close on itself or extend to infinity. We shall test and confirm this expectation with an explicit flat-space example in Sec.~\ref{reconstruction-completion}. 

Next, we recall our conclusion that the radiation-gauge MP fails to be absolutely integrable over a surface crossing a string. This suggests that the MP does not possess an expansion in harmonics on such a surface; expressed as integrals of the MP against the harmonics, the coefficients in the expansion do not exist. Therefore, it seems to us unlikely that a mode-by-mode CCK reconstruction can recover the string singularity of the half-string and full-string solutions. Rather, the reconstruction scheme will recover the regular side of a half-string solution (or, equivalently, the no-string solution to one side of the discontinuous surface). Which of the two regular sides is recovered will depend on whether one computes the Hertz potential by integrating from one asymptotic domain or the other. We will confirm these expectations with our explicit flat-space example of Sec.~\ref{reconstruction-completion}, and we conjecture here that they apply more generally, for generic orbits in Kerr spacetime.


\section{Self-force in a locally deformed radiation gauge: the half-string case}\label{LL-method}

In this section, we describe our formalism of locally deformed radiation gauges, in which we slightly alter the gauge Rad$'$ (i.e., the completed radiation gauge) to make it LL. By design, this slightly altered gauge will fall within the Barack-Ori class of gauges, and will thus be amenable to the standard mode-sum method of calculating the GSF, with the standard Lorenz-gauge regularization parameters. We begin by describing the deformation to an LL gauge, and we then describe the formulation of a practical mode-sum formula in that gauge. Throughout this section we specialize the discussion to the case of a Kerr background (although many of our intermediate results will apply more generally). The formulation of the mode-sum scheme will also involve specifying a coordinate system, which we take to be Boyer-Lindquist coordinates, denoted $(t,r,\theta,\varphi)$. For the most part we shall continue to use $\ell^{\alpha}$ to denote either one of the two principal null vectors of $g_{\alpha\beta}$; the form of all expressions will be insensitive to our particular choice. We specialize to the ingoing radiation gauge only when calculating explicit regularization parameters.

\subsection{Locally Lorenz gauges}\label{LL-definitions}

To define what we mean by an LL gauge, we first recall the form of the globally Lorenz MP near the particle, given in our Fermi-like coordinates in Eq.\ \eqref{LL_sing}. In any coordinates, the expression reads \cite{Poisson-Pound-Vega:11} 
\beq\label{Lor-singularity}
h_{\alpha\beta}^{\rm Lor}=\frac{2\mu}{s} (g_{\alpha\beta}+2 \tilde u_{\alpha}\tilde u_{\beta}) +O(1).
\eeq
Here $s$ is the geodesic distance to $\Gamma$, and $\tilde u_{\alpha}$ can be any smooth extension of the four-velocity $u_\alpha$ off $\Gamma$. The terms $O(1)$ are finite but not necessarily continuous on $\Gamma$. By an LL gauge, we mean any gauge in which the MP possesses the same leading-order singularity structure as $h_{\alpha\beta}^{\rm Lor}$; that is,
\beq\label{LL-singularity}
h_{\alpha\beta}^{\rm LL}=\frac{2\mu}{s} (g_{\alpha\beta}+2 \tilde u_{\alpha}\tilde u_{\beta}) +o(s^{-1}).
\eeq
The terms $o(s^{-1})$ may diverge at the particle, but not as strongly as does the leading-order singularity. In particular, we shall need to allow logarithmic divergences, which potentially arise in the radiation gauge at sub-leading order, as our analysis in the previous section suggests.

Our goal is to locally transform $h^{{\rm Rad}'}_{\alpha\beta}$ to some $h_{\alpha\beta}^{\rm LL}$. We could instead locally deform $h^{{\rm Rad}'}_{\alpha\beta}$ to give it the singularity structure of any gauge within Gralla's class, which would equally well allow us to use the mode-sum formula in its standard form. The advantage of working specifically in an LL gauge, besides the familiarity of its singularity structure, is that we have already established the leading-order gauge transformation relating the completed radiation gauges to the Lorenz gauge. Because $h^{\rm LL}_{\alpha\beta}$ is identical to $h^{\rm Lor}_{\alpha\beta}$ at leading order, the gauge transformation $\xi_{\alpha}=\xi_{\alpha}^{{\rm Rad}'\to {\rm LL}}$ must satisfy the same equations as did $\xi_{\alpha}=\xi_{\alpha}^{{\rm Rad}'\to {\rm Lor}}$. Those equations, \eqref{Fermi_eq} and \eqref{trace-free-condition}, in arbitrary coordinates read
\beq \label{gaugetransformation2}
\ell^{\beta}(\xi_{\alpha;\beta}+\xi_{\beta;\alpha})=\frac{2\mu}{s}(\ell_\alpha+2\tilde u_\alpha \tilde u_\beta \ell^\beta)+o(s^{-1})
\eeq
and
\beq \label{gaugetransformation3}
\xi^\alpha{}_{;\alpha}=\frac{2\mu}{s}+o(s^{-1}).
\eeq

Finding an LL gauge is simply a matter of solving Eqs.~\eqref{gaugetransformation2} and \eqref{gaugetransformation3} for $\xi^{\alpha}$. If we start from an MP $h^{{\rm Rad}'}_{\alpha\beta}$ in any of the three classes of radiation gauges, then the corresponding half-, full-, and no-string gauge vectors $\xi_\alpha=\xi^0_\alpha+Z_\alpha+o(1)$ found in Sec.~\ref{Fermi-analysis} will transform the MP to an LL gauge. In Sec.~\ref{Fermi-analysis}, the terms $Z_\alpha+o(1)$ in the transformation are, in principle, uniquely determined by the subleading behavior of the particular Lorenz gauge and completed radiation gauge that are being related. In the present context, however, the terms $Z_\alpha+o(1)$ may be chosen arbitrarily: $\xi^0_\alpha$ is fixed by the leading-order singularities in $h^{{\rm Rad}'}_{\alpha\beta}$ and $h^{\rm LL}_{\alpha\beta}$, but the LL gauge is completely unspecified beyond leading order. Different choices of $Z_\alpha$, and of the subleading terms of $o(1)$ in the transformation, correspond to different choices of LL gauge. 

In our analysis, we take the stance that since the GSF is gauge-dependent, we must clearly specify the gauge in which we calculate it. A numerical reconstruction and completion procedure will yield an MP in a particular gauge ${\rm Rad}'$. From this starting point, we choose a \emph{specific} gauge vector that brings the MP to a corresponding LL gauge, thereby locally identifying our choice of LL gauge. To be precise, suppose we wish to calculate the GSF in Boyer-Lindquist coordinates. We take the gauge vector to be 
\beq\label{LL-choice}
\xi^{\rm Rad'\to LL}_\alpha=\xi^{0}_\alpha
\eeq
with $\xi^{0}_\alpha$ given, \emph{in Boyer-Lindquist coordinates}, by Eq.~\eqref{xi0-coords}. $\xi^{0}_\alpha$, as given in Eq.~\eqref{xi0-coords}, depends on the choice of reference point $x'$ about which $\xi_\alpha$ is expanded. We identify $x'(x)$ as the point on $\Gamma$ with the same Boyer-Lindquist time as $x$: $x'=x_p(\tau(t))$, where $\tau(t)$ is the proper time on $\Gamma$ at coordinate time $t$. Explicitly,
\beq
 x^{\alpha'}(t)=(t,r_p(t),\th_p(t),\varphi_p(t)), 
\eeq 
and
\beq
\delta x^{\alpha'}=(0,r-r_p(t),\th-\th_p(t),\vf-\vf_p(t)).
\eeq
Equation~\eqref{LL-choice} then uniquely specifies an LL counterpart to each given completed-radiation-gauge MP. In Sec.~\ref{alt-gauge} we will discuss the effect of making alternative choices.

The gauge vector is ``specific'' in terms of its local behavior only; we have additional freedom in how we extend it away from the local neighborhood of the particle. To deal with this freedom, we take our specific choice $\xi_\alpha=\xi^{0}_\alpha$ everywhere in a neighborhood $\mathcal{N}_1$ of the particle, and we then let $\xi_\alpha$ go smoothly to zero at the boundary of a slightly larger region $\mathcal{N}_2\supsetneq\mathcal{N}_1$, leaving us with a uniquely specified LL gauge in $\mathcal{N}_1$ and the numerically determined gauge Rad$'$ everywhere outside $\mathcal{N}_2$. For GSF purposes, this degree of specificity is sufficient: the GSF is calculated in a specific LL gauge, and gravitational waves, for example, are calculated in the specific completed radiation gauge. We will refer to the gauge as an ``LL half-string", ``LL full-string", or ``LL no-string" gauge, as appropriate. 

\subsection{Mode-sum formula for the GSF in an LL gauge}
Recall that our goal is to construct a gauge within the Barack-Ori class, meaning that the generator $\hat\xi_\alpha\equiv\xi_{\alpha}^{\rm Lor\to LL}$ of the gauge transformation from $h_{\alpha\beta}^{\rm Lor}$ to $h_{\alpha\beta}^{\rm LL}$ must be continuous. From the local singularity  structures of Eqs.~\eqref{Lor-singularity} and \eqref{LL-singularity}, it follows that the generator satisfies 
\beq \label{gtrans}
\hat\xi_{\alpha;\beta}+\hat\xi_{\beta;\alpha}=o(s^{-1})
\eeq
near $\Gamma$. This alone is not sufficient to ensure that $\hat\xi_\alpha$ is a transformation within the Barack-Ori class; it does not rule out, for example, jump discontinuities. Therefore, when constructing our LL gauges, we shall require more of $\hat\xi_{\alpha}$, demanding that it be continuous.
 
With this demand satisfied, the LL gauge falls within the Barack-Ori class, and the GSF is given by the same mode-sum formula~\eqref{mode-sum_form} as in the Lorenz gauge, with the same parameter values. We write the formula here with more specificity as
\beq \label{MS1}
F_{\alpha}^{\rm LL}=\sum_{\ell=0}^{\infty}\left[
(\tilde F^{\rm LL}_{\alpha})^\ell_{\pm}-A_{\alpha}^{\pm}L-B_{\alpha}-C_{\alpha}/L\right] - D_{\alpha}.
\eeq
Here $L\equiv\ell+1/2$ and the parameters $A_\alpha^{\pm}$, $B_\alpha$, $C_\alpha$, and $D_\alpha$ (which are $\ell$-independent but depend on the position and velocity of the particle at the point where the GSF is evaluated) take their Lorenz-gauge values \cite{Barack-Ori:03b,Barack:09}. The meaning of the label $\pm$ will be explained below. The quantities $(\tilde F^{\rm LL}_{\alpha})^\ell_{\pm}$ are the multipole modes of the full force in the LL gauge, evaluated at the particle limit; their exact definition is as follows. 

For an arbitrary MP $h_{\alpha\beta}$, we define the {\it full force} to be the vector field
\beq\label{full_force}
\tilde F_{\alpha} \equiv -\frac{1}{2}\mu\tilde P_{\alpha}{}^\beta(2\tilde\nabla_{\!\delta}h_{\beta\gamma}-\tilde \nabla_{\!\beta}h_{\gamma\delta})\tilde u^\gamma \tilde u^\delta,
\eeq
which is a generalization of the full force defined in Eq.~\eqref{full_force1}. Here we have introduced not only an extension $\tilde u^\alpha$ of the four-velocity off $\Gamma$, but also extensions $\tilde P_{\alpha}{}^\beta$ and $\tilde\nabla_{\!\alpha}$ of the projection operator $P_{\alpha}{}^\beta\equiv \delta_\alpha^\beta+u_\alpha u^\beta$ and covariant derivative off the worldline. [The extension of $u^\alpha$ here need not be related to that of $u_\alpha$ in Eq.~\eqref{Lor-singularity}.] Extending $u^\alpha$ and $P_\alpha{}^\beta$ is necessary in order to define the full force as a field. Extending the covariant derivative, allowing it to differ from the derivative compatible with $g_{\alpha\beta}$ off $\Gamma$, is optional but sometimes useful. The LL-gauge full force $\tilde F^{\rm LL}_{\alpha}$ is the one associated with the LL-gauge MP $h_{\alpha\beta}^{\rm LL}$. Prior to calculating explicit quantities, we will leave the choice of extension arbitrary.

Given $\tilde F^{\rm LL}_{\alpha}$, the $\ell$ modes $(\tilde F^{\rm LL}_{\alpha})^\ell_{\pm}$ are constructed by expanding each coordinate component of this field (artificially considered as a scalar field) in spherical-harmonic functions on a surface of constant Boyer-Lindquist time $t$ and radius $r$, then adding up all azimuthal numbers $m$ for given multipole number $\ell$, and finally evaluating the result at the particle's limit.  This limit will generally be direction-dependent, and one 
must ensure that it is taken from the same direction as was used to derive the regularization parameters. In the mode-sum formula \eqref{MS1} the limit is taken from one of the radial directions, $r\to r_p^{\pm}$, holding $t,\theta,\varphi$ fixed. $(\tilde F^{\rm LL}_{\alpha})^\ell_{\pm}$ and $A_{\alpha}^{\pm}$ denote the corresponding one-sided values (the values of the parameters $B_{\alpha}$, $C_{\alpha}$ and $D_{\alpha}$ turn out not to depend on the direction).

Let us continue with our formulation. The essential step now is to rewrite Eq.~\eqref{MS1} in terms of the modes of the full force in the Rad$'$ gauge, $(\tilde F^{{\rm Rad}'}_{\alpha})^\ell$. Without this step, Eq.~\eqref{MS1} has no utility: in a numerical implementation using metric reconstruction and completion, we calculate the modes of the MP in the Rad$'$ gauge, not in the LL gauge. To determine the difference between the modes of the full force in the two gauges, we treat $h_{\alpha\beta}$ as a perturbation of a background metric $\tilde g_{\alpha\beta}$ that is compatible with the derivative $\tilde\nabla_{\!\alpha}$. Under a gauge transformation generated by a vector $\xi_\alpha$, we then have $h_{\alpha\beta}\to h_{\alpha\beta}+2\tilde\nabla_{\!(\alpha}\xi_{\beta)}$, and so $\tilde F_{\alpha}\to \tilde F_{\alpha}+\d_{\xi} \tilde F_{\alpha}$, with 
\beq \label{dxiF}
\d_{\xi} \tilde F_{\alpha}=-\mu\left[
\tilde P_{\alpha}{}^{\lambda}\tilde u^\mu \tilde u^\nu \tilde\nabla_{\!\mu}\tilde\nabla_{\!\nu}\xi_{\lambda}+\tilde R_{\alpha \mu\lambda\nu}\tilde u^{\mu} \xi^{\lambda} \tilde u^{\nu}
\right].
\eeq
This result is easily derived by substituting $h_{\alpha\beta}+2\tilde\nabla_{\!(\alpha}\xi_{\beta)}$ into Eq.~\eqref{full_force} and utilizing the Ricci identity, noting that here $\tilde R_{\alpha \mu\lambda\nu}$ is the Riemann tensor defined by the commutator $[\tilde\nabla_{\!\alpha},\tilde\nabla_{\!\beta}]$. In the special case of a continuous transformation, Eq.~\eqref{dxiF} can be evaluated directly on $\Gamma$, where it reduces to
\beq \label{dxiF-continuous}
\d_{\xi} \tilde F_{\alpha}=-\mu\left[
P_{\alpha}{}^{\lambda}\frac{D^2\xi_{\lambda}}{d\tau^2}+R_{\alpha \mu\lambda\nu}u^{\mu} \xi^{\lambda} u^{\nu}
\right].
\eeq
Equation~\eqref{dxiF-continuous} is the formula one obtains by considering how the acceleration of a worldline changes under an infinitesimal translation~\cite{Barack-Ori:01}. We refer the reader to Appendix~\ref{dF-arbitrary} for a summary of useful properties and local expansions of $\d_\xi\tilde F_{\alpha}$, which we shall appeal to as necessary.

Let $\d_{\xi} \tilde F^{\rm Rad'\to LL}_{\alpha}$ be the change in the full force induced by transforming to the LL gauge, and denote its $\ell$ modes by $(\d_{\xi} \tilde F^{{\rm Rad}'\to {\rm LL}}_{\alpha})_{\pm}^\ell$, where we allow for a directional dependence corresponding to $r\to r_p^{\pm}$. We can rewrite Eq.~\eqref{MS1} as
\begin{eqnarray} \label{MS2}
F_{\alpha}^{\rm LL}=\sum_{\ell=0}^{\infty}\left[(\tilde F^{{\rm Rad}'}_{\alpha})_{\pm}^\ell+(\d_{\xi} \tilde F^{\rm Rad'\to LL}_{\alpha})_{\pm}^\ell
\right. \nonumber\\
\left.
-A^{\pm}_{\alpha}L-B_{\alpha}-C_{\alpha}/L 
\right] -D_{\alpha},
\end{eqnarray}
where both $(\tilde F^{{\rm Rad}'}_{\alpha})_{\pm}^\ell$ and $(\d_{\xi} \tilde F^{\rm Rad'\to LL}_{\alpha})_{\pm}^\ell$ must be calculated via the same directional limit to the particle as were the regularization parameters, and all terms must be defined with the same extension of $u^\alpha$, $P_\alpha{}^\beta$, and $\nabla_\alpha$. 

We assume, tentatively, that $(\d_{\xi} \tilde F^{{\rm Rad}'\to {\rm LL}}_{\alpha})_{\pm}^\ell$ admits a large-$\ell$ asymptotic expansion of a form similar to that of $(\tilde F_{\alpha}^{\rm Lor})^{\ell}_{\pm}$, namely
\beq\label{asym}
(\d_{\xi} \tilde F^{{\rm Rad}'\to {\rm LL}}_{\alpha})_{\pm}^\ell= \delta A_{\alpha}^{\pm} L+\d B^\pm_{\alpha} +\d C^\pm_{\alpha}/L +O(1/L^2),
\eeq
where $\delta A^{\pm}_{\alpha}$, $\d B^\pm_{\alpha}$ and $\d C^\pm_{\alpha}$ are $\ell$-independent parameters [we will verify this form with an explicit calculation in Sec.~\ref{RP-general}, showing that the parameter values are in fact zero through $O(1/L)$]. With this assumption, Eq.~(\ref{MS2}) becomes 
\begin{eqnarray} \label{MS3}
F_{\alpha}^{\rm LL}=\sum_{\ell=0}^{\infty}\left[(\tilde F^{{\rm Rad}'}_{\alpha})_{\pm}^\ell-(A^{\pm}_{\alpha}-\d A^{\pm}_{\alpha})L-(B_{\alpha}-\d B^\pm_{\alpha})
\right.
\nonumber\\
\left.
-(C_{\alpha}-\d C^\pm_{\alpha})/L 
\right] -(D_{\alpha}-\d D^\pm_{\alpha}),\nonumber\\
\end{eqnarray}
where 
\beq\label{dD}
\d D^\pm_{\alpha}\equiv\sum_{\ell=0}^{\infty} \left[(\d_{\xi} \tilde F^{\rm Rad'\to LL}_{\alpha})_{\pm}^\ell-\delta A^{\pm}_{\alpha} L-\d B^\pm_{\alpha} -\d C^\pm_{\alpha}/L \right].
\eeq
Since the argument in the last sum is $O(L^{-2})$ at large $\ell$, the sum should be convergent. And since we started with a convergent sum in Eq.\ (\ref{MS2}), the sum in Eq. (\ref{MS3}) should therefore also be convergent. 

Equation (\ref{MS3}) is a mode-sum formula for the GSF in an LL gauge. It requires as input three ingredients (all of which must be given with the same extension): (i) the modes $(\tilde F^{{\rm Rad}'}_{\alpha})^\ell$, which are to be derived from the MP obtained numerically via CCK reconstruction and completion; (ii) the standard, Lorenz-gauge regularization parameters $\{A^{\pm}_{\alpha}, B_{\alpha},  C_{\alpha}, D_{\alpha}\}$, given in Refs.~\cite{Barack-Ori:03b,Barack:09} for generic orbits in Kerr and for various choices of extension; and (iii) the new parameters $\{\delta A^{\pm}_{\alpha},\d B^{\pm}_{\alpha},\d C^{\pm}_{\alpha},\d D^{\pm}_{\alpha}\}$ associated with the particular LL gauge chosen. The latter can be obtained analytically via a local analysis, as we demonstrate in Sec.~\ref{RP-general}. For any admissible choice of LL gauge and for any extension, we find that $\{\delta A^{\pm}_{\alpha},\d B^\pm_{\alpha},\d C^\pm_{\alpha}\}$ all vanish. However, there is generically a nonzero $\d D^\pm_{\alpha}$ correction that must be included in the mode-sum formula. 

But before proceeding to the calculation of the parameters, there remains an important issue to clarify: Which of the radiation-gauge types discussed in the previous section (half-string, full-string, no string) are suitable as input for the mode-sum formula (\ref{MS3})? As we argued above, the CCK reconstruction probably cannot be used to compute the full-string MP, making this class of solutions  irrelevant in practice. The full-force modes $(\tilde F^{{\rm Rad}'}_{\alpha})_{\pm}^\ell$ could be derived from either ``half'' of a no-string MP, by taking the corresponding limits $r\to r^{\pm}$. However, deforming a no-string gauge to a gauge within the Barack-Ori class would be highly nontrivial. Imagine beginning in a no-string gauge Rad$'$. It contains jump discontinuities across a surface $\mathcal{S}$ intersecting the particle, and these discontinuities occur both at leading and subleading orders in $h^{\rm Rad'}_{\alpha\beta}$. The transformation generated by $\xi^0_\alpha$ will remove the discontinuity at $O(s^{-1})$ in $h^{\rm Rad'}_{\alpha\beta}$, but removing the discontinuity at the next order would require including a very precise choice of discontinuous vector $Z^\pm_\alpha(\tau,x^a)$ in the gauge transformation, of the form displayed in \eqref{nostring}. Making that choice would require more complete knowledge of the gauges imposed on $h^{\rm Rad}_{\alpha\beta}$ and $h^{\rm Cmpl}_{\alpha\beta}$ in any particular numerical calculation. If the appropriate $Z^\pm_\alpha(\tau,x^a)$ is not known, then nonremovable discontinuities will remain at the particle, the LL gauge will not be within the Barack-Ori class, and the mode-sum formula \eqref{MS1} can not be expected to apply. Therefore, we conclude that like the full-string case, the no-string case is not relevant to our LL formalism.

Rather, the full-force modes $(\tilde F^{{\rm Rad}'}_{\alpha})_{\pm}^\ell$ should be derived from a half-string MP, with the limit $r\to r^{\pm}$ taken from the side opposite the string. Gauge vectors $\xi_{\alpha}^{\rm Lor\to LL}$ associated with half-string solutions {\em are} continuous, because the corresponding vector $\xi_{\alpha}^{\rm Rad'\to LL}=\xi^0_\alpha$ accounts explicitly for the full discontinuity in $h_{\alpha\beta}^{\rm Rad'}$ at the relevant order. Hence, an LL gauge derived from a half-string radiation gauge belongs to the Barack-Ori class, as required. A CCK reconstruction (and completion) presumably yields only the ``regular half'' of a half-string solution, so fixing the string direction (by fixing the half-string gauge) dictates the direction from which the limit $r\to r_p^{\pm}$ should be taken in computing $(\tilde F^{{\rm Rad}'}_{\alpha})_{\pm}^\ell$  and $A^{\pm}_{\alpha}$ in Eq.\ \eqref{MS3}: for a string extending over $r>r_p$ take $r\to r_p^{-}$; for a string extending over $r<r_p$ take $r\to r_p^{+}$. The sub- and superscripts ``$\pm$'' in Eq.~\eqref{MS3} now have the dual purpose of denoting (i) the ``$+$'' or ``$-$'' LL half-string gauge in which the force is valid and (ii) the directional limit from which the quantities in the formula are calculated.

\subsection{Regularization parameters }\label{RP-general}
We now describe the calculation of $\d A^\pm_\alpha$, $\d B^\pm_\alpha$, $\d C^\pm_\alpha$, and $\d D^\pm_\alpha$ for arbitrary geodesic orbits in Kerr. When presenting the explicit final results we will, for simplicity, specialize to specific classes of orbits and a specific choice of extension. Section\ \ref{alt-extension} discusses the effect of alternative choices of extension and derives some general properties of the parameter values that apply to all orbits and any extension. Section\ \ref{alt-gauge} comments on how the parameter values are influenced by alternative choices of the LL gauge.

We assume we are given either the modes $(\tilde F^{{\rm Rad}'}_{\alpha})_{+}^\ell$ or the modes $(\tilde F^{{\rm Rad}'}_{\alpha})_{-}^\ell$ in a half-string gauge Rad$'$, and we calculate the GSF in an LL counterpart related to Rad$'$ by the gauge vector $\xi_\alpha^{\pm}=\xi^{0\pm}_\alpha$ given in Eq.~\eqref{xi0-coords}, where the $\pm$ corresponds to  $(\tilde F^{{\rm Rad}'}_{\alpha})_{\pm}^\ell$.

From this starting point, our calculation follows closely the method of Refs.\ 
\cite{Barack-Ori:02,Barack-Ori:03a,Barack-Ori:03b}, as reviewed in \cite{Barack:09}, and in the text below we will refer to these works where details are incomplete. 

We are interested in calculating the GSF at the point $x'$ on $\Gamma$ with Boyer-Lindquist coordinates $x^{\alpha}_p=(t,r_p,\theta_p,\varphi_p)$. We introduce new polar coordinates $(\tilde\theta,\tilde\varphi)$, so that the particle is located at the pole ($\tilde\theta=0$) of the new system, and $\tilde\varphi$ is chosen so that the particle's velocity at $x_p$ (projected onto the 2-sphere) points along the $\tilde\varphi=0$ longitude line. [The purpose of the transformation to the $(\tilde\theta,\tilde\varphi)$ system is to simplify the multipole decomposition below: in that system, the value of each $\ell$-mode of the full force at the particle has a sole contribution from the axially-symmetric, $m=0$ azimuthal mode.] We then introduce locally Cartesian coordinates $\hat x=\rho\cos\tilde\varphi$, $\hat y=\rho\sin\tilde\varphi$, where $\rho=\rho(\tilde\theta)$ is some smooth function with the property $\rho=\tilde\theta+O(\tilde\theta^2)$ near the particle. In terms of these variables, we have $\delta\theta=\hat x+O(s^2)$ and $\delta\varphi=\hat y/\sin\theta_p+O(s^2)$ \cite{Barack:09}.  Then, at leading order, we can write $\delta_\xi \tilde F_\alpha^{\pm}(\delta x';x')$ as $\delta_\xi \tilde F_\alpha^{\pm}(\delta r,\hat x,\hat y;x_p)$; recall that we have chosen $\delta t=0$. 

The $\ell$ modes of $\delta_\xi F_\alpha^{\pm}$ are calculated via \cite{Barack-Ori:03b,Barack:09}
\begin{align}\label{dF-modes0}
(\delta_\xi \tilde F_\alpha)^\ell_{\pm} = \frac{L}{2\pi}&\lim_{\delta r\to0^{\pm}}\int_{-1}^1 d(\cos\tilde\theta) \mathsf{P}_\ell(\cos\tilde\theta)\nonumber\\
	& \times \int_0^{2\pi}d\tilde\varphi\,\delta_{\xi}\tilde F_\alpha^{\pm}(\delta r,\hat x,\hat y),
\end{align}
where $\mathsf{P}_\ell$ is the Legendre polynomial. At first glance this equation seems to require global information about $\xi_\alpha$, since it involves integrating over a sphere that extends far away from the particle. However, our local expression for $\xi_\alpha$ suffices; terms that vanish
at the particle may affect the individual $\ell$ modes, but they do not alter the sum of modes evaluated at the particle.

Equation~\eqref{dF-modes0} can be concretely evaluated only once an extension is chosen. It can be slightly simplified, however, by appealing to the general properties of the full force. In Appendix~\ref{dF-arbitrary}, we show that (a) the piece of $\xi_\alpha$ parallel to $u_\alpha$, $\xi_{\alpha\parallel}$, does not contribute to $\delta_{\xi}\tilde F_\alpha^{\pm}$ at leading order, and (b) if $\xi_{\alpha\perp}$ is bounded, then $\delta_{\xi}\tilde F_\alpha^{\pm}$ is as well. Given these properties, we can write Eq.~\eqref{dF-modes0} as
\begin{align}\label{dF-modes}
(\delta_\xi \tilde F_\alpha)^\ell_{\pm} = \frac{L}{2\pi}&\int_{-1}^1 d(\cos\tilde\theta) \mathsf{P}_\ell(\cos\tilde\theta)\nonumber\\
	& \times \int_0^{2\pi}d\tilde\varphi\,\lim_{\delta r\to0^{\pm}}\delta_{\xi_\perp}\tilde F_\alpha^{\pm}(\delta r,\hat x,\hat y),
\end{align}
where we have brought the limit inside the integral using the fact that the integrand is bounded.
 
The above is valid for any extension. For our explicit calculations, we take that extension to be a ``rigid" one, with $\tilde u^\alpha(x)\equiv u^{\alpha'}$ and $\tilde\Gamma^\alpha_{\beta\gamma}(x)\equiv\tilde\Gamma^{\alpha'}_{\beta'\gamma'}$. The explicit formula for $\delta_{\xi_\perp}\tilde F_\alpha^{\pm}$ in that case is given in Eq.~\eqref{dF-rigid-extension}; its important property is that it includes no derivatives of the $\delta x'$ dependence in $\xi_\alpha$. That property allows us to avoid directly evaluating the integral in Eq.~\eqref{dF-modes}, writing instead $(\delta_{\xi_\perp} \tilde F_\alpha)^\ell=\delta_{(\xi_\perp)^\ell} \tilde F_\alpha$ and obtaining the $\ell$ modes of $\delta_\xi \tilde F_\alpha$ directly from those of $\xi_{\alpha\perp}^{\pm}$.

Our job is thus reduced to  
finding the $\ell$ modes of the vector $\xi_{\alpha\perp}^{\pm}$ given in Eq.~\eqref{xi_perp}. These modes are calculated from 
\begin{align}
(\xi_{\alpha\perp})^\ell_{\pm} = \mu\frac{L}{\pi}Q_{\alpha\beta}&\int_{-1}^1 d(\cos\tilde\theta) \mathsf{P}_\ell(\cos\tilde\theta)\nonumber\\
	& \times \int_0^{2\pi}d\tilde\varphi\lim_{\delta r\to0^{\pm}}\frac{\delta x^{\beta}}{\epsilon_0\pm z_0},
\end{align}
where $Q_{\alpha\beta}$, $s_0$ and $z_0$ are given in Eqs.\ \eqref{Q-body}, \eqref{s0-body} and \eqref{z0-body}, respectively. We now note that at $\delta t=\delta r=0$, the numerator and denominator of the integrand both scale linearly with $\rho$. It follows that the integrand is independent of $\rho$. The integral over $\cos\tilde\theta$ therefore reduces to $2\delta^\ell_0$, leaving us with
\begin{align}
(\xi_{\alpha\perp})^\ell_{\pm} = \frac{\mu}{\pi}\delta^\ell_0 \int_0^{2\pi}d	\tilde\varphi\frac{Q_{\alpha\theta}\cos	\tilde\varphi+Q_{\alpha\varphi}\sin	\tilde\varphi/\sin\theta_p}{R^\pm(x_p,\tilde\varphi)},\label{xi0ell}
\end{align}
where $R^\pm$ is $(\epsilon_0\pm z_0)/\rho$ evaluated at $\delta t = 0 = \delta r$.

The integral in Eq.\ \eqref{xi0ell} is elementary for any orbit \cite{Barack:09}. For example, specializing to {\em equatorial} orbits ($\theta_p\equiv \pi/2$), we find
\begin{align}
R^\pm &= r_p\left[1+(P_{\varphi\varphi}/r_p^2-1)\sin^2\tilde\varphi\right]^{1/2}\nonumber\\
	&\quad \mp \left(u_\varphi+\frac{\ell_\varphi}{\ell_\alpha u^\alpha}\right)\sin\tilde\varphi,\label{Rpm}
\end{align}
and
\beq
(\xi_{\alpha\perp})^\ell_{\pm} = \pm \frac{\mu}{r_p}\delta^\ell_0 Q_{\alpha\varphi}\frac{2c}{b-c^2}\left(1-\frac{1}{\sqrt{1+b-c^2}}\right),\label{xi0ellEvaluated}
\eeq
where $b\equiv P_{\varphi\varphi}/r_p^2-1$ and $c\equiv \frac{1}{r_p}[u_\varphi+\ell_\varphi/(\ell_\alpha u^\alpha)]$ are the factors appearing in Eq.~\eqref{Rpm}, and we have used the fact that $P_{\theta\theta}=r_p^2$ for equatorial orbits.

Given $(\xi_{\alpha\perp})^\ell_{\pm}$, calculating $(\delta_{\xi^0} \tilde F_\alpha)^\ell_{\pm}$ is a straightforward matter of substituting Eq.~\eqref{xi0ellEvaluated} into Eq.~\eqref{dF-rigid-extension}. Since $\xi_{\alpha\perp}$ contains a single $\ell$ mode, recalling Eq.\ \eqref{asym} we can then read off 
\beq\label{dAdBdC}
\delta A^\pm_\alpha = \delta B^\pm_\alpha = \delta C^\pm_\alpha = 0.
\eeq
Equation \eqref{dD}, in turn, gives
\beq
\delta D_{\alpha}^{\pm}=\sum_{\ell} (\d_{\xi^0} \tilde F_{\alpha})^\ell_{\pm}
=(\d_{\xi^0} \tilde F_{\alpha})^{\ell=0}_{\pm}
=\d_{(\xi_\perp)^{\ell=0}_{\pm}} \tilde F_{\alpha}.
\eeq

We proceed to show explicit results for $\delta D^\pm_\alpha$ in a few special cases.

\subsubsection{Arbitrary geodesic orbit in Schwarzschild geometry}

Specializing first to the Schwarzschild background, let $M$ denote the black hole mass, and ${\cal E}$ and ${\cal L}$ stand for the particle's specific energy and angular momentum---two conserved quantities that can be used to parametrize the geodesic orbit. Without loss of generality we set the particle to move on the equatorial plane. In the expressions below we use $r$ in lieu of $r_p$ for simplicity, introduce $f\equiv 1-2M/r_p$, and denote by $\dot r$ the derivative of $r_p$ with respect to proper time. The latter can be written in terms of ${\cal E}$, ${\cal L}$, and $r$ using $\dot{r}=\pm[{\cal E}^2-f(1+{\cal L}^2/r_p^2)]^{1/2}$. The four-velocity is $u^{\alpha}=({\cal E}/f,\dot{r},0,{\cal L}/r_p^2)$, and the principal null vector is $\ell^{\alpha}=(f^{-1},1,0,0)$---here we assume that the reconstructed part of the MP is given in a radiation gauge, $h_{\alpha\beta}^{\rm Rad}\ell^{\beta}=0$.

We note that $(\xi_{\alpha\perp})^\ell_{\pm}$ as written in Eq.~\eqref{xi0ellEvaluated} is not defined at $a=0$, where $b-c^2$ vanishes. However, the limit $a\to0$ is well defined, and it yields
\beq
(\xi^0_{\alpha\perp})^\ell_{\pm}= \pm \frac{\mu\mathcal{L}}{r_p^2}\delta^\ell_0 Q_{\alpha\varphi},
\eeq
where now
\beq
Q_{\alpha\varphi} = \left(\frac{-f\mathcal{L}}{\mathcal{E}-\dot r}\,,\,\frac{\mathcal{L}}{\mathcal{E}-\dot r}\,,\,0\,,\,r_p^2\right).
\eeq
This result may be verified by an independent calculation in Schwarzschild (using the results of Appendix~\ref{Schw-example}, for example). From it we find, with the aid of computer algebra,
\begin{eqnarray}
\d D_t^{\pm}&=&\pm \frac{\mu^2 \mathcal{L}^2 C_t(\mathcal{E},r,\rdot)}{r^7(\mathcal{E}-\rdot)^3},\quad\quad
\d D_r^{\pm}=\pm \frac{\mu^2 \mathcal{L}^2 C_r(\mathcal{E},r,\rdot)}{r^7(\mathcal{E}-\rdot)^3f},
\nonumber\\
\d D_\th^{\pm}&=&0,\quad\quad
\d D_\vf^{\pm}=\pm \frac{2\mu^2 \mathcal{L} C_\vf(\mathcal{E},r,\rdot)}{r^4(\mathcal{E}-\rdot)^2},
\end{eqnarray}
where 
\begin{eqnarray}
C_t(\mathcal{E},r,\rdot)&=&
2 r f [r^2 (1 - \mathcal{E}^2)+  M r (3 \mathcal{E}^2 -4)+ 4 M^2]
\nonumber\\
&&+  [3 r^2 (1 - \mathcal{E}^2)+ 4 M r (\mathcal{E}^2-4) +20 M^2 ]r \mathcal{E} \rdot
\nonumber\\
&&+ [r^2 (9 \mathcal{E}^2-1) + 6 M r (1 - 2 \mathcal{E}^2) - 8 M^2] r \rdot^2 
\nonumber\\
&&- 3 (3r-4 M) r^2 \mathcal{E} \rdot^3 
\nonumber\\
&&+ (3r-4 M)r^2  \rdot^4 ,
\end{eqnarray}
\begin{eqnarray}
C_r(\mathcal{E},r,\rdot)&=&
r^3 (-2 + \mathcal{E}^2 + \mathcal{E}^4) - 6 M r^2 ( \mathcal{E}^2-2) 
\nonumber\\
&& + 8 M^2 r (\mathcal{E}^2-3)+ 16 M^3 
\nonumber\\
&& - [r^2 (1 + 3 \mathcal{E}^2)- 8 M r+ 12 M^2  ]  r \mathcal{E} \rdot 
\nonumber\\
&& + r ( 3 r^2 \mathcal{E}^2 - 2 M r  +4 M^2 ) \rdot^2 
\nonumber\\
&& -  r^3 \mathcal{E} \rdot^3 ,
\end{eqnarray}
\begin{eqnarray}
C_\vf(\mathcal{E},r,\rdot)&=&
 r^2 (\mathcal{E}^2-1)- M r ( 3 \mathcal{E}^2-4) - 4 M^2 
\nonumber\\
&&+  [r (\mathcal{E}^2-1)+ 4 M ] r E  \rdot 
\nonumber\\
&& -   (2 r \mathcal{E}^2 + M) r\rdot^2 
\nonumber\\
&&+  r^2 \mathcal{E} \rdot^3 .
\end{eqnarray}

\subsubsection{Circular geodesic orbit in Schwarzschild geometry}

In the special case of circular motion [with $\rdot=0$, $\mathcal{E}=f(1-3M/r_p)^{-1/2}$ and $\mathcal{L}=(Mr_p)^{1/2}(1-3M/r_p)^{-1/2}$], the above expressions for $\d D_{\alpha}$ simplify to 
\begin{eqnarray}
\d D_r^{\pm} &=& \pm \frac{3\mu^2 M^2}{r_p^{5/2}(r_p-3M)^{3/2}},\label{dDr}
\nonumber\\
\d D_t^{\pm}&=&\d D_\th^{\pm} = \d D_\vf^{\pm}=0 
\quad
\text{(circular orbit)}.
\end{eqnarray}

\subsubsection{Circular equatorial orbits in Kerr geometry}

We now generalize to Kerr but immediately specialize to circular equatorial orbits, for simplicity. We denote by $M$ and $aM$ the mass and spin of the black holes, and we introduce 
\beq
\Delta\equiv r_p^2-2Mr_p+a^2,\quad\quad v\equiv \sqrt{M/r_p}.
\eeq
The specific energy and angular momentum are given in terms of the Boyer-Lindquist orbital radius as 
\begin{align}
\mathcal{E} &= \frac{1-2v^2+av^3/M}{\sqrt{1-3v^2+2av^3/M}},\\
\mathcal{L} &= r_pv\frac{1-2av^3/M+a^2v^4/M^2}{\sqrt{1-3v^2+2av^3/M}}.
\end{align}

We find
\beq
\delta D^\pm_\alpha=\pm\mathcal{Q}_\alpha\frac{2\mu^2 c}{r(b-c^2)}\left(1-\frac{1}{\sqrt{1+b-c^2}}\right),
\eeq
where 
\begin{align}
b &= r_p^{-3} \left[\mathcal{L}^2 r_p + a^2 (2 M + r_p)\right],\\
c &= \frac{a^2 \mathcal{E} \mathcal{L} + \mathcal{E} \mathcal{L} r_p^2 -  a \mathcal{L}^2 -a\Delta}{r_p\Bigl(a^2 \mathcal{E} -  a \mathcal{L} + \mathcal{E}r_p^2\Bigr)},
\end{align}
and
\begin{align}
\mathcal{Q}_t &= \mathcal{Q}_\th = \mathcal{Q}_\vf = 0,\\
\mathcal{Q}_r &= \frac{3 M}{r_p^3} \frac{v r_p^2 - a (r_p-M)-a^2 v }{r_p - 3 M + 2 a v}.
\end{align}

As with $(\xi_{\alpha\perp})^\ell_{\pm}$, $\d D^\pm_\alpha$ as written is not defined at $a=0$. But its limit as $a\to0$ is well defined, and that limit agrees, of course, with the Schwarzschild result displayed in Eq.~\eqref{dDr}.

\subsection{Alternative extensions}\label{alt-extension}

The values of the parameter corrections derived above correspond to our
particular choice of an off-worldline extension for the full force.
Since other extensions may prove beneficial in practice, it is useful to
discuss how our results may change if a different extension were chosen.
Below we will establish that two important features are insensitive to the
choice of extension: the vanishing of the ``large-$\ell$'' parameter
corrections $\delta A_{\alpha}$, $\delta B_{\alpha}$ and $\delta
C_{\alpha}$; and the fact that $\delta D_{\alpha}^{+}=-\delta
D_{\alpha}^{-}$ (while the actual values $\delta D_{\alpha}^{\pm}$ do
depend on the extension). We discuss the two features in turn.

\subsubsection{Extension-independence of $\delta A_{\alpha}=\delta
B_{\alpha}=\delta C_{\alpha}=0$ }

That Eq.\ \eqref{dAdBdC} applies for any smooth extension follows from a
simple general consideration (which we will prove momentarily): the sum of $\ell$ modes $(\delta_\xi \tilde
F)^{\ell}_{\pm}$ is guaranteed to converge. Therefore the large-$\ell$
expansion of $(\delta_\xi \tilde F)^{\ell}_{\pm}$ [Eq.\ \eqref{asym}]
cannot contain terms of orders $L^1$, $L^0$ or $L^{-1}$, and it follows
that $\delta A_{\alpha}=\delta B_{\alpha}=\delta C_{\alpha}=0$, and 
\beq\label{dD2}
\delta D_{\alpha}^{\pm}=\sum_{\ell} (\d_{\xi^0} \tilde F_{\alpha})^\ell_{\pm}.
\eeq
This result is insensitive to the extension chosen.

To show that the sum of modes converges, we appeal to a general formula for such situations. For a function $f:S^2\to\mathbb{R}$ that is discontinuous at the north pole, the function's spherical harmonic expansion at the pole is equal to the average of the function around an infinitesimal latitude
line surrounding the pole~\cite{Sansone:77}; that is, $\sum_\ell f^\ell=\lim_{\tilde\theta\to0}\frac{1}{2\pi}\int_0^{2\pi}f(\tilde\theta,\tilde\vf)d\tilde\vf$, where $\tilde\theta$ is the angle from the pole. This formula is analogous to the statement that a Fourier expansion of a function at a jump discontinuity yields the average across the jump. It is true so long as the average is of bounded variation as a function of $\tilde\theta$. In our case, it implies that 
\beq \label{Legendre-sum0}
\sum_\ell(\delta_\xi\tilde F^\alpha)_\pm^\ell = \lim_{\tilde\theta\to0}\frac{1}{2\pi}\int_0^{2\pi}\lim_{\d r\to 0^\pm}\delta_{\xi_\perp}\tilde F^\alpha(\d r,\hat x,\hat y) d\tilde\varphi,
\eeq
if the integral is of bounded variation. Clearly it is: On the regular side of the particle, $\xi_{\alpha\perp}$ is a simple (bounded) rational function of $\tilde\theta$, implying it is of bounded variation. From the results of Appendix~\ref{dF-arbitrary}, $\lim_{\d r\to 0^\pm}\delta_{\xi_\perp}\tilde F^\alpha(\d r,\hat x,\hat y)$ is bounded for any smooth extension if $\xi_{\alpha\perp}$ is bounded; and since it is constructed from $\xi_{\alpha\perp}$ via derivatives and multiplication by smooth functions, it will also be of bounded variation. Therefore the sum converges for any choice of smooth extension, and our result is established.

\subsubsection{Parity and the extension-independence of $\delta
D_{\alpha}^{+}=-\delta D_{\alpha}^{-}$}\label{deltaD}

In all three examples considered in the previous subsection we have
found (in our particular extension) $\delta D_{\alpha}^{+}=-\delta
D_{\alpha}^{-}$. Here we establish this result in full generality: for any point along
{\em any} geodesic orbit in Kerr, and for {\em any} smooth extension,
one has $\delta D_{\alpha}^{+}=-\delta D_{\alpha}^{-}$.

That this is true follows from the relationship between the parities of
the ``$+$'' and ``$-$'' solutions. Recall that for a no-string gauge, the
components of the gauge vector $\xi_{\alpha\perp}$ have odd parity under $\delta x^{\alpha'}\to-\delta x^{\alpha'}$, inherited from the odd parity of $\xi_a(x^b)$. It
follows that the half-string gauge vectors relate to one another
according to
$\xi^{0+}_{\alpha\perp}(\delta x^{\alpha'})=-\xi^{0-}_{\alpha\perp}(-\delta x^{\alpha'})$, except at the surface of discontinuity 
$\mathcal{S}$. This relationship is most easily visualized on a small sphere of
constant geodesic distance from the particle, with half the sphere in
the regular half of the ``$+$'' solution and half in the regular half of
the ``$-$'' solution. At antipodal points, the $\pm$ gauge vectors point
in opposite directions with equal magnitudes. (See the lower panel of
Fig.\ \ref{fig:parity} in the next section for an illustration.)

Now return to Eq.~\eqref{dF-modes} with this parity relation in mind.
 From the relationship between the gauge vectors, we have the
corresponding relation $\delta_{\xi^{0+}} \tilde F_\alpha(x',\delta
x')=-\delta_{\xi^{0-}} \tilde F_\alpha(x',-\delta x')$, which follows
from the results of Appendix~\ref{dF-arbitrary}, where we show that
$\delta_\xi \tilde F_\alpha$ has the same parity as $\xi_\alpha$ under
$\delta x^{\alpha'}\to-\delta x^{\alpha'}$. In terms of the variables in
Eq.~\eqref{dF-modes}, the relation becomes $\delta_{\xi^{0+}}\tilde
F_\alpha(\delta r,\hat x,\hat y)=-\delta_{\xi^{0-}}\tilde
F_\alpha(-\delta r,-\hat x,-\hat y)$. Hence, we immediately arrive
at the desired conclusion
\begin{align}
\lim_{\delta r\to0^+}\int d\cos\tilde\theta d\tilde\varphi
\mathsf{P}_\ell(\cos\tilde\theta)\delta_{\xi^{0+}}\tilde F_\alpha(\delta
r,\hat x,\hat y)
\nonumber\\
=-\lim_{\delta r\to0^-}\int d\cos\tilde\theta d\tilde\varphi
\mathsf{P}_\ell(\cos\tilde\theta)\delta_{\xi^{0-}}\tilde F_\alpha(\delta
r,-\hat x,-\hat y)
\nonumber\\
=-\lim_{\delta r\to0^-}\int d\cos\tilde\theta d\tilde\varphi
\mathsf{P}_\ell(\cos\tilde\theta)\delta_{\xi^{0-}}\tilde F_\alpha(\delta
r,\hat x,\hat y),
\end{align}
where the first equality follows from the odd parity of
$\delta_{\xi}\tilde F_\alpha(\delta r,\hat x,\hat y)$ under $(\delta
r,\hat x,\hat y)\to(-\delta r,-\hat x,-\hat y)$, and the second follows
from the invariance of the integral under the change of integration
variables $(\hat x,\hat y)\to(-\hat x,-\hat y)$ (which corresponds to a
rotation $\tilde\varphi\to	\tilde\varphi+\pi$). Hence, $(\delta_{\xi^0} \tilde
F_\alpha)^\ell_{+}=-(\delta_{\xi^0} \tilde F_\alpha)^\ell_{-}$. Since
[recalling Eq.\ \eqref{dD2}] we have simply $\delta
D_{\alpha}^{\pm}=\sum_{\ell} (\d_{\xi^0} \tilde F_{\alpha})^\ell_{\pm}$,
we immediately find $\delta D_{\alpha}^{+}=-\delta D_{\alpha}^{-}$.

This concludes the proof: we have shown that the correction $\delta D$
in the ``$+$'' solution is equal to minus the corrections $\delta D$ in
the ``$-$'' solution, for generic orbits in Kerr and regardless of the
choice of extension. We now have the mode sum formulas
\beq\label{MS4}
F_{\alpha}^{{\rm LL}\pm} = \sum_{\ell=0}^{\infty}\left[(\tilde F^{\rm
Rad'}_{\alpha})^\ell_\pm-A^\pm_{\alpha}L-B_{\alpha}-C_{\alpha}/L\right] -\d
D^\pm_{\alpha},
\eeq
where $\d D^\pm_{\alpha}=-\d D^\mp_{\alpha}$. These two formulas enable
us to calculate the GSF either from inside the particle's orbit, with
modes of the full force in the ``$-$'' half-string gauge, or from outside
the orbit, with modes in the ``$+$'' gauge.
Section~\ref{reconstruction-completion} clarifies how those two gauges
are constructed in practice.

\subsection{Alternative choices of locally Lorenz gauge}\label{alt-gauge}

In our construction of the LL gauge, we made a specific
choice: given a particular half-string gauge, the LL gauge
is related to it by the gauge vector $\xi_\alpha=\xi^0_\alpha$. Adding

terms of $o(1)$ to $\xi_\alpha$ has no impact on the GSF in the LL
gauge, so such terms need not be considered. However, adding an $O(1)$
term does affect the GSF, and we could have made the alternative choice
$\xi_\alpha=\xi^0_\alpha+Z_\alpha(\tau)$, with any $Z_\alpha(\tau)$.

Suppose we had done so. Then Eq.~\eqref{MS4} would have become (omitting
here the $\pm$  for simplicity)
\begin{align}
F_{\alpha}^{\rm LL}=\sum_{\ell=0}^{\infty}\left[(\tilde F^{{\rm
Rad'}}_{\alpha})^\ell-A_{\alpha}L-B_{\alpha}-C_{\alpha}/L\right] +\d
D^{\rm new}_{\alpha},\nonumber\\
\end{align}
where the new $\d D_\alpha$ parameter is
\begin{align}\label{dD-new}
\d D^{\rm new}_{\alpha} &= \sum_{\ell=0}^{\infty} (\d_\xi \tilde
F_{\alpha})^\ell\nonumber\\
                         &= \sum_{\ell=0}^{\infty} \left[(\d_{\xi^0}
\tilde F_{\alpha})^\ell+(\d_Z \tilde F_{\alpha})^\ell\right].
\end{align}
The first term is the $\delta D_\alpha$ that we have already calculated,
and the second term is the change to that result due to the nonzero
$Z_\alpha$. From this, one can see that the freedom to choose $Z_\alpha$
allows us to alter $\d D_\alpha$ almost arbitrarily. The question then
arises of whether we have made the best choice in setting $Z_\alpha$ to
zero. For example, we might try to choose a $Z_\alpha$ for which $\d
D^{\rm new}_\alpha=0$. To do so, we note that $\d_Z \tilde F_{\alpha}$
is smooth at the worldline, allowing us to write $\sum_\ell (\d_Z \tilde
F_{\alpha})^\ell$ simply as
\beq
\d_Z \tilde F_{\alpha} =
-\mu\left(P_{\alpha}^{\lambda}\frac{D^2Z_{\lambda}}{d\tau^2}+R_{\alpha
\mu\lambda\nu}u^{\mu} Z^{\lambda} u^{\nu}\right),
\eeq
where here all quantities are evaluated on the worldline. Finding a
$Z_\alpha$ for which $\d D^{\rm new}_{\alpha}=0$ then requires solving
the ODE
\beq\label{Zode}
\mu\left(P_{\alpha}^{\lambda}\frac{D^2Z_{\lambda}}{d\tau^2}+R_{\alpha
\mu\lambda\nu}u^{\mu} Z^{\lambda} u^{\nu}\right) = \delta D_\alpha,
\eeq
with $\delta D_\alpha$ given by $\sum_{\ell=0}^{\infty}(\d_{\xi^0}
\tilde F_{\alpha})^\ell$.

However, we reiterate our stance that since the GSF is gauge dependent,
when we calculate the GSF we must specify which LL gauge we are
calculating it in. For that reason, there is no apparent advantage to
knowing that there might exist an LL gauge in which $\d D^{\rm
new}_\alpha=0$ vanishes; finding such a gauge would still require us to
calculate $\sum_\ell(\d_{\xi^0} \tilde F_{\alpha})^\ell$ analytically,
and it would only add the extra step of solving the ODE \eqref{Zode} to
obtain $Z_\alpha$.


\section{Self-force in an undeformed radiation gauge: the no-string case}\label{force-no-string}
We now move on to our second method of calculating the GSF: a direct formulation in a radiation gauge, without local deformation. Our method begins with the formula~\eqref{Delta z} for the shift of a small object's center of mass under a gauge transformation. We apply the formula to find the change in position due to the transformation from Lorenz to completed radiation gauge. By taking two derivatives of the change in position and combining it with the acceleration in the Lorenz gauge, we obtain an equation of motion from which we can read off the total GSF. This procedure and its foundations are outlined in more detail in Appendix~\ref{Rad-force}. 

Our focus will be on the no-string case. The half- and full-string cases we relegate to Appendix~\ref{force-with-string}, due to their undesirable parity and singularity properties described in Sec.~\ref{Fermi-analysis}. There are several motivations for developing a direct formulation in the no-string gauge. First, a no-string MP has the obvious benefit of being regular on both sides of the particle. In Kerr, that will translate into an MP that is regular both at the event horizon and at asymptotic infinity. To take advantage of this, we must appeal to a direct formulation, because, as discussed in the previous section, the gauge's discontinuities prevent the LL formalism from being easily applied.

A second reason for considering the no-string gauge is its parity. It is parity-regular in the sense of Gralla. Therefore one might hope that Gralla's invariance results hold true despite the gauge's irregularities, making the Quinn-Wald-Gralla and standard mode-sum formulas applicable. In what follows, we show that the Quinn-Wald-Gralla angle-averaging formula~\eqref{Gralla-average} does remain valid. We find that the standard one-sided-limit form of the mode-sum formula fails, but an alternative form, requiring no corrections to parameters, does apply: the GSF is given by the Lorenz-gauge mode-sum formula, but with both the full modes and the Lorenz-gauge regularization parameters replaced by their average values calculated from two opposite sides of the particle. This is precisely the formula one would obtain by taking the average of the two half-string formulas~\eqref{MS4} and using the result that $\d D^+_{\alpha}+\d D^-_{\alpha}=0$.

In this section we seek results valid in any algebraically special vacuum background. However, when deriving our mode-sum formula, for concreteness we specialize to Kerr spacetime.

\subsection{Changes in position and force relative to globally Lorenz gauge}\label{force-no-string-calc}
The gauge vector $\xi^{\rm Lor\to Rad'}_\alpha$ that brings a global Lorenz gauge to a global no-string gauge is given by $\xi_\alpha = -\xi^0_\alpha-Z_\alpha+o(1)$, where $\xi^0_\alpha$ and $Z_\alpha$ are found in \eqref{xi0_t-no}--\eqref{xi0_A-no} and \eqref{Z-no}. Each term contains a jump discontinuity across the plane described by $p_ax^a=0$. Substituting $\xi_\alpha$ into Eq.~\eqref{Delta z} gives
\beq
\Delta z_1^a = \frac{3}{4\pi}\lim_{s\to0}\int n^a n^b(\xi^0_b+Z_b)d\Omega,
\eeq
where the integral is over a sphere of radius $s$ around the particle, and $d\Omega=\sin\theta d\theta d\phi$, where $\theta$ and $\phi$ are defined in the usual way from $x^a=(\sin\theta\cos\phi,\sin\theta\sin\phi,\cos\theta)$. [Here we use $\phi$ rather than $\varphi$ to distinguish the coordinates on this sphere from those on the Boyer-Lindquist coordinate sphere of constant $(t,r)$.] Since $\xi^0_b$ has odd parity on the sphere and $n^a n^b$ has even, the first term in the integral vanishes; the discontinuity in $\xi^0_b$ is immaterial. Therefore 
\begin{align}
\Delta z_1^a = \frac{3}{4\pi}\left(Z^+_b\int_{\tfrac{1}{2}S^2} n^a n^bd\Omega+Z^-_b\int_{\tfrac{1}{2}S^2} n^a n^bd\Omega\right),
\end{align}
where each of the integrals is over precisely half the sphere. Using the identity $\int n^a n^bd\Omega=\frac{4\pi}{3} \delta^{ab}$ and the even parity of the integrand, we arrive at the following simple result:
\beq
\Delta z_1^a(\tau) = \frac{1}{2}\left[Z^{a}_+(\tau)+Z^{a}_-(\tau)\right].\label{Delta z-nostring}
\eeq
In words, the shift in position is simply the average of the translations $Z_\pm^a$ that act from opposite sides of the particle. The odd parity of $\xi^0_a$ allows us to write this result in terms of the full gauge vector as
\beq
\Delta z_1^a(\tau) = -\frac{1}{2}\lim_{x^b\to0}\left[\xi^{a}_+(\tau,x^b)+\xi^{a}_-(\tau,-x^b)\right],\label{DzLim}
\eeq
with $x^b$ chosen to ensure that each of the two terms is evaluated in the region $\pm p_a x^a>0$, where it is regular. With this coordinated choice of limit to the particle, the singular pieces of $\xi^a_+$ and $\xi_-^a$ cancel. If the limit were not coordinated in this way, it would be ill-defined, since $\xi^a_+$ and $\xi_-^a$ do not separately have unique limits at the particle.  We now note that $\tilde P^{\alpha\beta}\xi_\beta(x+\delta x)$, with $x\in\Gamma$, has the same parity (at leading order) under $\delta x\to-\delta x$ as does $\xi_a$ under $x^a\to-x^a$, as shown in Appendix~\ref{trans_to_global}. Hence, the change in position in arbitrary coordinates can be expressed as
\beq
\Delta z_1^{\alpha} = -\frac{1}{2}\lim_{\delta x\to 0}\left[\tilde P^{\alpha\beta}\xi_\beta^++\tilde P^{\alpha\beta}\xi_\beta^-\right],
\eeq
where we have multiplied Eq.~\eqref{DzLim} by $e^\alpha_a$ and used the fact that $e^\alpha_a\xi^a=\tilde P^{\alpha\beta}\xi_\beta+O(s\xi)$ for any smooth extension. The $\xi_\alpha^+$ term is evaluated at $x^\alpha+\delta x^\alpha$, and the $\xi_\alpha^-$ term at $x^\alpha-\delta x^\alpha$, with $\delta x^\alpha$ chosen to ensure that each term is evaluated in the corresponding regular half of spacetime.

From the shift in position, an equation of motion can be found simply by taking two derivatives along the worldline, leading to
\begin{align}
\mu\frac{D^2\Delta z_1^{\alpha}}{d\tau^2} &= -\frac{1}{2}\mu\lim_{\delta x\to 0}\left[\tilde P^{\alpha\beta}\tilde u^\mu \tilde \nabla_\mu\left(\tilde u^\nu\tilde \nabla_\nu\xi_\beta^+\right)\right.\nonumber\\
	&\qquad\qquad \left.+\tilde P^{\alpha\beta}\tilde u^\mu \tilde \nabla_\mu\left(\tilde u^\nu\tilde \nabla_\nu\xi_\beta^-\right)\right]\nonumber\\
	&= -\mu R^\alpha{}_{\mu\beta\nu}u^\mu \Delta z_1^\beta u^\nu \nonumber\\
	&\quad +\frac{1}{2} \lim_{\delta x\to 0}\left[\delta_{\xi^+}\tilde F^\alpha-\mu\tilde P^{\alpha\beta}(\tilde u^\mu\tilde \nabla_\mu \tilde u^\nu)\tilde \nabla_\nu \xi^+_\beta\right. \nonumber\\
	&\quad \left.+ \delta_{\xi^-}\tilde F^\alpha-\mu\tilde P^{\alpha\beta}(\tilde u^\mu\tilde \nabla_\mu\tilde u^\nu)\tilde \nabla_\nu \xi^-_\beta\right].\label{a1}
\end{align}
The first equality holds for any smooth extensions of $u^\mu$, $P^{\alpha\beta}$, and $\nabla$ off the worldline, as can be seen from the results of Appendix~\ref{dF-arbitrary}. The second equality then follows from expressing $\tilde P^{\alpha\beta}\tilde u^\mu\tilde  \nabla_\mu\left(\tilde u^\nu\tilde \nabla_\nu\xi_\beta^\pm\right)$ in terms of $\delta_{\xi^\pm}\tilde F^\alpha$ using Eq.~\eqref{dF-arb-extension}. We now use Eq.~\eqref{a-term}, which shows that the terms involving $(\tilde u^\mu\tilde \nabla_\mu \tilde u^\nu)$ in Eq.~\eqref{a1} cancel one another, because the contribution of $Z^\pm_\alpha$ to those terms vanishes in the limit, and the contribution from $\xi^{0}_\alpha$ is odd under $\delta x\to-\delta x$. Our final result for the equation of motion is
\begin{align}
\mu\frac{D^2\Delta z_1^{\alpha}}{d\tau^2} &= -\mu R^\alpha{}_{\mu\beta\nu} u^\mu \Delta z_1^\beta u^\nu \nonumber\\
	&\quad +\frac{1}{2}\lim_{\delta x\to 0}\left[\delta_{\xi^+}\tilde F^\alpha + \delta_{\xi^-}\tilde F^\alpha\right].
\end{align}

As we described in Sec.~\ref{gauges} and Appendix~\ref{Rad-force}, the change in acceleration induced by $\xi^\alpha$ comes in the form of a background geodesic-deviation term plus a self-force term. The self-force term reads 
\beq\label{dF-nostring1}
\Delta F^\alpha = \frac{1}{2}\lim_{\delta x\to 0}\left[\delta_{\xi^+}\tilde F^\alpha + \delta_{\xi^-}\tilde F^\alpha\right],
\eeq
which is simply the average of the change in the full force as calculated from two opposite sides of the particle. This result is valid for any choice of extension. Using the fact that the contributions from $\xi^0_\alpha$ cancel, we can write it in the equivalent form
\beq\label{dF-nostring2}
\Delta F^\alpha = -\frac{1}{2}\left(\delta_{Z^+}\tilde F^\alpha + \delta_{Z^-}\tilde F^\alpha\right),
\eeq
where all quantities are evaluated on $\Gamma$.

The total GSF in the no-string gauge is then the sum of the GSF in the original Lorenz gauge plus the change due to the transformation:
\beq
F^\alpha = F^\alpha_{\rm Lor}+\Delta F^\alpha.\label{F-nostring}
\eeq
 
\subsection{Mode-sum formula}\label{MS-NS}
To obtain a formula useful for numerical implementation, we wish to recast Eq.~\eqref{F-nostring} in the form of a mode sum. For concreteness, we specialize to a Kerr background in Boyer-Lindquist coordinates, and we let $\delta x^\mu=\delta^\mu_r\delta r$. The first step is to write
\begin{align}
F^\alpha &=  F^\alpha_{\rm Lor}+\frac{1}{2}\lim_{\delta x\to 0}\left[\delta_{\xi^+}\tilde F^\alpha + \delta_{\xi^-}\tilde F^\alpha\right]\nonumber\\
	&= \frac{1}{2}\sum_\ell\left[(\tilde F^\alpha_{\rm Lor})^\ell_+ - A_+^\alpha L-B^\alpha-C^\alpha/L \right]\nonumber\\
	&\quad +\frac{1}{2}\sum_\ell\left[(\tilde F^\alpha_{\rm Lor})^\ell_- - A_-^\alpha L-B^\alpha-C^\alpha/L \right]\nonumber\\
	&\quad +\frac{1}{2}\sum_\ell\left[(\delta_{\xi}\tilde F^\alpha)^\ell_+ + (\delta_{\xi}\tilde F^\alpha)^\ell_-\right],
\end{align}
where the notation follows that of Sec.~\ref{LL-method}. We have made two simple manipulations in deriving this result: written $F^\alpha_{\rm Lor}$ as the average of its two one-sided-limit mode sums $F^\alpha_{\rm Lor}=\sum_\ell\left[(\tilde F^\alpha_{\rm Lor})^\ell_\pm - A_\pm^\alpha L-B^\alpha-C^\alpha/L \right]$, choosing the same (arbitrary) extension used for $\d_\xi \tilde F^\alpha$; and decomposed $\delta_{\xi^\pm}\tilde F^\alpha$ into modes. The second step is to note that the combination $(\tilde F^\alpha_{\rm Lor})_\pm^\ell+(\delta_{\xi}\tilde F^\alpha)_\pm^\ell$ is $(\tilde F^\alpha)^\ell_\pm$, the mode of the full force in the no-string gauge. Utilizing that fact, as well as the fact that $A^\alpha_+=-A^\alpha_-$~\cite{Barack-Ori:03b,Barack:09}, we arrive at the simple two-sided-average formula
\begin{align}\label{mode-sum-no-string}
F^\alpha 
	&= \sum_\ell\left[\frac{1}{2}(\tilde F^\alpha)_+^\ell+\frac{1}{2}(\tilde F^\alpha)_-^\ell - B^\alpha - C^\alpha/L \right]-D^\alpha,
\end{align}
where the choice of extension is arbitrary and $B^\alpha$, $C^\alpha$, and $D^\alpha$ are the standard Lorenz-gauge parameters corresponding to that extension. 

Equation~\eqref{mode-sum-no-string} is the primary practical result of our analysis of the no-string gauge. In numerical calculations it will give us the capability of calculating the GSF from perfectly regular MP information on both sides of the particle, and it requires no corrections to the standard regularization parameters. It is worth mentioning that although we have found this GSF formula by examining the undeformed no-string gauge, it \emph{also applies in an LL no-string gauge}. Specifically, it applies in the LL gauge related to the no-string gauge by the gauge transformation $\xi_\alpha=\xi^0_\alpha$. We can see this by noting that $\xi^0_\alpha$, due to its odd parity, does not alter the GSF. Therefore, the GSF is identical to that in the undeformed gauge: $F_{\rm LL}^\alpha = F^\alpha$. The GSF calculated from the mode-sum~\eqref{mode-sum-no-string} can be interpreted equally well as the GSF in the undeformed no-string gauge or as the GSF in its LL counterpart.

\subsection{Comparison with Gralla's invariance results}
Having established a practical mode-sum scheme, we now turn to the more formal question of whether Gralla's invariance results apply. Gralla proved two main results in Ref.~\cite{Gralla:11}: (i) the Quinn-Wald-Gralla formula~\eqref{Gralla-average} is valid for any MP within his class of gauges, and (ii) the mode-sum formula is likewise invariant. We now discuss each in turn.

\subsubsection{Self-force in a Quinn-Wald-Gralla form}\label{force-averaging}
We begin by confirming that Gralla's first invariance result does hold in the no-string gauge: the GSF is given by a simple average of the full force (with a particular extension) around the particle. Returning to Eq.~\eqref{Delta z-nostring}, we note that the right-hand side is easily found to agree with $-\frac{1}{4\pi}\lim_{s\to 0}\int \xi^a d\Omega$, meaning we can write
\beq
\Delta z_1^a = -\frac{1}{4\pi}\lim_{s\to 0}\int \xi^a d\Omega.
\eeq
In other words, the shift in position satisfies the same simple averaging formula as it would if $\xi_\alpha$ were truly parity-regular, rather than only nearly so (in the sense described in Sec.~\ref{no-string-Fermi}); the fact that $\xi_a$ contains discontinuities off the particle does not spoil the result. From here, reaching the desired conclusion is straightforward. In concord with Gralla's assumptions, when defining the full force off the worldline, we allow all metric-related quantities to take their natural values, and we extend $u^\alpha$ off $\Gamma$ via parallel propagation along geodesics in the spatial surfaces $\Sigma$. We then adopt Fermi coordinates $(\tau, y^a)$---\emph{not} our Fermi-\emph{like} coordinates $(\tau,x^a)$---in which parallel propagation of $u^\alpha$ is equal to a rigid coordinate extension through order $s$, because the coordinates are locally inertial. In these coordinates we write the components of $\Delta z_1^\alpha$ as
\begin{align}\label{Dz-averaging}
\Delta z_1^\alpha = -\frac{1}{4\pi}\lim_{s\to 0}\int \tilde P^{\alpha\beta}\xi_\beta d\Omega,
\end{align}
with $\tilde P^{\alpha\beta}=g^{\alpha\beta}+\tilde u^\alpha \tilde u^\beta$. Since the Christoffel symbols vanish on $\Gamma$ in Fermi coordinates (unlike in our Fermi-like coordinates), we can write
\begin{align}\label{Dzddot-averaging}
\mu\frac{D^2\Delta z_1^\alpha}{d\tau^2} =\mu\partial_\tau^2\Delta z_1^\alpha = -\frac{\mu}{4\pi}\lim_{s\to 0}\int \tilde P^{\alpha\beta}\partial^2_{\tau}\xi_\beta d\Omega.
\end{align}
We now express $P^{\alpha\beta}\partial^2_{\tau}\xi_\beta$ in terms of the full force. This is accomplished using the general expression~\eqref{dF-rigidu-extension} for the full force with a rigid extension of $u^\alpha$ and natural extension of the Christoffel symbols. In that equation, we replace primed indices with barred ones; the reference point on the worldline is now the point $\bar x=x_p(\tau)$. We also replace the derivative $\delta_\mu$ with the ordinary partial derivative $\delta^a_\mu\partial_a=\delta^a_\mu\frac{\partial}{\partial y^a}$; in Fermi coordinates, the coordinate differences $\delta x^{\alpha'}$ are simply the coordinates $y^a$ themselves. Analogously, the derivative $\hat\partial_{\mu'}$, which moves the point along the worldline, is replaced by $\delta_\mu^\tau\partial_\tau$. Making these substitutions and again using the fact that the Christoffel symbols vanish on the worldline, we find 
\begin{align}
\mu\tilde P^{\alpha\beta}\partial^2_{\tau}\xi_\beta &= \mu\tilde P^{\alpha\beta}\tilde u^\mu\tilde u^\nu \partial_\mu\partial_\nu \xi_\beta+O(s^2\xi)\nonumber\\
		&= -\delta_\xi \tilde F^\alpha -\mu R^{\bar\alpha}{}_{\bar\mu\bar\gamma\bar\nu}u^{\bar\mu}\xi^\gamma u^{\bar\nu} \nonumber\\
		&\quad +\mu P^{\bar\alpha\bar c}\Gamma^{\bar a}_{\bar\tau\bar\tau,\bar b}y^b\partial_a\xi_c+o(1).
\end{align}
[The fact that the second Christoffel term in Eq.~\eqref{dF-rigidu-extension} vanishes can be derived from the concrete expressions for the background metric in Fermi coordinates in Ref.~\cite{Poisson-Pound-Vega:11}; it can also be seen from the fact that no nonvanishing component of the Riemann tensor or its derivatives can have a lone spatial index.] Upon substitution into Eq.~\eqref{Dzddot-averaging}, the final term in this result vanishes, since $y^b\partial_a\xi_c=y^b\partial_a\xi^0_c+O(s)$ has odd parity. Making use of this and Eq.~\eqref{Dz-averaging},  we find that Eq.~\eqref{Dzddot-averaging} gives
\beq
\mu\frac{D^2\Delta z_1^\alpha}{d\tau^2} = -\mu R^{\alpha}{}_{\mu\beta\nu}u^{\mu}\Delta z_1^\beta u^{\nu}+\frac{1}{4\pi}\lim_{s\to 0}\int \delta_\xi \tilde F^\alpha d\Omega,\label{Dz1ddot-sphere}
\eeq
where now all tensors are evaluated on the worldline (except the term within the integral).

Combining Eqs. \eqref{Dz1ddot-sphere} and \eqref{motion-in-Lorenz} for $\mu\frac{D^2\Delta z_1^\alpha}{d\tau^2}$ and $\mu\frac{D^2z_{1{\rm Lor}}^\alpha}{d\tau^2}$ immediately yields our desired result:
\beq
\mu\frac{D^2z_1^\alpha}{d\tau^2} = -\mu R^{\alpha}{}_{\mu\beta\nu}u^{\mu} z_1^\beta u^{\nu}+\frac{1}{4\pi}\lim_{s\to 0}\int \tilde F^\alpha d\Omega,
\eeq
where $z_1^\alpha=z_{1{\rm Lor}}^\alpha+\Delta z_1^\alpha$ is the perturbative correction to the position in the no-string gauge, and the full force is defined with a parallelly propagated extension of $u^\alpha$ off $\Gamma$ and a natural extension of $\tilde\nabla$. 

This confirms Gralla's conjecture that the GSF in a completed radiation gauge is given by the Quinn-Wald-Gralla formula despite irregularities away from the particle. Our analysis has shown that this is true specifically in a no-string gauge; Appendix~\ref{force-with-string} shows that it is not, generically, valid in a completed radiation gauge with a string.

\subsubsection{Generalization of Gralla's parameter-invariance result}
We now move to Gralla's second result: the invariance of the mode-sum formula. Unlike the Quinn-Wald-Gralla formula, this result \emph{does not} hold in the no-string gauge, as we have already seen in Sec.~\ref{MS-NS}. 

More precisely, the mode-sum formula in the one-sided-limit form 
\beq
F^\mu=\sum_\ell\left[(\tilde F^\mu)^\ell_\pm-A^\mu_\pm L-B^\mu\right],
\eeq
with Lorenz-gauge parameter values, is not valid in the no-string gauge. (Here for compactness we make use of the fact that $C^\mu=D^\mu=0$.) But the mode-sum formula in the alternative, two-sided-average form 
\beq
F^\mu=\sum_\ell\left[\tfrac{1}{2}(\tilde F^\mu)^\ell_++\tfrac{1}{2}(\tilde F^\mu)^\ell_--B^\mu\right],
\eeq
again with the Lorenz-gauge value of $B^\mu$, \emph{is} valid. In the Lorenz gauge, the two formulas are guaranteed to be equivalent by the fact that the two sums $\sum_\ell[(\tilde F_{\rm Lor}^\mu)^\ell_+-A^\mu_+ L-B^\mu]$ and $\sum_\ell[(\tilde F_{\rm Lor}^\mu)^\ell_--A^\mu_- L-B^\mu]$ are equal. In the no-string gauge, on the other hand, the two sums are not equal, and only the two-sided average yields the correct GSF.

The failure of the one-sided-limit formula can be traced to the gauge's discontinuity away from the particle. Gralla's proof of the formula's validity is based on evaluating the sum $\sum_\ell (\delta_\xi\tilde F^\alpha)^\ell$ at the particle in terms of local quantities (e.g., our vector $Z^\alpha$) and showing it evaluates to the correct $\Delta F^\alpha$. That allowed him to write the total GSF as 
\begin{align}
F^\mu &= F^\mu_{\rm Lor}+\Delta F^\mu \nonumber\\
	&= \sum_\ell\left[(\tilde F^\mu_{\rm Lor})^\ell_\pm+(\delta_\xi\tilde F^\mu_{\rm Lor})^\ell_\pm-A^\mu_\pm L-B^\mu\right]\nonumber\\
	&= \sum_\ell\left[(\tilde F^\mu)^\ell_\pm-A^\mu_\pm L-B^\mu\right].
\end{align}

Here we follow Gralla's method of proof and discover precisely where it fails in the no-string gauge. First, we note that in this gauge $\sum_\ell (\delta_\xi\tilde F^\alpha)^\ell$ is not defined at the particle. But in the one-sided-limit formula, we require only the limit from inside or outside the particle's radial position, $\big(\!\lim_{r\to r_p^\pm}\delta_\xi\tilde F^\alpha\big)^{\!\ell}=(\delta_\xi\tilde F^\alpha)_\pm^\ell$, which is well defined. Following Gralla, we evaluate $\sum_\ell(\delta_\xi\tilde F^\alpha)_\pm^\ell$ by appealing to the same general formula we utilized in Sec.~\ref{alt-extension} for the sum of harmonic modes at a point of discontinuity, which led to Eq.~\eqref{Legendre-sum0}. With the particle placed at the pole $\tilde\theta=0$ (as described in Sec.~\ref{RP-general}), the formula in the present case becomes
\beq \label{Legendre-sum}
\sum_\ell(\delta_\xi\tilde F^\alpha)_\pm^\ell = \frac{1}{2\pi}\int_0^{2\pi}\lim_{\tilde\theta\to0}\left(\delta_{\xi_\perp}\tilde F^\alpha\right)_{r=r_p^\pm} d\tilde\varphi,
\eeq
where we have brought the limit inside the integral using the boundedness of the integrand. Near the particle, $\delta_\xi\tilde F^\alpha|_{r=r_p^\pm}$ contains two contributions: the continuous term $-\delta_{Z^\pm}\tilde F^\alpha$, and the discontinuous one $-\delta_{\xi^{0\pm}}\tilde F^\alpha$. Therefore Eq.~\eqref{Legendre-sum} becomes
\beq\label{Legendre-sum2} 
\sum_\ell(\delta_\xi\tilde F^\alpha)_\pm^\ell = -\delta_{Z_\pm}\tilde F^\alpha-\frac{1}{2\pi}\int_0^{2\pi}\lim_{\tilde\theta\to0}\delta_{\xi^0}\tilde F^\alpha|_{r=r_p^\pm} d\tilde\varphi.
\eeq

Compare this with the result for the change in GSF, which reads, according to Eq.~\eqref{dF-nostring2}, $\Delta F^\alpha=-\frac{1}{2}(\delta_{Z_+}\tilde F^\alpha+\delta_{Z_-}\tilde F^\alpha)$. Clearly $\Delta F^\alpha$ is not equal to the right-hand side of Eq.~\eqref{Legendre-sum2}. Each of the two vectors $Z^\alpha_\pm$ can be changed arbitrarily while remaining in the class of no-string gauges; ergo, there need not be any relation between them, nor one between either of them and $\xi_0$. In that sense, because of the jump discontinuity in the gauge, the sum of modes from the ``$+$'' side is blind to the gauge on the ``$-$'' side, and there is no reason for it to yield the correct GSF, which depends on the MP all around the particle. Indeed, the second term on the right-hand side of Eq.~\eqref{Legendre-sum2} is simply $-\sum_\ell(\delta_{\xi^0}\tilde F^\alpha)_\pm^\ell=-\delta D^\alpha_\pm$, the (generically nonzero) gauge correction to $D^\alpha$ in the LL half-string gauge.

How is this outcome altered, and the correct GSF recovered, when there is no jump discontinuity in the gauge? Suppose our gauge were in the Gralla class. The gauge vector could still be written in the form $\xi_\alpha(\tau,x^a)=-Z_\alpha^+(\tau)\theta(z)-Z_\alpha^-(\tau)\theta(-z)-\xi^0_\alpha(n^a)+o(1)$ near the particle, but $Z_\alpha^+$ and $Z_\alpha^-$ would be identical, and $\xi^0_\alpha$ would be smooth off the particle. We would then have $\Delta F^\alpha=-\delta_{Z^-}\tilde F^\alpha=-\delta_{Z^+}\tilde F^\alpha$. The first term in Eq.~\eqref{Legendre-sum2} would then be the correct GSF. We will show momentarily that the second term would vanish. These two facts together lead to the desired conclusion that $\sum_\ell(\delta_\xi\tilde F^\alpha)_\pm^\ell=-\delta_{Z^\pm}\tilde F^\alpha=\Delta F^\alpha$. 

The fact that $\int_0^{2\pi}\lim_{\tilde\theta\to0}\delta_{\xi^0}\tilde F^\alpha|_{r=r_p^\pm} d\tilde\varphi$ vanishes for any smooth, parity-regular $\xi^0_\alpha(n^a)$ brings an important point to the discussion. For a transformation within the class of sufficiently regular gauges, the change in GSF depends on the parity of $\delta_\xi\tilde F^\alpha$ in a three-dimensional region around the particle. Contrastingly, the change in the mode sum of the full force, $\sum_\ell(\delta_\xi\tilde F^\alpha)_\pm^\ell$, depends on the parity of $\delta_\xi\tilde F^\alpha$ in the two-dimensional plane tangent to the sphere $(t_p,r_p)$ at the particle. That can be seen from Eq.~\eqref{Legendre-sum}: if the integrand $\lim_{\tilde\theta\to0}\delta_\xi\tilde F^\alpha|_{r=r_p^\pm}$ has odd parity in the aforementioned plane, then the integral vanishes; only even-parity pieces of $\lim_{\tilde\theta\to0}\delta_\xi\tilde F^\alpha|_{r=r_p^\pm}$ contribute to the integral. 

This distinction between three-dimensional and two-dimensional parities is the root cause of the differing values of $\int_0^{2\pi}\lim_{\tilde\theta\to0}\delta_{\xi^0}\tilde F^\alpha|_{r=r_p^\pm} d\tilde\varphi$ in the smooth versus discontinuous cases. If $\xi^0_\alpha(n^a)$ is smooth and has odd three-dimensional parity, then $\lim_{\tilde\theta\to0}\delta_{\xi^0}\tilde F^\alpha|_{r=r_p^\pm}$ inherits odd two-dimensional parity, ensuring that $\int_0^{2\pi}\lim_{\tilde\theta\to0}\delta_{\xi^0}\tilde F^\alpha|_{r=r_p^\pm} d\tilde\varphi$ vanishes. On the other hand, if $\xi^0_\alpha(n^a)$ possesses a jump discontinuity at $r=r_p$, then $\lim_{\tilde\theta\to0}\delta_{\xi^0}\tilde F^\alpha|_{r=r_p^\pm}$ need not have any particular parity.

\begin{figure}[t]
\begin{center}
\includegraphics[width=\columnwidth]{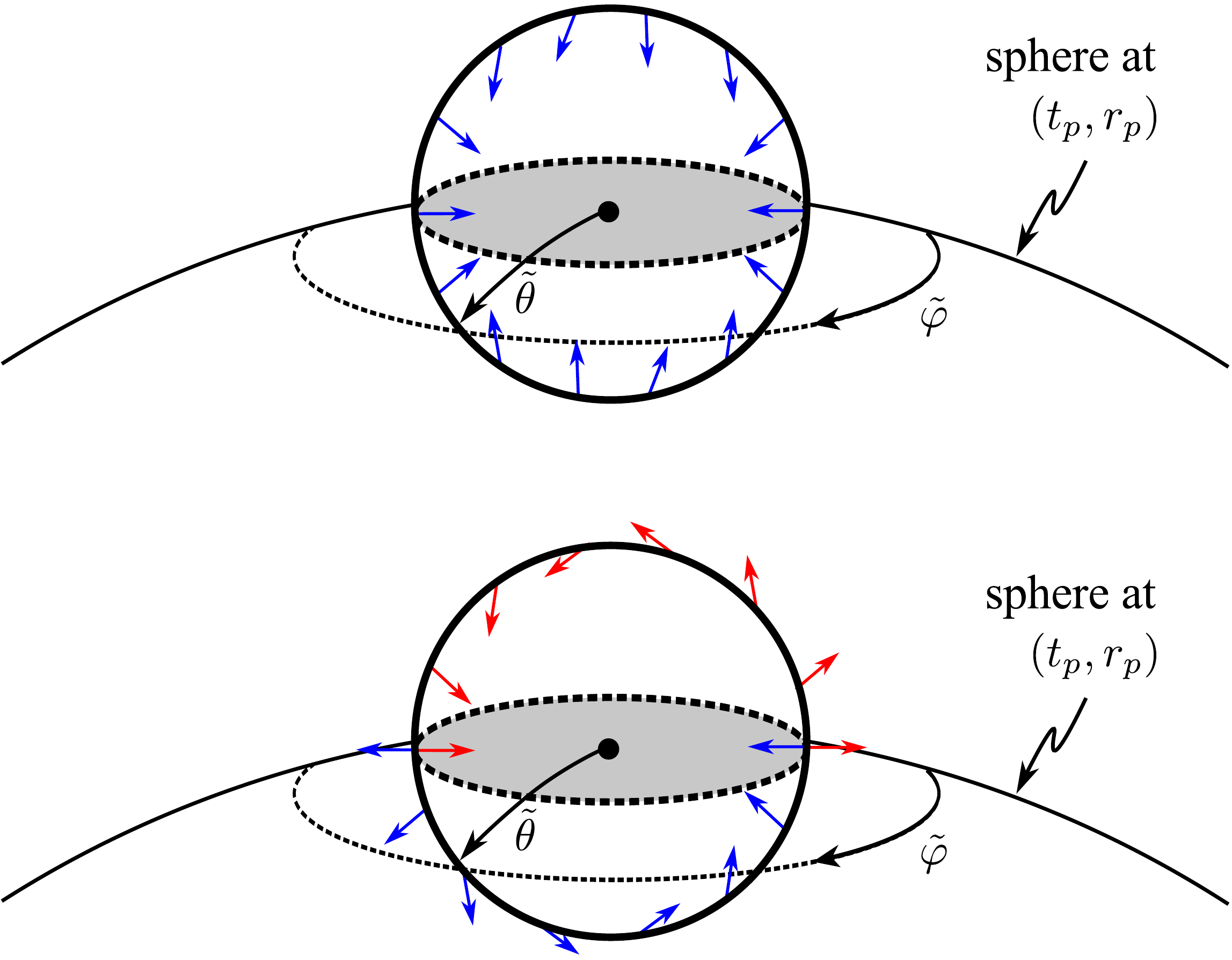}
\caption{\label{fig:parity}(Color online) Parity of vector fields around the particle. The particle, indicated by the black ball, sits at the north pole $\tilde\theta=0$ of the Boyer-Lindquist coordinate sphere defined by $(t,r)=(t_p,r_p)$. It is surrounded by a much smaller sphere of radius $s$. The shaded disc is tangent, at $\tilde\theta=0$, to the large sphere. Upper panel: a smooth vector field with odd parity is shown on the surface of the smaller sphere. Its restriction to the shaded disc inherits odd parity under reflection through the center of the disk. Lower panel: a discontinuous vector field with odd parity is shown. The field exhibits a jump discontinuity across the disc. Although it possesses odd parity on the small sphere, its limit to the disc, either from above (shown in red) or from below (in blue), does not inherit that parity.}
\end{center}
\end{figure}

Figure \ref{fig:parity} illustrates this inheritance of parity (or lack thereof) in two sample vector fields around the particle, one continuous and one with a jump discontinuity. We may also see algebraically how it occurs in quantities relevant to our calculations. Consider the function $s_0/\rho$, where $\rho(\tilde\theta)$ is the local distance introduced on the Boyer-Lindquist sphere in Sec.~\ref{RP-general}. This function is smooth off the particle and has even parity under $\delta x^{\mu'}\to-\delta x^{\mu'}$, where $\delta x^{\mu'}$ is the coordinate distance from the particle in Boyer-Lindquist coordinates. Its limit to the tangent plane is
\begin{align}
\lim_{\tilde\theta\to0}&\lim_{r\to r_p}\frac{1}{\rho}\left[P_{\alpha'\beta'}\delta x^{\alpha'}\delta x^{\beta'}\right]^{1/2} \nonumber\\
&= \left[P_{\theta'\theta'}(\hat n^x)^2+2P_{\theta'\vf'}\hat n^x\hat n^y+P_{\vf'\vf'}(\hat n^y)^2\right]^{1/2},
\end{align}
where $\hat n^x\equiv \displaystyle\lim_{\tilde\theta\to0}\frac{\hat x}{\tilde\theta}=\cos\tilde\vf$ and $\hat n^y\equiv  \displaystyle\lim_{\tilde\theta\to0}\frac{\hat y}{\tilde\theta\sin\theta_p}=\sin\tilde\vf$ are ``unit vectors'' pointing outward from the particle in the plane tangent to $(t,r)=(t_p,r_p)$ at $\tilde\theta=0$. One can see that the limiting field has inherited an even parity under $(\hat n^x,\hat n^y)\to(-\hat n^x,-\hat n^y)$. Now contrast this with the case of a field with a jump discontinuity. Consider $\frac{1}{\rho}(s_0+q_{\mu'}\delta x^{\mu'})\theta(r-r_p)+\frac{1}{\rho}(s_0-q_{\mu'}\delta x^{\mu'})\theta(r_p-r)$. It has even three-dimensional parity, but its limit to the tangent plane,
\begin{align}
\lim_{\tilde\theta\to0}&\lim_{r\to r_p^\pm}\frac{1}{\rho}(s_0\pm q_{\mu'}\delta x^{\mu'})\nonumber\\
 &= \left[P_{\theta'\theta'}(\hat n^x)^2+2P_{\theta'\vf'}\hat n^x\hat n^y+P_{\vf'\vf'}(\hat n^y)^2\right]^{1/2}\nonumber\\
&\quad\pm \left(q_{\theta'}\hat n^x+q_{\vf'}\hat n^y\right),
\end{align}
has no definite parity under $(\hat n^x,\hat n^y)\to(-\hat n^x,-\hat n^y)$.

These facts make clear why the one-sided limit formula is incorrect in a no-string gauge. They also provide a heuristic explanation of why the two-sided-average formula is correct: it incorporates the three-dimensional parity of the full force, which is lost in the one-sided limit. In that sense, the two-sided-average formula is the more general of the mode-sum formulas, and it is invariant in a broader class of gauges.


\section{Metric reconstruction in a flat-space example}\label{reconstruction-completion}
Although we have derived practical formulas for calculating the GSF from the modes of a half-string or no-string MP, an important question remains: are these perturbations calculable from a mode-by-mode CCK reconstruction and completion procedure? In this section, we consider that question within a toy model of a static point mass in flat spacetime. This problem has been considered in the past by Keidl~\emph{et al.}~\cite{Keidl-etal:07}, but in that earlier work the focus was on metric reconstruction using closed-form, four-dimensional quantities. Here we are interested specifically in how the reconstruction plays out when working with a decomposition in harmonics, because a standard numerical procedure of reconstruction proceeds at the level of individual modes. The distinction is an important one. As we have stressed at various points in earlier sections, the string singularities in the half- and full-string MPs are not absolutely integrable over a two-dimensional spatial surface intersecting the string, suggesting they do not possess harmonic expansions.

For the purpose of developing an analogy with numerical calculations in black hole spacetimes, we situate the static particle off the origin, at a radial position $r_p$. We then perform the reconstruction and completion of the MP by working mode by mode in a decomposition into spherical harmonics. The spacetime naturally divides into two regions: the region inside the sphere $\mathcal{S}$ of radius $r=r_p$ centered at the origin; and the region outside that sphere. (We shall sometimes refer to these two regions as half-spacetimes, irrespective of their disparate volumes.) We show that when either region contains a string, a reconstructed MP cannot be found anywhere in that region as a sum over modes. In the regions with no string, we explicitly reconstruct the regular half of each half-string MP (and hence both halves of the no-string MP, with the sphere $\mathcal{S}$ being the surface of discontinuity). Some subtleties arise in the completion of the no-string MP, but we show that in practice those subtleties will not affect calculations of GSF. We conjecture that the same basic conclusions hold true for dynamic particles in Schwarzschild and Kerr.

Our goal in the toy problem is to solve the linearized EFE
\begin{equation}
\delta G_{\alpha\beta}[h_{\mu\nu}]=8\pi T_{\alpha\beta},\label{EFE}
\end{equation}
where $\delta G_{\alpha\beta}$ is the linearized Einstein tensor, and $T_{\mu\nu}$ is the static point particle stress-energy
\begin{equation}\label{T_ab}
T^{\alpha\beta} = \mu u^\alpha u^\beta \delta^3(\vec x - \vec x_p).
\end{equation}
Here the coordinates are $(t,\vec x)$, the four-velocity is $u^\alpha=\delta^\alpha_t$, and the particle sits at a spatial position $\vec x_p$. To solve the EFE using the CCK procedure, we split the perturbation into two pieces,
\begin{equation}
h_{\alpha\beta}=h^{\rm Rad}_{\alpha\beta}+h^{\rm Cmpl}_{\alpha\beta},
\end{equation}
where $h^{\rm Rad}_{\alpha\beta}$ is the perturbation reconstructed from the curvature scalars, and $h^{\rm Cmpl}_{\alpha\beta}$ is whatever is required to complete the solution. Given a reconstructed perturbation $h^{\rm Rad}_{\mu\nu}$, $h^{\rm Cmpl}_{\mu\nu}$ must satisfy
\begin{equation}
\delta G_{\alpha\beta}[h^{\rm Cmpl}_{\mu\nu}]=8\pi T_{\alpha\beta}-\delta G_{\alpha\beta}[h^{\rm Rad}_{\mu\nu}].\label{EFE_comp}
\end{equation}

For our explicit calculations, we will adopt an ingoing radiation gauge for the reconstructed MP. We use the complex null tetrad $\ell^\mu$, $n^\mu$, $m^\mu$, and $m^{\mu*}$, where the legs that appear explicitly in the calculation are
\begin{align}
\ell^\mu = (1,1,0,0),\quad m^\mu =\frac{1}{r\sqrt 2}(0,0,1,i\csc\theta), 
\end{align}
and $m^{\mu*}$, the complex conjugate of $m^\mu$. We work in polar coordinates, which for later convenience we write as $(x^a,\theta^A)$, where $x^a=(t,r)$ and $\theta^A=(\theta,\varphi)$. [The reader should take note of the change in notation from the Fermi-like coordinates $x^a=(x^A,z)$ used elsewhere in this paper.] The particle's spatial position in these coordinates is $(r_p,\theta^A_p)$. For vector and tensor harmonics, we adopt the definitions of Martel and Poisson~\cite{Martel-Poisson:05}. In these definitions, the even-parity harmonics are 
\begin{align}
Y^{\ell m}_A &= D_A Y_{\ell m},\\
Y^{\ell m}_{AB} &= D_AD_B Y_{\ell m}+\frac{1}{2}\ell(\ell+1)\Omega_{AB}Y_{\ell m},
\end{align}
where $Y_{\ell m}(\theta^A)$ are the ordinary scalar spherical harmonics, $\Omega_{AB}=(1,\sin^2\theta)$ is the metric of a unit two-sphere, and $D_A$ is the covariant derivative compatible with it. We shall not require the odd-parity harmonics. We use the standard conventions~\cite{Goldberg-etal:67} for the spin-weight-$s$ harmonics ${}_sY_{\ell m}(\theta^A)$ and for the spin-raising and -lowering angular derivatives $\eth$ and $\bar\eth$.  

Our presentation here will be brisk, skipping over many details. To avoid a swathe of review material and lists of equations, we assume some degree of familiarity with spin-weighted harmonics, the Teukolsky equation, and the CCK formalism. A more self-contained and detailed presentation of the calculation will appear, alongside a more thorough discussion of metric reconstruction and completion in black hole spacetimes, in a forthcoming paper~\cite{Barack-etal:14}.

\subsection{Reconstruction}
The CCK metric reconstruction procedure involves three steps: (1) solving the Teukolsky equation for $\psi_0$ or $\psi_4$, (2) finding a Hertz potential $\Psi$ that satisfies both the Teukolsky equation and a certain differential equation with $\psi_0$ or $\psi_4$ as a source, and (3) operating on the Hertz potential with another differential operator to obtain an MP in a (traceless) radiation gauge.

\subsubsection{Half-string, full-string, and no-string Hertz potentials}
We begin by expanding the Weyl scalar $\psi_0$ in terms of spin-weighted spherical harmonics,
\beq
\psi_0 = \sum_{\ell\geq2,m}\psi^{\ell m}_0(r) {}_2Y_{\ell m}(\theta^A),
\eeq
where we assume no dependence on $t$. After substituting this into the spin-2 Teukolsky equation, we arrive at the radial equation 
\beq
\left(\partial^2_r+\frac{6}{r}\partial_r-\frac{(\ell+3)(\ell-2)}{r^2}\right)\psi_0^{\ell m} = -8\pi T_0^{\ell m},\label{Teukolsky}
\eeq
where $T_0^{\ell m}$ is the radial coefficient in the expansion $T_0 = \sum_{\ell\geq2,m}T_0^{\ell m}(r) {}_2Y_{\ell m}(\theta^A)$ of the Teukolsky source. That source is constructed from derivatives of the stress-energy tensor, which in our case simplify to 
\beq
T_0 = -\frac{1}{2r^2}\eth^2 (T_{\alpha\beta}\ell^\alpha\ell^\beta).
\eeq
The expansion of $T_0$ in spin-weighted harmonics can be found by expanding the stress-energy tensor in ordinary scalar harmonics, using
\begin{equation}
\delta^3(\vec x - \vec x_p) = \frac{1}{r_p^2}\delta(r-r_p)\sum_{\ell\geq0,m}Y^*_{\ell m}(\theta^A_p)Y_{\ell m}(\theta^A).\label{delta_expansion}
\end{equation}
From the identity $\eth^2 Y_{\ell m} = \sqrt{\frac{(\ell+2)!}{(\ell-2)!}}{}_2Y_{\ell m}$, we find
\begin{align}
-8\pi T_0^{\ell m} &= \frac{4\pi \mu}{r_p^4}\sqrt{\frac{(\ell+2)!}{(\ell-2)!}}Y^*_{\ell m}(\theta^A_p)\delta(r-r_p)\nonumber\\
					&\equiv S_{\ell m}\delta(r-r_p).
\end{align}
Solving Eq.~\eqref{Teukolsky} is now a simple matter. We impose regularity at the origin and infinity, and we find
\beq
\psi^{\ell m}_0 = -\frac{S_{\ell m} r_p}{2\ell+1}\!\left[\!\left(\frac{r}{r_p}\right)^{\!\!\ell-2}\!\!\theta(r_p-r)+\left(\frac{r_p}{r}\right)^{\!\!\ell+3}\theta(r-r_p)\right].
\eeq

That completes step (1) of the reconstruction. Step (2) is to find the Hertz potential $\Psi$ by solving the equation $(\ell^\alpha \partial_\alpha)^4\Psi^*=\psi_0$, which for a static solution reduces to
\beq
\partial_r^4\Psi^*=\psi_0.\label{Hertz-eq}
\eeq
We again assume an expansion in spin-weighted harmonics,
\begin{equation}
\Psi^* = \sum_{\ell\geq2,m} \Psi^*_{\ell m}(r)\,{}_2Y_{\ell m}(\theta^A).
\end{equation}
Let us assume that we can bring the four derivatives in Eq.~\eqref{Hertz-eq} inside the sum. Solving mode by mode in the two regions $r<r_p$ and $r>r_p$, we find the general solution
\beq\label{Psi}
\Psi^*_{\ell m} = r_p^2S_{\ell m}\begin{cases} H^<(r) & \text{for } r<r_p,\\
								H^>(r) & \text{for } r>r_p,
						\end{cases}
\eeq
where 
\begin{align}
H^<(r) &= k\frac{r^{\ell+2}}{r_p^{\ell-1}}+\sum_{j=0}^3 b^<_{j\ell m} r^j,\\
H^>(r) &= k\frac{r_p^{\ell+2}}{r^{\ell-1}}+\sum_{j=0}^3 b^>_{j\ell m} r^j.
\end{align}
The common constant is $k=\frac{-1}{2\ell+1}\frac{(\ell-2)!}{(\ell+2)!}$, and the $b^\lessgtr_{j\ell m}$ terms are homogeneous solutions to Eq.~\eqref{Hertz-eq}. 

Equation \eqref{Psi} satisfies $\partial_r^4\Psi^*_{\ell m}=\psi^{\ell m}_0$ for all $r\neq r_p$, regardless of the choice of $b^\lessgtr_{j\ell m}$. If in addition we demand that $\Psi^*_{\ell m}$ satisfies that equation at $r=r_p$, then we place constraints on the constants $b^\lessgtr_{j\ell m}$. Specifically,
\begin{align}
b_{0\ell m}^> - b_{0\ell m}^< &= \frac{r_p^3}{6(\ell+2)(\ell-1)},\label{b0}\\
b_{1\ell m}^> - b_{1\ell m}^< &= -\frac{r_p^2}{2\ell(\ell+1)},\\
b_{2\ell m}^> - b_{2\ell m}^< &= \frac{r_p}{2\ell(\ell+1)},\\
b_{3\ell m}^> - b_{3\ell m}^< &= -\frac{1}{6(\ell+2)(\ell-1)}.\label{b3}
\end{align}
Since we have four equations and eight free parameters, we can solve these equations in multiple ways. However, regardless of which solution we choose, we immediately come up against two major problems. First, recall that if the reconstructed MP is to satisfy the vacuum EFE away from $\mathcal{S}$, the Hertz potential must satisfy not only $\partial_r^4\Psi^*=\psi_0$, but also the spin-(-2) Teukolsky equation. The reader can easily verify that if any of the parameters $b^\lessgtr_{j\ell m}$ is nonzero, then the $\ell m$-modes of $\Psi$ fail to satisfy the radial Teukolsky equation in the entire half-spacetime $r\lessgtr r_p$. A valid $\Psi$ could still be obtained if the sum over modes satisfied the Teukolsky equation, but that possibility brings us to a second, irresolvable problem: the sum of modes diverges everywhere in the half-spacetime $r\lessgtr r_p$ where $b^\lessgtr_{j\ell m}\neq0$. Suppose we solve Eqs.~\eqref{b0}--\eqref{b3} by choosing all $b^>_{jlm}=0$. For large $\ell$, we then have $b^<_{j\ell m}\sim 1/\ell^2$ and $S_{\ell m}\sim \ell^2 Y^*_{\ell m}(\theta^A_p)$, leading to
\begin{equation}
\Psi^*_{\ell m}\,{}_2Y_{\ell m} \sim Y^*_{\ell m}(\theta^A_p){}_2Y_{\ell m}\sim \frac{1}{\ell^2}\eth^2\left[Y^*_{\ell m}(\theta^A_p)Y_{\ell m}\right].
\end{equation} 
We can eliminate the sum over $m$ by using 
\begin{equation}\label{sum_m}
\sum_m Y^*_{\ell m}(\theta^A_p)Y_{\ell m}(\theta^A)=\frac{2\ell+1}{4\pi}\mathsf{P}_\ell(\cos\gamma),
\end{equation}
where $\gamma$ is the angle between $\vec x$ and $\vec x_p$. For $|\cos\gamma|\neq1$, the Legendre polynomial scales as $\mathsf{P}_\ell(\cos\gamma)\sim1/\ell^{1/2}$, and each derivative of it introduces one additional power of $\ell$. So we arrive at a sum of the form 
\begin{equation}\label{Psi-divergence}
\sum_{\ell m}\Psi^*_{\ell m}\,{}_2Y_{\ell m} \sim \sum_\ell \frac{1}{\ell}\eth^2 \mathsf{P}_\ell\sim \sum_\ell \ell^{1/2},
\end{equation} 
which manifestly diverges. This shows that for a solution with $b^>_{j\ell m}=0$ and $b^<_{j\ell m}\neq0$, the sum of modes diverges in the entire half-spacetime $r<r_p$. Likewise, for a solution with $b^<_{j\ell m}=0$ and $b^>_{j\ell m}\neq0$, the sum of modes diverges in the entire half-spacetime $r>r_p$. For a solution with some $b^<_{j\ell m}\neq0$ and some $b^>_{j\ell m}\neq0$, the sum diverges everywhere.

We conclude that it is impossible to construct a global solution to $\partial_r^4\Psi^*=\psi_0$ by working at the level of individual modes. The only way to obtain a convergent sum is to work in the two separate vacuum regions $r<r_p$ and $r>r_p$. In each region, we can obtain a valid solution to $\partial_r^4\Psi^*=\psi_0$ and to the Teukolsky equation by choosing all $b^\lessgtr_{j\ell m}=0$.

Suppose we work in one of the two regions, say in $r>r_p$ with $b^>_{j\ell m}=0$. If we were to evaluate the sum over modes of the Hertz potential in that region and then analytically continue the result into the region $r<r_p$, we would find that the Hertz potential acquires a string in the $r<r_p$ region. For brevity's sake, and since we wish to continue to work mode by mode, we do not perform that calculation here, but an analogous result will be obtained at the level of the MP in Sec.~\ref{recon-MP}. Consequently, the modes $\Psi_{\ell m}$ in $r>r_p$ with all $b^>_{j\ell m}=0$ can be identified as the modes of the regular half of a half-string Hertz potential $\Psi^+$. Analogously, the modes $\Psi_{\ell m}$ in $r<r_p$ with all $b^<_{j\ell m}=0$ can be identified as those of the regular half of a half-string potential $\Psi^-$. The impossibility of working mode by mode globally can reasonably be ascribed to the fact that the string in the Hertz potential does not possess a convergent harmonic expansion. If we worked with four-dimensional solutions rather than with harmonics, the half-string potentials could be found directly in their singular half-spacetimes, as seen in the four-dimensional calculation in Ref.~\cite{Keidl-etal:07}. 

Instead of working with two different Hertz potentials in their respective domains of regularity, we may work with both regions simultaneously by gluing together the regular halves of $\Psi^+$ and $\Psi^-$, creating a no-string potential $\Psi=\Psi^+\theta(r-r_p)+\Psi^-\theta(r_p-r)$. The modes of this potential are given by Eq.~\eqref{Psi} with all $b^\lessgtr_{j\ell m}$ set to zero. As in the half-string case, these modes represent a solution to $\partial_r^4\Psi^*=\psi_0$ (and to the Teukolsky equation) only \emph{off the sphere} $\mathcal{S}$. The potential contains a discontinuity across $\mathcal{S}$, but that will prove to be of little consequence, because in the reconstruction procedure we will consistently work off the sphere. 

In what follows, we will use the no-string potential, for the simple reason of compactness. Once the no-string MP is reconstructed, each of the half-string MPs will be read off straightforwardly.

\subsubsection{Reconstructed metric perturbations}\label{recon-MP}
We now move to the final step in the reconstruction procedure. For the moment we work with a generic Hertz potential $\Psi^*_{\ell m} = r_p^2S_{\ell m}H(r)$, without specifying $H(r)$. The MP is reconstructed from the Hertz potential using the CCK formula, which in our toy problem reduces to\footnote{We have multiplied the right-hand side of the standard formula by an overall factor of 2. The missing factor of 2 is a longstanding error in the CCK formalism, as noted by Keidl \emph{et al.}~\cite{Keidl-etal:07}, who corrected it by altering the relation between $\Psi$ and $\psi_0$ rather than by altering the reconstruction formula as we have.}
\begin{align}
h^{\rm Rad}_{\alpha\beta} &= -\left\lbrace\frac{1}{r^2}\ell_\alpha\ell_\beta\bar\eth^2
			+2m^*_\alpha m^*_\beta\!\left(\!\partial_r+\frac{1}{r}\right)\!\!\left(\!\partial_r-\frac{3}{r}\right)\right.\nonumber\\
			&\quad\left.+\sqrt{2}\ell_{(\alpha}m^*_{\beta)}\!\left[\partial_r\left(\frac{1}{r}\bar\eth\!\right)\!+\frac{1}{r}\bar\eth\left(\!\partial_r-\frac{3}{r}\right)\!\right]\!\right\rbrace\!\Psi^*+{\rm c.c.},
\end{align}
where ``c.c." denotes the complex conjugate of the entire first term. From this we obtain an MP  
\begin{align}
h^{\rm Rad}_{ab} &= \sum_{\ell\geq2, m}h^{\ell m}_{ab}Y^{\ell m},\label{hrec_ab}\\
h^{\rm Rad}_{aA} &= \sum_{\ell\geq2, m}h^{\ell m}_{a}Y_A^{\ell m},\\
h^{\rm Rad}_{AB} &= \sum_{\ell\geq2, m}h^{\ell m}Y_{AB}^{\ell m},\label{hrec_AB}
\end{align}
with the summands found to be (after some simplification)
\begin{align}
h^{\ell m}_{ab} &= 2r_p^2 S_{\ell m}\ell_a \ell_b \frac{1}{r^2}H(r)\sqrt{\frac{(\ell+2)!}{(\ell-2)!}},\label{htilde_ab}\\
h^{\ell m}_{aA} &= 2r_p^2 S_{\ell m}\ell_a (\partial_r-2/r)H(r)\sqrt{\frac{(\ell+2)(\ell-1)}{\ell(\ell+1)}} ,\\
h^{\ell m}_{AB} &= 4r_p^2 S_{\ell m} (r^2\partial^2_r-2r\partial_r)H(r)\sqrt{\frac{(\ell-2)!}{(\ell+2)!}}.\label{htilde_AB}
\end{align}
Each of these modes is constructed from the spin-weighted harmonic mode of the Hertz potential with the same $\ell$.

Equations \eqref{htilde_ab}--\eqref{htilde_AB} are valid for any $H(r)$. In a moment we will take $H(r)$ in its no-string form, but before we do so, let us consider what would happen if we were to use the Hertz potential~\eqref{Psi} with some nonzero $b^\lessgtr_{j\ell m}$ in $H(r)$. Following the analysis that led to Eq.~\eqref{Psi-divergence}, we would find from Eqs.~\eqref{hrec_ab} and \eqref{htilde_ab} that 
\begin{equation}
h^{\rm Rad}_{ab}\sim\sum_{\ell}\ell^3 \mathsf{P}_\ell(\cos\gamma)\sim\sum_\ell\ell^{5/2}
\end{equation}
in the half-spacetime containing the nonzero $b^\lessgtr_{j\ell m}$. This sum manifestly diverges, even more strongly than did the Hertz potential with nonzero $b^\lessgtr_{j\ell m}$. We conclude again that to perform a reconstruction using an expansion in harmonics, we must work pointwise in the separate regions $r<r_p$ and $r>r_p$, with all $b^\lessgtr_{j\ell m}$ set to zero in whichever region we work in. So let us now do just that. 

(a) \emph{No-string case.}
First consider the no-string case, where $H(r)=H^<(r)\theta(r_p-r)+H^>(r)\theta(r-r_p)$ with all $b^\lessgtr_{j\ell m}=0$ in Eq.~\eqref{Psi}. We work strictly off the sphere $\mathcal{S}$, ignoring any delta functions that arise from differentiating the Heaviside functions. Including such delta functions in $h^{\rm Rad}_{\alpha\beta}$ would simply force $h^{\rm Cmpl}_{\alpha\beta}$ to include terms that exactly cancel them. By ignoring them, we make our calculation equivalent to reconstructing the regular half of each half-string MP and then gluing the two together at $\mathcal{S}$. 

With that idea in mind, we find
\begin{align}
h^{\ell m}_{ab} &= \ell_a\ell_b C_{\ell m}(r),\label{hab_mode}\\
h^{\ell m}_{a} &= -\ell_a r C_{\ell m}(r)\left[\frac{1}{\ell}\theta(r-r_p)-\frac{1}{\ell+1}\theta(r_p-r)\right],\\
h^{\ell m} &= \frac{2r^2C_{\ell m}(r)}{\ell(\ell+1)}.\label{hAB_mode}
\end{align}
The common function of $r$ is
\begin{equation}
C_{\ell m}(r)=\frac{8\pi\mu}{2\ell+1}Y^*_{\ell m}(\theta^A_p)\frac{r_<^\ell}{r_>^{\ell+1}},
\end{equation}
where $r_<=\text{min}(r,r_p)$ and $r_>=\text{max}(r,r_p)$. We can see that the modes of $h_{ab}$ and $h_{AB}$ are continuous at $r=r_p$, while the modes of $h_{aA}$ are discontinuous.

The sum of modes can be evaluated analytically using the generating function
\begin{equation}
\sum_{\ell\geq2}\frac{r_<^\ell}{r_>^{\ell+1}}\mathsf{P}_\ell(\cos\gamma) = \frac{1}{R}-\frac{1}{r_>}-\frac{r_< \cos\gamma}{r_>^2},
\end{equation}
where 
\begin{equation}
R=\sqrt{r^2+r_p^2-2rr_p\cos\gamma}
\end{equation}
is the distance from the particle. The end result is
\begin{align}
h^{\rm Rad}_{ab} &= 2\mu\ell_a\ell_b \left(\frac{1}{R}-\frac{1}{r_>}-\frac{r_< \cos\theta}{r_>^2}\right),\label{hRadab}\\
h^{\rm Rad}_{a\theta} &= -2\mu\ell_a\sin\theta\Bigg\lbrace\frac{r_p}{r}\left[1-\frac{r^2+rR}{R(R+r-r_p \cos\theta)}\right]\theta^+\nonumber\\
&\quad+\frac{r}{r_p}\left[-\frac{r}{2r_p}+\frac{rr_p}{R(R+r_p-r\cos\theta)}\right]\theta^-\Bigg\rbrace,\label{hRadaA}\\
h^{\rm Rad}_{AB} &= 2\mu r^2\Bigg[\Omega_{AB}\left(\frac{1}{R}-\frac{1}{r_>}\right)\nonumber\\
&\quad
+\d^\th_A\d^\th_B\frac{2(R-r_>+r_<\cos\theta)(r_<-r_>\cos\theta)}{r_< r_> R\sin^2\theta}\nonumber\\
&\quad-\d^\vf_A\d^\vf_B\frac{2\cos\theta(R-r_>+r_<\cos\theta)}{r_> r_<}\Bigg],\label{hRadAB}
\end{align}
and $h^{\rm Rad}_{a\varphi}=0=h^{\rm Rad}_{\theta\varphi}$, where $\theta^\pm=\theta[\pm(r-r_p)]$. In these expressions, we have placed the particle at the pole $\theta=0$, making $\gamma=\theta$. While the MP components appear to contain terms with $\ell=0,1$ symmetry, those terms are exactly cancelled by pieces of, e.g., $1/R$.

With this result for the reconstructed MP, we can return to, and reconfirm, the endpoint of our local analysis in Sec.~\ref{Fermi-analysis}: the local form of the MP near the particle in a completed radiation gauge. To obtain the local form near the particle, we first transform to Cartesian coordinates centered at the particle's position, making $R$ the new radial coordinate. Expanding for $R\ll r_p$ and picking off the leading-order term allows us to directly compare to the results of Sec.~\ref{Fermi-analysis}. As expected, we find that \emph{our reconstructed no-string MP precisely recovers the local no-string MP} shown in Table I. This also validates our assumption in Sec.~\ref{Fermi-analysis} that the completion term in the MP does not contribute to the leading-order local singularity.

(b) \emph{Half-string case.}
From the no-string MP one can read off the two half-string MPs. The individual modes in the regular half-spacetime of each MP can be found from Eqs.~\eqref{hab_mode}--\eqref{hAB_mode}. Once the modes are summed in the regular half-spacetime, the result can be analytically continued to the half-spacetime containing the string. To condense the discussion, we give the result for $h^{\rm Rad-}_{\alpha\beta}$ only. Taking the evaluated sum from Eqs.~\eqref{hRadab}--\eqref{hRadAB}, we find
\begin{align}
h^{\rm Rad-}_{ab} &= 2\mu\ell_a\ell_b \left(\frac{1}{R}-\frac{1}{r_p}-\frac{r \cos\theta}{r_p^2}\right),\\
h^{\rm Rad-}_{a\theta} &= 2\mu\ell_a\frac{r\sin\theta}{r_p}\left[\frac{r}{2r_p}-\frac{rr_p}{R(R+r_p-r\cos\theta)}\right],\\
h^{\rm Rad-}_{AB} &= 2\mu r^2\Bigg[\Omega_{AB}\left(\frac{1}{R}-\frac{1}{r_p}\right)\nonumber\\
&\quad
+\d^\th_A\d^\th_B\frac{2(R-r_p+r\cos\theta)(r-r_p\cos\theta)}{r r_p R\sin^2\theta}\nonumber\\
&\quad-\d^\vf_A\d^\vf_B\frac{2\cos\theta(R-r_p+r\cos\theta)}{r r_p}\Bigg],
\end{align}
and $h^{\rm Rad-}_{a\varphi}=0=h^{\rm Rad-}_{\theta\varphi}$. Again, for these expressions the particle is at the pole $\theta=0$. The MP is regular for $r<r_p$, but for all $r>r_p$ it diverges on the half-string at $\theta=0$. As with the no-string MP, after expanding for $R\ll r_p$ this MP agrees with the local result displayed in Table I.


\subsection{Completion}
We now seek to complete the MP by adding the field $h^{\rm Cmpl}_{\alpha\beta}$. The typical picture of reconstruction (in the flat-space or Schwarzschild case) is that it provides the whole of the $\ell>1$ contribution to the MP, leaving $h^{\rm Cmpl}_{\alpha\beta}$ to contain only $\ell=0,1$ contributions. This picture is inspired by Wald's theorem~\cite{Wald:73} stating that for vacuum perturbations of Kerr, $\psi_0$ or $\psi_4$ contain all the gauge-invariant information about the MP except corrections to the Kerr mass and spin parameters.\footnote{Wald also identified gauge-invariant perturbations corresponding to linearized terms in C-metrics and Kerr-NUT metrics, but they can be excluded due to their singularities~\cite{Wald:73,Keidl-etal:10}.} Here we show that problems with this description can arise in the no-string case, but they are obviated by an appropriate choice of gauge. We also briefly discuss the situation in the half-string case, but we leave the main analysis to the sequel~\cite{Barack-etal:14}.

\subsubsection{No-string case}
Recall that at the end of Sec.~\ref{metric-summary}, we noted that gluing together the regular halves of two half-string gauge vectors yields a valid solution to the EFE, but gluing together the regular halves of two half-string MPs does not. In the foregoing calculation, we have done the gluing at the level of the MP, not at the level of a gauge vector. We now show that this leads to the necessity of including $\ell>1$ terms in $h^{\rm Cmpl}_{\alpha\beta}$. However, we also show that a suitable gauge can be chosen to confine these modes to the sphere $\mathcal{S}$, allowing us to preserve the usual picture of adding only $\ell=0,1$ terms in $h^{\rm Cmpl}_{\alpha\beta}$ off $\mathcal{S}$.  

Concretely, we wish to find a field $h^{\rm Cmpl}_{\alpha\beta}$ satisfying Eq.~\eqref{EFE_comp}, subject to staticity as well as regularity conditions at the origin and infinity. Substituting Eqs.~\eqref{hrec_ab}--\eqref{hrec_AB} with \eqref{hab_mode}--\eqref{hAB_mode} into Eq.~\eqref{EFE_comp}, using the tensor-harmonic decomposition of the Einstein tensor given in Ref.~\cite{Martel-Poisson:05}, and noting the harmonic expansion~\eqref{delta_expansion} for the stress-energy tensor, we find that $h^{\rm Cmpl}_{\mu\nu}$ must satisfy
\begin{align}
\delta G_{tt}[h^{\rm Cmpl}_{\mu\nu}] &= \frac{8\pi\mu}{r_p^2}\delta(r-r_p)\sum_{\ell<2,m}Y^*_{\ell m}(\theta_p^A)Y_{\ell m},\label{EFEtt_comp}\\
\delta G_{tr}[h^{\rm Cmpl}_{\mu\nu}] &= \frac{4\pi\mu}{r_p^2}\delta(r-r_p)\sum_{\ell\geq2,m}Y^*_{\ell m}(\theta^A_p)Y_{\ell m},\label{EFEtr_comp}\\
\delta G_{tA}[h^{\rm Cmpl}_{\mu\nu}] &= 4\pi\mu\delta'(r-r_p)\sum_{\ell\geq2,m}\frac{Y^*_{\ell m}(\theta^B_p)}{\ell(\ell+1)}Y_A^{\ell m}.\label{EFEtA_comp}
\end{align}
All other components of $\delta G_{\alpha\beta}[h^{\rm Cmpl}_{\mu\nu}]$ vanish. Equation~\eqref{EFEtt_comp} is simply the $\ell=0,1$ piece of the original EFE $\delta G_{\alpha\beta}[h_{\mu\nu}]=8\pi T_{\alpha\beta}$; $\delta G_{tt}[h^{\rm Rad}_{\mu\nu}]$ has exactly canceled the $\ell\geq2$ terms in $8\pi T_{\alpha\beta}$. The nonzero sources in the $tr$ and $tA$ equations arise from the fact that the reconstructed MP does not solve the $\ell\geq2$ piece of the original EFE.

What must be added to $h^{\rm Rad}_{\alpha\beta}$ to satisfy the $\ell\geq2$ pieces of the EFE? To answer that, we expand $h^{\rm Cmpl}_{\mu\nu}$ in terms of even-parity tensor harmonics:
\begin{align}
h^{\rm Cmpl}_{ab} &= \sum_{\ell\geq0, m}j^{\ell m}_{ab}Y^{\ell m},\\
h^{\rm Cmpl}_{aA} &= \sum_{\ell\geq1, m}j^{\ell m}_{a}Y_A^{\ell m},\\
h^{\rm Cmpl}_{AB} &= r^2\sum_{\ell\geq2, m}j^{\ell m}Y_{AB}^{\ell m}+r^2\sum_{\ell\geq0, m}K^{\ell m}\Omega_{AB}Y_{\ell m};
\end{align}
since no odd-parity harmonics appear in Eqs.~\eqref{EFEtt_comp}--\eqref{EFEtA_comp}, coefficients of odd-parity harmonics would vanish. Rather than substituting this expansion into Eqs.~\eqref{EFEtr_comp} and \eqref{EFEtA_comp} and directly solving for the coefficients $j^{\ell m}_{ab}$, $j^{\ell m}_{a}$, $j^{\ell m}$, and $K^{\ell m}$, which would require specifying a gauge, we instead solve the linearized Einstein equation in the form presented in Ref.~\cite{Martel-Poisson:05}, where the Einstein tensor is written in terms of certain gauge-invariant quantities $\tilde j^{\ell m}_{ab}$ and $\tilde K^{\ell m}$, to be defined below. More specifically, we seek time-independent solutions regular at $r=0$ and $r\to\infty$. With those specifications, the solution to Eqs.~\eqref{EFEtt_comp}--\eqref{EFEtA_comp} reads
\begin{align}
\tilde j^{\ell m}_{tt} = \tilde j^{\ell m}_{rr} = \tilde K^{\ell m}=0,\label{K}\\
\tilde j^{\ell m}_{tr} = \frac{8\pi\mu}{\ell(\ell+1)}Y^*_{\ell m}(\theta^A_p)\delta(r-r_p)\label{jtr}
\end{align}
for all $\ell\geq2$. The sum over modes can be evaluated to find
\begin{equation}\label{RW}
\sum_{\ell\geq2,m}\tilde j^{\ell m}_{tr}Y_{\ell m} = -2\mu\left[1+\frac{3}{2}\cos\gamma+\ln\sin^2\frac{\gamma}{2}\right]\delta(r-r_p).
\end{equation} 
It is worth observing that this delta function term diverges logarithmically at $\gamma=0$, where the particle sits. 

These results show that in terms of its gauge-invariant content, the completion of the $\ell\geq2$ modes of the MP is restricted to a distribution on the sphere $\mathcal{S}$. To determine how that content is expressed in any given gauge, we must use the definitions of $\tilde j^{\ell m}_{ab}$ and $\tilde K^{\ell m}$ in terms of components of the MP. For time-independent fields, the relationships are~\cite{Martel-Poisson:05}
\begin{align}
\tilde j^{\ell m}_{tt} &= j^{\ell m}_{tt},\\
\tilde j^{\ell m}_{tr} &= j^{\ell m}_{tr}-\partial_rj^{\ell m}_{t},\\
\tilde j^{\ell m}_{rr} &= j^{\ell m}_{rr}-2\partial_rj^{\ell m}_{r}+r^2\partial^2_r j^{\ell m}+2r\partial_rj^{\ell m},\\
\tilde K^{\ell m} &= K^{\ell m}-\frac{2}{r}j^{\ell m}_r+r\partial_r j^{\ell m}+\frac{1}{2}\ell(\ell+1)j^{\ell m}.
\end{align}
If we were to impose the ingoing radiation gauge condition on the $\ell\geq2$ modes of $h^{\rm Cmpl}_{\alpha\beta}$, we would find that those modes contain a nonvanishing trace, unlike the reconstructed MP (cf. the analysis of Price \emph{et al}.~\cite{Price-Shankar-Whiting:07}, which showed that the radiation gauge condition and trace-free condition are incompatible in the presence of matter). We would also find that the trace term introduces a Dirac-delta-type half-string into the MP; although the gauge-invariant content of the MP is restricted to a singularity on $\mathcal{S}$, the gauge condition extends that singularity out to infinity along a string. We \emph{can}, however, find at least one choice of gauge that keeps the content confined to $\mathcal{S}$ even in the individual components of the MP: the Regge-Wheeler gauge, in which $\tilde j^{\ell m}_{ab}=j^{\ell m}_{ab}$, $\tilde K^{\ell m}=K^{\ell m}$, and $j^{\ell m}_a=0=j^{\ell m}$. Hence we adopt the Regge-Wheeler gauge for the $\ell\geq2$ completion terms.

This leaves only the $\ell=0,1$ piece of the MP, which is the static solution to Eq.~\eqref{EFEtt_comp}, subject to regularity at $r=0$ and $r\to\infty$. We can, for example, take the $\ell=0$ and 1 terms from the Lorenz-gauge solution to Eq.~\eqref{EFE}, given by $h_{\alpha\beta}=\frac{2\mu}{R}(g_{\alpha\beta}+2\d^t_\alpha \d^t_\beta)$, where $R$ is the distance from the particle. The $\ell=0,1$ piece of the solution then reads
\begin{align}
h^{\rm Cmpl,\ell=0,1}_{ab} &= 2\mu\delta_{ab}p(r),\label{habLow}\\ 
h^{\rm Cmpl,\ell=0,1}_{AB} &= 2\mu r^2\Omega_{AB}p(r),\label{hABLow}
\end{align}
and $h^{\rm Cmpl,\ell=0,1}_{aA}=0$, where
\begin{align}
p(r) &= \left(\frac{1}{r_p}+\frac{r\cos\gamma}{r_p^2}\right)\theta(r_p-r)\nonumber\\
&\quad +\left(\frac{1}{r}+\frac{r_p\cos\gamma}{r^2}\right)\theta(r-r_p).
\end{align}
We can see directly, and calculations confirm, that the terms for $r>r_p$ carry gauge-invariant content. The $2\mu/r$ terms contribute an Arnowitt-Deser-Misner mass $\mu$; the $2\mu r_p\cos\gamma/r^2$ terms contribute a mass dipole moment, created by the displacement of the center of mass from the origin.\footnote{The mass dipole can, of course, be removed by a global coordinate transformation that puts the particle at the origin. But it cannot be removed via a gauge transformation, because the distance $r_p$ is a background quantity unrelated to the perturbative quantity $\mu$.} Contrasting with this, in the region $r<r_p$ the $\ell=0,1$ terms are pure gauge. They contribute nothing to the curvature inside $\mathcal{S}$, but they are required to satisfy the junction condition at $r=r_p$, as determined by Eq.~\eqref{EFEtt_comp}. Put another way, they are pointwise gauge for $r<r_p$, but distributionally, due to the presence of the Heaviside function multiplying them, they contribute to the curvature at $r=r_p$.

With this, the completion process has come to an end. The final MP is
\beq
h_{\alpha\beta}=h^{\rm Rad}_{\alpha\beta}+h^{\rm Cmpl,\ell>1}_{\alpha\beta}+h^{\rm Cmpl,\ell=0,1}_{\alpha\beta}.
\eeq
The reconstructed term $h^{\rm Rad}_{\alpha\beta}$ is given in a traceless ingoing radiation gauge in Eqs.~\eqref{hRadab}--\eqref{hRadAB}, the $\ell>1$ completion term is given [according to Eq.~\eqref{RW}] in the Regge-Wheeler gauge by
\beq
h^{\rm Cmpl,\ell>1}_{tr}=-2\mu\left[1+\frac{3}{2}\cos\gamma+\ln\sin^2\frac{\gamma}{2}\right]\delta(r-r_p),
\eeq
with all other components vanishing, and the $\ell=0,1$ completion term is given in Eqs.~\eqref{habLow} and \eqref{hABLow}. It is easy to show (mode by mode) that this completed solution is related to the Lorenz gauge solution by a gauge transformation that is discontinuous across $\mathcal{S}$.

Several important lessons can be gleaned from this completion process. First, while something must be added to complete the $\ell\geq2$ piece of the solution, that ``something'' is restricted to the sphere $\mathcal{S}$. Off the sphere, the reconstruction procedure immediately yields the correct $\ell\geq2$ piece of a no-string vacuum solution. Second, whatever $\ell=0,1$ terms are to be added must simply satisfy the $\ell=0,1$ piece of the EFE. Those $\ell=0,1$ terms are  pure gauge inside the sphere, while outside the sphere they contain gauge invariant monopole and dipole moments.  

\subsubsection{Half-string case}
In the regular half of each half-string MP, the completion can be immediately performed on the basis of the no-string results. We have shown that for $r>r_p$, the reconstructed no-string solution is equal (up to gauge) to the Lorenz-gauge solution once we complete the no-string MP with the mass and mass dipole terms in Eqs.~\eqref{habLow} and \eqref{hABLow}. It follows that the half-string MP $h^+_{\alpha\beta}$ in its regular region $r>r_p$ can be completed by the addition of those same mass and mass dipole terms. Similarly, we have shown that for $r<r_p$, the reconstructed no-string solution is equal up to gauge to the Lorenz-gauge solution. It follows that the half-string MP $h^-_{\alpha\beta}$ in its regular region $r<r_p$ is already complete in the form obtained by reconstruction. 

With the tools at hand in the present paper, we cannot easily complete the half-string MPs globally, because we cannot work mode by mode in the irregular region of each solution, making it difficult to directly solve the EFE for $h^{\rm Cmpl}_{\alpha\beta}$. But for a calculation of the GSF, we need only have the modes of a half-string solution in its regular half, so a global solution is not required.

\subsection{Comments on reconstruction and completion in black-hole spacetimes}
Recapitulating our findings,
\begin{itemize}
\item If a regular MP is reconstructed mode by mode either inside or outside the particle's orbit, that MP is the regular half of an (uncompleted) half-string solution (or equivalently, one half of an uncompleted no-string solution).   
\item If a regular MP is reconstructed mode by mode both inside and outside the particle's orbit \emph{and} taken to be part of the same global MP, it must be interpreted as an (uncompleted) no-string solution.
\item Outside the sphere intersecting the particle at a given instant of time, the ``$+$'' half-string solution (or the no-string solution) can be completed by adding a stationary vacuum MP determined by the system's physical multipole moments.
\item Inside that sphere, the completion of the ``$-$'' half-string solution (or the no-string solution) is pure gauge.
\end{itemize}
Though the situation is more complicated \emph{on} the sphere, that need not concern us. In practical calculations of the GSF, we will always be calculating modes off the particle and then taking limits to it from one or both sides of the sphere. Therefore, for practical purposes, we may simply ignore the fact that completing the no-string solution requires delta functions on the sphere, and we may likewise ignore the fact that we cannot easily obtain a completed half-string solution on the irregular side of the sphere.

In practice, these calculations will be performed in black hole spacetimes, and we must extrapolate our results to those contexts. We conjecture that the first two of the bulleted points above will remain true at each given instant of coordinate time in Kerr, in terms of Boyer-Lindquist coordinate spheres. However, since the mode-by-mode reconstruction in Kerr must be performed in the frequency domain, for non-circular orbits the ``position'' of the particle (and therefore the surface of discontinuity $\mathcal{S}$ or the endpoint of the half-string) will become smeared over space, which could complicate matters.

The last two bulleted points, pertaining to the completion of the MP, are perhaps more delicate. In Schwarzschild, the completion has been performed in the past in a manner analogous to that described here, by adding a solution to the $\ell=0$ and 1 pieces of the EFE~\cite{Shah-etal:11}. But in Kerr, the situation is perplexed by the fact that we cannot separate the EFE into $\ell$ modes. 

These issues will be discussed in more detail in a future paper~\cite{Barack-etal:14}, where we will present a new MP completion procedure in Kerr. (See also Ref.~\cite{Shah-etal:12}.)  


\section{Summary and concluding remarks}

Let us summarize. We have analyzed the local form of the MP near a point particle in the class of (either ingoing or
outgoing) completed radiation gauges. Our analysis assumed very little about the
particle's motion or the background geometry, and it is applicable, in
particular, for arbitrary motion in Kerr spacetime. We found that each completed
radiation gauge belongs to one of three categories, according to whether
the MP is singular along half a radial ray, singular along a full
radial ray, or discontinuous along a surface through the particle. Table
\ref{metric} summarized the local form of the singularity (in the
language of local Fermi-like coordinates) for gauges in each of the
three categories.

We then argued that full-string solutions are too singular to be
considered via the standard mode-by-mode CCK reconstruction procedure.
On the other hand, we argued, the procedure can be used to construct the
regular sides of half-string solutions, and (equivalently) the
no-string solution on either side of the discontinuous surface. We
illustrated this with an explicit application of the CCK method to
(analytically) reconstruct the MP for a static particle in flat space.

Equipped with the above understanding, we moved on to consider the
definition of the GSF in the various types of completed radiation gauges. We
showed that none of the completed radiation gauges falls within the Barack-Ori class of gauges, which are related to the Lorenz gauge by a continuous transformation. However, we also showed that completed radiation gauges
of all three types (half-, full-, and no-string) fall within the class of gauges for which the GSF is well defined in terms of a certain 
integral over a small sphere around the particle. (We generalized this class beyond, e.g., the Gralla-Wald class~\cite{Gralla-Wald:11}, to allow irregularities on surfaces extending off the particle.)

We then devised two practical mode-sum schemes for the GSF, each using as input
an MP in a completed radiation gauge. The first scheme was tailored to the half-string case and formulated within the
Barack-Ori class of gauges. The GSF was expressed in an LL
gauge belonging to that class, which was obtained via a local gauge deformation of a half-string MP in a completed radiation gauge. We derived a mode-sum formula for the GSF in the LL gauge, given
in Eq.~\eqref{MS4}. The formula retains the standard Lorenz-gauge form, but
with modified regularization parameters. We showed how to derive these
parameters, and for a specific (analytically given) choice of gauge deformation, we gave the explicit parameter values for arbitrary
geodesic motion in Schwarzschild and for circular and equatorial
geodesics in Kerr. For its numerical input, the mode-sum formula requires
the $\ell$-modes of the full force associated [via
Eq.~\eqref{full_force}] with a half-string radiation gauge solution. One can use
here either of the $\pm$ half-string solutions, together with the
appropriate $\delta D_{\alpha}^{\pm}$.

Our second scheme was formulated directly in a no-string completed radiation
gauge, without a gauge deformation. To achieve this, we appealed to the general formulation of the GSF involving integration over a sphere around the particle. The mode-sum formula for the GSF in a no-string gauge is given in Eq.~\eqref{mode-sum-no-string}. It has
the Lorenz-gauge form, this time without any modification to the
regularization parameters, but the input full modes of the force must be
computed by averaging the two values obtained from either radial side
of the particle, and likewise the parameters must be replaced by their average values. In effect, our no-string mode-sum formula is precisely
the average of the two half-string formulas. Our derivation (along with our discussion of GSF in gauges with irregularities off the particle), provides a formal justification for taking this average, and gives a physical meaning to it. We have also derived a mode-sum formula directly in the half-string gauges, without local deformation, but in this case we consider the representation of the particle's motion to be less intuitively meaningful than that described by the LL formalism.
 
It should be noted that the final GSF value obtained using the ``$+$''
half-string solution should by no means agree with the final GSF value
obtained using the ``$-$'' solution, or with the one obtained using the
no-string solution (which is the average of the former two): the three
GSF values are given in different gauges. We recall here that a complete
gauge-invariant description of the motion involves the GSF as well as
the associated MP, both given in the same gauge. Our GSF schemes would
have not been complete without a prescription for obtaining the MP in
the relevant gauges. In the case of the half-string scheme, the
prescription is simple: take the CCK-reconstructed (and completed)
half-string radiation-gauge MP, and add to it the gauge perturbation
$2\xi^0_{(\alpha;\beta)}$, where $\xi^0_\alpha$ is given in Eqs.~\eqref{xi0-coords}--\eqref{xi0xi-hs}; this perturbation can be attenuated in any
convenient way to suppress its support away from the particle. This will
produce an LL MP in a gauge corresponding to the one in which
the GSF is given. In the case of the no-string scheme, the situation is a bit more subtle: the force is given in the same gauge as the original reconstructed (and completed) MP, making it simpler in one sense, but the MP in that gauge has a discontinuity across (and
perhaps also a delta function on) a surface through the particle, which might complicate calculations of some gauge-invariant aspects of
the motion.

Let us also stress the issue of {\em off-worldline extension} relevant
to the definition of the various quantities that go into the mode-sum
formulas. As we have explained, the values of both the multipole modes of the
full force and the regularization parameters are sensitive to the way we
define the underlying vector fields through extensions of their values
off the particle's (zeroth-order) worldline. The freedom in choosing the extension can
be used to one's benefit (e.g., to try minimize the degree of coupling
between the original Teukolsky-mode multipoles arising from the CCK
reconstruction and the spherical-harmonic multipoles that go into the
mode-sum formula), but one must be careful to use the same extension for
both the full modes and the parameters. We emphasize that the particular
parameter values we give explicitly in the half-string case come with a specific
extension label, and should only be used in conjunction with full modes
computed in that same extension. If a different extension is deemed more
practical, one would need to rederive the parameters corresponding to
that extension. Our discussion provides sufficient detail to enable the
calculation of the parameters in any such extension. In the no-string case, our results are independent of the extension, and the Lorenz-gauge parameter values appearing in our final mode-sum formula may be used with any of the extensions for which they have been calculated~\cite{Barack-Ori:03b,Barack:09}.

Modulo the choice of extension, both our mode-sum schemes, Eqs.~\eqref{MS4} and \eqref{mode-sum-no-string}, can be implemented
immediately using the existing computational infrastructure developed by Shah {\it
et al.}~\cite{Shah-etal:12}, which implements the CCK reconstruction.
This infrastructure currently exists only for orbits (in Kerr) that are both
circular and equatorial, but a generalization to more generic orbits should
be feasible. The choice of off-worldline extension will need to be considered carefully, and in the half-string case, corresponding parameters will need to be computed. In the no-string case, where only Lorenz-gauge parameters are required, no calculations of new parameters will be necessary (unless an extension is chosen for which Lorenz-gauge parameters are currently unavailable). One would need, in that case, to reconstruct the regular half-string solutions on either side of the particle, but this should not result in 
doubling the computational cost, since the computationally expensive
part of the procedure---namely, obtaining homogeneous solutions to the
Teukolsky equation---is shared by the two half-string reconstructions.
The only major remaining open issue is that of the {\em completion} of the
CCK-reconstructed MP. We have described some facets of this problem in our analysis in flat spacetime, and we will address the issue thoroughly in a forthcoming paper \cite{Barack-etal:14}, where we will prescribe a procedure for completing the
solutions for generic bound orbits in Kerr spacetime.

\begin{acknowledgments}
We thank Amos Ori, John Friedman, James Vickers, Soichiro Isoyama, and Abhay Shah for helpful discussions. The research leading to these results received funding from the European Research Council under the European Union's Seventh Framework Programme (FP7/2007-2013)/ERC grant agreement no.~304978. AP acknowledges additional support from the Natural Sciences and Engineering Research Council of Canada. CM acknowledges support from CONACyT through grant number 214690. LB acknowledges additional support from STFC in the UK through grant number PP/E001025/1.
\end{acknowledgments}


\appendix

\section{Motion as defined in matched asymptotic expansions}\label{Rad-force}
Deriving an equation of motion necessitates first defining the motion that the equation is meant to describe. The standard method of accomplishing that in the GSF context is matched asymptotic expansions~\cite{Mino-Sasaki-Tanaka:97,Gralla-Wald:08, Pound:10a, Pound:10b, Gralla:11, Pound:12a, Pound:12b, Gralla:12}. This appendix reviews the method and a particular way of using its results to obtain equations of motion in non-Lorenz gauges. It also describes a way to rigorously interpret the method's results in the case of gauges with irregularities off the particle.
 
\subsection{Center of mass}
In the method of matched asymptotic expansions, one assumes that the particle is actually a small, compact object. Let ${\sf g}_{\mu\nu}(x,\varepsilon)$ be the exact solution to the full, nonlinear Einstein equations for the spacetime including that small object, where $\varepsilon$ is used to count powers of $\mu$ but can be set to 1 at the end of the calculation. Also let $\mathcal{R}$ denote the other lengthscales of the system, which are much larger than $\mu$.
 
Suppose we work in the local Fermi-like coordinates $(\tau,x^a)$ centered on $\Gamma$, introduced in Sec.~\ref{local_coords}. We do not begin with any definite association between $\Gamma$ and the bulk motion of the small object, but we start by assuming that the object is only a small distance from $\Gamma$. At distances $s\gg\mu$, far from the object, one can expand the exact metric as ${\sf g}_{\mu\nu}=g_{\mu\nu}+\varepsilon h^{(1)}_{\mu\nu}+\varepsilon^2 h^{(2)}_{\mu\nu}+O(\varepsilon^3)$, which is the form of the expansion assumed throughout the earlier sections of this paper. We call this the outer expansion. In this expansion the first-order perturbation, $h^{(1)}_{\mu\nu}\equiv h_{\mu\nu}$, is that of a point particle moving on $\Gamma$ in the background $g_{\alpha\beta}$~\cite{Gralla-Wald:08}. 
 
At distances $s\sim\mu$, near the object, the outer expansion fails, because in that region the metric is dominated not by $g_{\mu\nu}$, but by the gravity of the small object. The method of matched asymptotic expansions overcomes that problem by adopting a second expansion near the object. Rather than taking the limit of small mass and size by keeping external distances fixed and sending the mass and size to zero, we take the limit by keeping the mass and size of the object fixed while sending other distances to infinity. This second limit is achieved by writing the metric components in terms of scaled spatial variables $\bar x^a = x^a/\mu$. Holding these scaled variables fixed while expanding for small $\mu$, we have
\beq
{\sf g}_{\mu\nu}(\tau,\bar x^a,\varepsilon) = g^{(0)}_{\mu\nu}(\tau,\bar x^a)+\varepsilon g^{(1)}_{\mu\nu}(\tau,\bar x^a)+O(\varepsilon^2),
\eeq
where $g^{(0)}_{\mu\nu}$ is the metric of the small body were it isolated. We call this the inner expansion.
 
The motion of the small object is defined by examining the metric in a buffer region $\mu\ll s\ll \mathcal{R}$ around the body. Because $s\gg\mu$, we can expect the outer expansion to be valid here; because $s\ll\mathcal{R}$, we can expect the inner expansion to also be valid here; and because they are both expansions of the same metric ${\sf g}_{\mu\nu}$, the two expansions must agree. This allows us to extract information about the outer expansion from information about the inner expansion in the buffer region. The first thing we infer is that the existence of an inner expansion requires the outer expansion to have the local form~\cite{Gralla-Wald:08,Pound:10a}
\begin{align}
h_{\mu\nu} \sim 1/s,\qquad h^{(2)}_{\mu\nu} \sim 1/s^2\label{h2}
\end{align}
near $\Gamma$. Any terms more singular would correspond to negative powers of $\varepsilon$ in the inner expansion.
 
Furthermore, we note that while the buffer region is asymptotically small from the perspective of the outer expansion, it corresponds to asymptotic infinity from the perspective of the inner expansion. Using that fact, we can define multipole moments of the inner expansion, and those multipole moments become the kernels of the outer expansion. As an example, we note that the Arnowitt-Deser-Misner mass of $g^{(0)}_{\mu\nu}(\tau,\bar x^a)$ in the inner expansion defines the point particle mass $\mu$ in the outer expansion~\cite{Gralla-Wald:08}. 
 
For the particular purpose of defining the object's motion, we will be interested in the mass dipole moment of the object's unperturbed metric:
\beq
M^a = \frac{3}{8\pi}\lim_{\bar s\to\infty}\int g^{(0)}_{\tau\tau}(\tau,\bar x^a) n^a dS,\label{dipole_rtoinfty}
\eeq
where the integration is over a sphere of radius $\bar s$ around the object, and $n^a$ is the unit vector $x^a/s=\bar x^a/\bar s$ normal to the sphere. Using this formula, we can meaningfully define the object's motion. Per unit mass, a mass dipole moment bears the interpretation of the position of the center of mass relative to the origin of the coordinates. Since we work in coordinates centered on the worldline $\Gamma$, the mass dipole per unit mass can be interpreted as the position relative to $\Gamma$. More explicitly, imagine the object's motion is described by a worldline $z^\alpha(\tau,\varepsilon)$ with the expansion
\beq
z^\alpha(\tau,\varepsilon) = z_0^\alpha(\tau) + \varepsilon z_1^\alpha(\tau)+O(\varepsilon^2),
\eeq
where $z_0^\alpha(\tau)$ are the coordinates on the geodesic $\Gamma$, and $z_1^\alpha(\tau)$ is a vector field on $\Gamma$. Then we define the leading-order correction $z^\alpha_1$ to the object's position as\footnote{An alternative method, called the self-consistent method, instead defines a mass dipole relative to the accelerated worldline $z^\alpha(\tau,\varepsilon)$, deriving an equation of motion for $z^\alpha$ by ensuring that mass dipole vanishes~\cite{Pound:10a}. That method is designed to maintain uniform accuracy on long timescales by avoiding an expansion of $z^\alpha(\tau,\varepsilon)$. Here, for simplicity, we work with the expanded worldline.}
\beq
z^\alpha_1\equiv \frac{M^\alpha}{\mu},
\eeq
where $M^\alpha\equiv e^\alpha_a M^a$. This was the method used by Gralla and Wald in the first rigorous derivation of the first-order GSF, and modifications of it have since been the basis for derivations of the second-order GSF~\cite{Pound:12a,Gralla:12}.
 
We can relate $M^\alpha$ to the perturbations in the outer expansion by appealing to the assumed agreement between the two expansions in the buffer region. In that region, we can expand $g^{(0)}_{\tau\tau}$ as 
\beq
g^{(0)}_{\tau\tau}(\tau,\bar s,n^a) = \frac{1}{\bar s}g^{(0,1)}_{\tau\tau}(\tau) +\frac{1}{\bar s^2}g^{(0,2)}_{\tau\tau}(\tau,n^a)+O(\bar s^{-3}). 
\eeq
The term $\frac{1}{\bar s}g^{(0,1)}_{\tau\tau}$ does not contribute to Eq.~\eqref{dipole_rtoinfty}, because it always takes the Schwarzschild form $\frac{2\mu}{\bar s}$. Therefore only the term $\frac{1}{\bar s^2}g^{(0,2)}_{\tau\tau}$ contributes. Written in terms of the unscaled variable $s$, this term becomes $\frac{\mu^2}{s^2}g^{(0,2)}_{\tau\tau}$, and we can see it must correspond to a $1/s^2$ term in $h^{(2)}_{\tau\tau}$ in the outer expansion. Therefore, noting Eq.~\eqref{h2}, we can write
\beq
\mu z_1^a=M^a = \frac{3}{8\pi}\lim_{s\to 0}\int h^{(2)}_{\tau\tau} n^a dS,\label{Ma}
\eeq
where now the integral is over a sphere of radius $s$.

\subsection{Equation of motion in sufficiently regular gauges}\label{motion-arbitrary-gauge}
In Ref.~\cite{Gralla-Wald:08}, a first-order self-forced equation of motion was found by solving the Einstein equation to sufficiently high order to establish a formula for $\partial_\tau^2M^a$. The result was\footnote{Throughout this paper, we assume the small object is nonspinning at leading order. Otherwise, a Papapetrou spin force would appear in Eq.~\eqref{motion-in-Lorenz-app}}
\beq
\mu\frac{D^2z_{1{\rm Lor}}^\alpha}{d\tau^2} = -\mu R^\alpha{}_{\mu\beta\nu}u^\mu z_{\rm 1Lor}^\beta u^\nu + F^\alpha_{\rm Lor}.\label{motion-in-Lorenz-app}
\eeq
The first term, $-R^\alpha{}_{\mu\beta\nu}u^\mu z_{\rm 1Lor}^\beta u^\nu$ describes the acceleration due to the background curvature. The second term, $F^\alpha_{\rm Lor}$, is the standard Lorenz-gauge GSF.
 
Using the result \eqref{motion-in-Lorenz-app} for the motion in the Lorenz gauge, we can find the motion in a different gauge by referring to how the mass dipole moment is altered by a gauge transformation. Under a gauge transformation generated by a first-order gauge vector $\xi_\alpha$, the second-order perturbation is altered according to $h^{(2)}_{\mu\nu}\to h^{(2)}_{\mu\nu}+\Delta h^{(2)}_{\mu\nu}$, where 
\begin{align}
\Delta h^{(2)}_{\mu\nu} &= h_{\mu\nu;\rho}\xi^\rho + 2h_{\rho(\mu}\xi^\rho{}_{;\nu)} + \xi^\rho\xi_{(\mu;\nu)\rho} + \xi^\rho{}_{;\mu}\xi_{\rho;\nu}\nonumber\\
                        &\quad +\xi^\rho{}_{;(\mu}\xi_{\nu);\rho},\label{dh2}
\end{align}
where, recall, we have defined $h_{\mu\nu}$ to be the first-order perturbation $h^{(1)}_{\mu\nu}$. We restrict our attention to gauge transformations preserving the form~\eqref{h2} for all $\tau$, to maintain compatibility with the existence of an inner expansion. Straightforward analysis of the transformation laws $\Delta h_{\alpha\beta}=2\xi_{(\alpha;\beta)}$ and \eqref{dh2} shows that this compatibility requirement is satisfied if we impose the following conditions on the asymptotic behavior of $\xi_\alpha$ in the limit $s\to0$:
\begin{enumerate}
\item[(SR1)] $\xi_\tau=f_1(\tau)\ln s+o(\ln s)$,\label{SR_condition1}
\item[(SR2)] $\xi_a=f_2(\tau,n^a)+o(1)$,\label{SR_condition2}
\item[(SR3)] $\tau$ derivatives do not increase the degree of singularity; e.g., $\partial_\tau\xi_\alpha=O(\xi_\alpha)$, 
\item[(SR4)] spatial derivatives increase the degree of singularity by at most one order of $s$; e.g., $\partial_a\xi_\alpha=O(\xi_\alpha/s)$.\label{SR_condition3}
\end{enumerate}
Here ``SR" stands for ``sufficiently regular". For the purpose of ascertaining the scaling with $s$ of the various terms in the gauge transformations of $h_{\mu\nu}$ and $h^{(2)}_{\mu\nu}$, the functions $f_1$ and $f_2$ must be twice differentiable almost everywhere, but they may otherwise be chosen arbitrarily.

Although it might be possible to formulate weaker conditions, it is easy to see that even slightly stronger singularities will generically violate~\eqref{h2}. For example, if $\xi_\tau\sim f_1(\tau,n^b)\ln s$, then terms of order $\frac{\ln s}{s}$ generically arise in $h_{\tau a}$, and if $\xi_a\sim f_2(\tau)\ln s$, then terms of order $\frac{\ln s}{s^2}$ generically arise in $h^{(2)}_{\alpha\beta}$. The conditions on the derivatives are less obvious, but careful inspection of Eq.~\eqref{dh2} reveals their necessity; without them, pathological behavior such as $\partial_\tau\sin(\tau/s)\sim 1/s$ and $\partial_x\sin(x/y^3)\sim 1/s^3$ could arise from the subleading terms in $\xi_\tau$ and $\xi_a$.

All the functions we consider in this paper will satisfy~(SR1)--(SR4). Given these conditions, a simple calculation shows that if we begin in the Lorenz gauge, where $h_{\mu\nu}=\frac{2\mu}{s}\delta_{\mu\nu}+O(1)$, the change in the time-time component of the second-order MP due to $\xi_\alpha$ is
\beq
\Delta h^{(2)}_{\tau\tau} = -\frac{2\mu}{s^2}n^a\xi_a +o(s^{-2}).\label{Delta htt}
\eeq
Of all the terms in Eq.~\eqref{dh2}, $h_{\mu\nu;\rho}\xi^\rho$ is the only one that contributes to this result. Referring to Eq.~\eqref{Ma}, we see that the change in mass dipole moment is
\beq
\Delta M^a = \frac{3}{8\pi}\lim_{s\to 0}\int \Delta h^{(2)}_{\tau\tau} n^a dS.\label{Delta M-app}
\eeq
Therefore $\Delta z_1^a=\Delta M^a/\mu$ reads
\beq
\Delta z_1^a = -\frac{3}{4\pi}\lim_{s\to 0}\int n^a n^b\xi_b d\Omega.  \label{Delta z-app}
\eeq
This is the final formula for the change in position under a gauge transformation. It forms the basis of our discussion in Sec.~\ref{gauges}, where it is reproduced as Eq.~\eqref{Delta z}.
 
Once the change in position is in hand, the change in the GSF can be calculated in a few short steps. First, we write the result covariantly using $\Delta z^\alpha = e^\alpha_a \Delta z_1^a$. Next, we calculate the acceleration of $\Delta z_1^\alpha$ by taking two covariant derivatives along the worldline, yielding 
\beq
\Delta a^\alpha\equiv\frac{D^2\Delta z_1^\alpha}{d\tau^2} = -\frac{3}{4\pi}\frac{D^2}{d\tau^2}\left[e^\alpha_a\lim_{s\to 0}\int n^a n^b\xi_b d\Omega\right],
\eeq
Finally, we add and subtract $R^\alpha{}_{\mu\beta\nu}u^\mu\Delta z_1^\beta u^\nu$, leading to the evolution equation
\beq
\mu\frac{D^2\Delta z_1^\alpha}{d\tau^2} = -\mu R^\alpha{}_{\mu\beta\nu}u^\mu\Delta z_1^\beta u^\nu + \Delta F^\alpha,\label{dzddot}
\eeq
where we have identified
\begin{align}\label{Df}
\Delta F^\alpha &\equiv \mu\Delta a^\alpha+\mu R^\alpha{}_{\mu\beta\nu}u^\mu\Delta z_1^\beta u^\nu\\
    &= -\frac{3}{4\pi}\frac{D^2}{d\tau^2}\left[e^\alpha_a\lim_{s\to 0}\int n^a n^b\xi_b d\Omega\right]\nonumber\\
    &\quad +\mu R^\alpha{}_{\mu\beta\nu}u^\mu\Delta z_1^\beta u^\nu     
\end{align}
as the change in the GSF under the transformation generated by $\xi_\alpha$. Our reason for adding zero in the form of Riemann terms is that doing so allows us to write the evolution equation for $z_{\rm 1Lor}^\alpha+\Delta z_1^\alpha$ in terms of a geodesic-deviation term plus a self-force term, as in Eq.~\eqref{motion-in-Lorenz-app}:
\begin{align}
\mu\frac{D^2}{d\tau^2}(z_{\rm 1Lor}^\alpha+\Delta z_1^\alpha) &= -\mu R^\alpha{}_{\mu\beta\nu}u^\mu(z_{\rm 1Lor}^\beta+\Delta z_1^\beta) u^\nu \nonumber\\
                                                    &\quad + F^\alpha_{\rm Lor} + \Delta F^\alpha.\label{zddot}
\end{align}
 
We now have our method of finding the GSF in a broad class of gauges, beginning in the Lorenz gauge and then transforming to the desired gauge. If the transformation satisfies conditions (SR1)--(SR4) enumerated above, \emph{and} it is such that the integral~\eqref{Delta z-app} yields a well-defined $C^2$ function of $\tau$ along $\Gamma$, then we say that the end gauge is \emph{sufficiently regular} to define the GSF. We calculate the change in GSF, $\Delta F^\alpha$, generated by such a transformation using Eq.~\eqref{Df}. The total GSF in the end gauge is then given by the GSF in the Lorenz gauge plus the change in the GSF. This is our approach to deriving expressions for the GSF in the radiation gauges in Sec.~\ref{force-no-string} and Appendix~\ref{force-with-string}.
 
Before proceeding, we comment on the difference between our class of sufficiently regular gauges and the Gralla-Wald class described in Ref.~\cite{Gralla-Wald:11}. The Gralla-Wald class is based on the coordinate freedom within a family of metrics satisfying certain smoothness conditions. Those smoothness conditions ensure the existence of an inner expansion, among other things. They also require the generator of a gauge transformation to be bounded at $\Gamma$, as well as precluding any terms involving $\ln s$, even bounded ones. However, weaker conditions can equally well ensure the existence of an inner expansion matchable to an outer one~\cite{Eckhaus:79}. Here we assume only that there exists a well-behaved inner expansion (i.e., one with no divergent powers of $\varepsilon$) that agrees with our outer expansion in the buffer region through sufficient order to make the identification $M^a=\mu z_1^a$. This allows us to impose only the relatively weak conditions (SR1)--(SR4) on the gauge vector.
 
\subsection{Mollified radiation gauges}\label{Colombeau}
We now consider whether any of the steps in the preceding derivation run aground in the completed radiation gauges, given those gauges' irregularities away from $\Gamma$. 
 
It is clear from the results of Sec.~\ref{Fermi-analysis} that the integral~\eqref{Delta z-app}, which gives the position $\Delta z^\alpha_1$ in a completed radiation gauge relative to that in a Lorenz gauge, is well defined in any of the half-, full-, or no-string gauges, since the divergences and discontinuities in $\xi_\alpha$ are well defined as distributions; the irregularities are integrable, as established in Sec.~\ref{Fermi-analysis}. Furthermore, the transformation from Lorenz to completed radiation gauge in each case preserves the local form~\eqref{h2} of the MP at all times. Therefore all three classes of completed radiation gauge are sufficiently regular (in the sense of the preceding section) to define the GSF. (We assume that the gauge vectors' dependence on $\tau$, which could not be fully determined in our leading-order local analysis, is at least $C^2$.)
 
But it is not immediately clear if that definition is sensible. Although the particular result \eqref{Delta z-app} for $\Delta z^\alpha_1$ is valid, the transformation of $h^{(2)}_{\alpha\beta}$ as a whole, given in Eq.~\eqref{dh2}, is not distributionally well defined: it contains products of $\xi_\alpha$ (and its derivatives), and multiplication is not defined between distributions.\footnote{Because of this fact, the specific route we took to arrive at Eq.~\eqref{Delta z-app} is not valid for a transformation to a radiation gauge. The terms in Eq.~\eqref{Delta htt} that are distributionally ill-defined all vanish in the limit $s\to0$, but in Eq.~\eqref{Delta M-app} the limit is taken only \emph{after} the ill-defined integral of those terms is evaluated. However, we can still arrive at Eq.~\eqref{Delta z-app} in a rigorous way by replacing Eq.~\eqref{Ma} with $\mu z_1^a=M^a = \frac{3}{8\pi}\int h^{(2,-2)}_{\tau\tau} n^a d\Omega$, where $h^{(2,-2)}_{\tau\tau}$ is the coefficient of $1/s^2$ in the small-$s$ expansion of $h^{(2)}_{\tau\tau}$. The notion of the coefficient of $s^n$ in $h^{(2)}_{\alpha\beta}$ remains well defined, despite $h^{(2)}_{\alpha\beta}$'s lack of distributional meaning, since the terms in Eq.~\eqref{dh2} remain defined almost everywhere in a pointwise sense.} Furthermore, we can question whether the half- or full-string positions can be meaningfully related to the mass dipole moment of an asymptotically flat inner background spacetime, since the string in the $1/s$ term in $h_{\mu\nu}$ disrupts the asymptotic flatness of $g^{(0)}_{\mu\nu}(\tau,\bar x^a)$.
 
Consequently, the notions of position and force we have ascribed to the particle might be spurious in these gauges. In this section we examine whether the definitions can be put on a sounder basis by interpreting them in terms of the Colombeau theory of nonlinear generalized functions. We answer in the affirmative for the no-string gauge. But we demonstrate that the answer is likely negative for the half-string and (most of the) full-string gauges.
 
Reference~\cite{Steinbauer-Vickers:06} reviews the application of Colombeau theory to general relativity. Reference~\cite{Grosser-etal:12} presents the more recent development of a diffeomorphism-invariant global version of the theory applicable to tensor distributions. Here, we only mention a few of the necessary ideas. A Colombeau algebra is a class $\mathcal{G}$ of nonlinear generalized functions the products of which are well defined, unlike products of distributions. An element $F$ of $\mathcal{G}$ is an equivalence class $F=[f_\eps]$ of smooth one-parameter families of functions $f_\eps$, where $0<\eps\leq1$. Each $f_\eps$ must not tend to infinity too rapidly when $\eps\to0$, in the sense that $f_\eps$ (and its derivatives to any given order) does not diverge more strongly than some inverse power of $\eps$. Two functions $f_\eps$ and $\hat f_\eps$ belong to the same equivalence class $F$ if they differ by functions that vanish faster than any positive power of $\eps$ (and whose derivatives to all orders likewise vanish) in the limit $\eps\to0$.\footnote{This definition applies in the so-called ``special algebra", which we use throughout this section.} Most importantly for us, a generalized function $F$ is said to be associated with (or weakly equivalent to) a distribution $T$ if for any (and therefore all) $f_\eps$ in $F$,
\beq
\lim_{\epsilon\to0}\int f_\epsilon \psi dV = \int T \psi dV\label{association}
\eeq
for all smooth test functions $\psi$.
 
We wish to interpret the fields $h^{\rm Rad'}_{\alpha\beta}$ and $\xi^{\rm Lor\to Rad'}_\alpha$ not as fundamental objects, but as distributions associated with nonlinear generalized functions $[h^{\eps}_{\alpha\beta}]$ and $[\xi^\eps_\alpha]$. Each element $h^\eps_{\alpha\beta}$ is a solution to the linearized EFE in a gauge that is ``close'' to Rad$'$ for small $\eps$ but is smooth away from the particle. We refer to the gauge of $h^\eps_{\alpha\beta}$ as a \emph{mollified radiation gauge}. The definition of perturbed position we have used is clearly well defined for each $h^\eps_{\alpha\beta}$: all the nonlinear terms in Eq.~\eqref{dh2} are manifestly well defined for each $\xi^\eps_\alpha$, and the inner expansion in Rad$'$ is asymptotically flat. Our goal is then to show that the position of the particle in the gauge Rad$'$ is associated with the position in the equivalence class of mollified gauges.
 
The next two sections explain how this goal can be attained in a no-string gauge, and the difficulties in doing the same in the gauges with a string. We satisfy ourselves with a moderately detailed sketch in each case rather than a rigorous proof, and we work at the level of components in the Fermi-like coordinates $(\tau,x^A,z)$ of Sec.~\ref{Fermi-analysis}.

\subsubsection{No-string gauges}
We assume, based on the local analysis of Sec.~\ref{Fermi-analysis} and the flat-space example of Sec.~\ref{reconstruction-completion}, that we can write the first-order MP in a no-string gauge as 
\beq
h^{\rm Rad'}_{\alpha\beta} = h^{\rm Lor}_{\alpha\beta} + 2\xi_{(\alpha;\beta)}
\eeq
with a gauge vector $\xi_\alpha$ that is smooth off the particle except at a $(2+1)d$ surface $\mathcal{S}$ intersecting the particle, across which it contains a jump discontinuity. $\xi_\alpha$ is manifestly well defined as a distribution, which implies that $h^{\rm Rad'}_{\alpha\beta}$ is as well, since it is constructed from derivatives and linear combinations of distributions. The distributional content in $h^{\rm Rad'}_{\alpha\beta}$ consists of a jump discontinuity across (and likely a Dirac delta function on) $\mathcal{S}$. In the Fermi-like coordinates at leading order we can locally approximate $\mathcal{S}$ by the plane $p_a(\tau)x^a=0$, and the gauge vector by
\beq\label{xi-no-app}
\xi_\alpha(\tau,x^b)=-\xi_\alpha^0(\tau,x^b)-Z_\alpha(\tau,x^a)+o(s^0),
\eeq
with $\xi_\alpha^0$ and $Z_\alpha$ given by Eqs.~\eqref{xi0_t-no}--\eqref{Z-no}. Everything here is perfectly sensible within linearized perturbation theory. 
 
However, it ceases to be sensible in second-order perturbation theory. If we attempt to write the second-order perturbation $h^{(2)}_{\alpha\beta}$ as
\beq
h^{(2)}_{\alpha\beta}=h^{\rm (2)old}_{\alpha\beta}+\Delta h^{(2)}_{\alpha\beta},
\eeq
where $\Delta h^{(2)}_{\alpha\beta}$ is given by Eq.~\eqref{dh2} and $h^{\rm (2)old}_{\alpha\beta}$ is the second-order perturbation in any gauge satisfying Eq.~\eqref{h2} and smooth off the particle, then $h^{(2)}_{\alpha\beta}$ is not well defined as a distribution, because the products of Heaviside and delta functions in $\Delta h^{(2)}_{\alpha\beta}$ are ill-defined.
 
To solve this problem, we upgrade the distribution $\xi_\alpha$ to a generalized function $[\xi^\eps_\alpha]$, such that the two are associated in the sense of Eq.~\eqref{association}. We want the MP $[h^\eps_{\alpha\beta}]=h^{\rm Lor}_{\alpha\beta}+2[\xi_{(\alpha;\beta)}^\eps]$ and the position perturbation $[(z^a_1)^\eps]=z^a_{\rm 1Lor}+[(\Delta z^a_1)^\eps]$ to likewise be associated with $h^{\rm Rad'}_{\alpha\beta}$ and $z^a_1+\Delta z_1^a$:
\begin{align}
\lim_{\eps\to0}\int h^\eps_{\alpha\beta}(\tau,x^a)\psi(x^a)d^3x &= \int h^{\rm Rad'}_{\alpha\beta}(\tau,x^a)\psi(x^a)d^3x,\label{h-association}\\
\lim_{\eps\to0}(\Delta z^a_1)^\eps(\tau) &= \Delta z^a_1(\tau).\label{z-association}
\end{align}
Here we treat individual components as distributions or generalized functions on $\mathbb{R}^3$ at each $\tau$. We note that the right-hand side of Eq.~\eqref{z-association} is calculated in Sec.~\ref{force-no-string} and given by Eq.~\eqref{Delta z-nostring}.

We can find a representative of $[\xi^\eps_\alpha]$ via the convolution of $\xi_\alpha$ with a \emph{mollifier} $\Phi$:
\beq
\xi^\eps_\alpha(\tau,x^a) = \frac{1}{\eps^3}\int \xi_\alpha(\tau,x^{a'})\Phi\left(\frac{x^{a'}-x^a}{\eps}\right)d^3x',\label{mollification}
\eeq
where $\Phi$ is a smooth function with unit volume (i.e., $\int\Phi(x^a)d^3x=1$), with rapidly decreasing derivatives at $x^a\to\pm\infty$, and with vanishing moments (i.e., $\int x^{a_1}\cdots x^{a_n}\Phi(x^a)d^3x=0$ for all $n>0$)~\cite{Steinbauer-Vickers:06}. For our purposes we restrict our attention to mollifiers with even parity, $\Phi(-x^a)=\Phi(x^a)$. One can easily see from Eq.~\eqref{mollification} that such mollifiers preserve the parity of $\xi_\alpha$, in the sense that if $\xi_\alpha(-x^a)=\pm \xi_\alpha(x^a)$ then $\xi^\eps_\alpha(-x^a)=\pm \xi^\eps_\alpha(x^a)$. 

Derivatives and products of generalized functions obey the natural rules $\partial_\beta[\xi^\eps_\alpha]=[\partial_\beta\xi^\eps_\alpha]$ and $[\xi^\eps_\alpha][\xi^\eps_\beta]=[\xi^\eps_\alpha\xi^\eps_\beta]$, allowing us to work with a given $\xi^\eps_\alpha$. Combining the rule for derivatives with Eq.~\eqref{mollification} immediately shows that Eq.~\eqref{h-association} is satisfied. Hence we must only validate Eq.~\eqref{z-association}.
 
Since each $\xi^\eps_\alpha$ at fixed $\eps$ is smooth, each satisfies conditions (SR1)--(SR4) at the particle (with $f_1\equiv0$ and $f_2$ independent of $n^a$). Therefore, given the smoothness off the particle, the steps leading to Eq.~\eqref{Delta z-app} for $(\Delta z_1^a)^\eps$ are manifestly well defined for each $\xi^\eps_\alpha$. Those steps culminate in the following expression for the object's position in the mollified gauge relative to that in the Lorenz gauge:
\beq\label{z1eps}
(\Delta z_1^a)^\eps = -\frac{3}{4\pi}\lim_{s\to 0}\int n^a n^b\xi^\eps_b d\Omega.
\eeq
Again using the smoothness of $\xi^\eps_\alpha$, we can write $\xi^\eps_a(\tau,x^b)=\xi^\eps_a(\tau,0)+O(s)$ and hence
\beq
(\Delta z_1^a)^\eps(\tau) = -\xi^\eps_a(\tau,0).
\eeq
The right-hand side can be evaluated directly from Eq.~\eqref{mollification}, after substituting Eq.~\eqref{xi-no-app} into the integrand. Given that $\xi_a^0(\tau,x^b)$ has odd parity and $\Phi(x^b)$ has even, $\xi^0_a$ contributes nothing to $\xi^\eps_a(\tau,0)$. The $o(s^0)$ terms in $\xi_a(\tau,x^b)$ contribute $o(\eps^0)$ terms; consider, for example, the integral $\frac{1}{\eps^3}\int s'\ln s'\Phi(x^{a'}/\eps)d^3x'=\int \eps s'\ln (\eps s')\Phi(x^{a'})d^3x'$. Again appealing to the even parity of the mollifier to evaluate the $Z^\pm_a$ terms, we have 
\beq
\xi^\eps_a(\tau,0)=-\frac{1}{2}Z^+_a(\tau)-\frac{1}{2}Z^-_a(\tau)+o(\eps^0)
\eeq
and our final result
\beq\label{Dz1-association}
\lim_{\eps\to0}(\Delta z_1^a)^\eps(\tau) =\frac{1}{2}Z^+_a(\tau)+\frac{1}{2}Z^-_a(\tau).
\eeq
This is precisely the result \eqref{Delta z-nostring} found for $\Delta z_1^a$ in the no-string gauge. Equation~\eqref{z-association} is therefore satisfied.

We conclude that the MP and GSF in a no-string gauge are associated, in a precise way, with the MP and GSF in an equivalence class of gauges that are smooth away from the particle. Since our only requirement of the mollifier was that it have even parity, there are a large number of such equivalence classes. Furthermore, we may note that since each $\xi^\eps_\alpha$ is smooth at the particle, each MP in any of the equivalence classes is within the Barack-Ori class of gauges.  

Rigorously proving these statements would require showing that they are valid covariantly and not merely in our Fermi-like coordinates. Since the key result~\eqref{Dz1-association} is established via a local analysis, this should not be prohibitively difficult. However, we believe our sketch here suffices, and we take the associations we have established to be a sound interpretation of the motion calculated in a no-string gauge.

\subsubsection{Half- and full-string gauges}
The half- and full-string cases turn out to be more troublesome. We assume that we can write the first-order MP as 
\beq
h^{\rm Rad'}_{\alpha\beta} = h^{\rm Lor}_{\alpha\beta} + 2\xi_{(\alpha;\beta)}
\eeq
with a gauge vector $\xi_\alpha$ that diverges on a $(1+1)d$ surface emanating from the particle, given locally by $x^A=0$ in the Fermi-like coordinates. The divergences in $\xi_\alpha$ are logarithmic or scale with inverse distance from the string, meaning they are integrable. Therefore, $\xi_\alpha$ is defined as a distribution, which implies that $h^{\rm Rad'}_{\alpha\beta}$ is as well. For concreteness we shall examine the half-string case; the full-string case proceeds analogously. In the Fermi-like coordinates we can then approximate the gauge vector by
\beq\label{xi-half-app}
\xi^\pm_\alpha(\tau,x^b)=-\xi_\alpha^{0\pm}(\tau,x^b)-Z^\pm_\alpha(\tau)+o(s^0),
\eeq
with $\xi_\alpha^{0\pm}$ given by Eqs.~\eqref{xi0_t-half}--\eqref{xi0_A-half}.

If we attempt to write the second-order perturbation $h^{(2)}_{\alpha\beta}$ as $h^{(2)}_{\alpha\beta}=h^{\rm (2)old}_{\alpha\beta}+\Delta h^{(2)}_{\alpha\beta}$, with $h^{\rm (2)old}_{\alpha\beta}$ in a gauge that is smooth off the particle, then $h^{(2)}_{\alpha\beta}$ contains terms that diverge with the fourth power of inverse distance from the string, which are not integrable, and it is not expressed as some linear operation on a distribution. So $h^{(2)}_{\alpha\beta}$ does not appear to be well defined as a distribution. Even if it can be written as a distribution, both it and $h_{\alpha\beta}$ are problematic because the singular string in them prevents the inner background $g^{(0)}_{\mu\nu}$ from being asymptotically flat.

We wish to overcome this problem by introducing a family of vectors $\xi^{\eps\pm}_\alpha$ that are smooth on the string and which satisfy Eqs.~\eqref{h-association} and \eqref{z-association}. As a first attempt to achieve this goal, we can consider mollifying $\xi^\pm_\alpha$ via Eq.~\eqref{mollification}. Each $\xi^{\eps\pm}_a$ would then be smooth at the particle, and we would find, as in the no-string case, $(\Delta z_1^a)^\eps(\tau) = -\xi^\eps_a(\tau,0)$ (for any choice of mollifier, regardless of its parity). But scrutinize the $z$ component of this equation. Since $\xi^{0\pm}_z=0$, we have $\xi^{\eps\pm}_z=Z^\pm_z+o(\eps^0)$. On the other hand, Appendix~\ref{force-with-string} shows that in the half-string gauge 
\beq\label{z1z}
\Delta z^z_{1\pm} = Z_z^\pm(\tau) \mp 2\mu.
\eeq
Therefore the perturbed position in the mollified gauges is not associated with that in the half-string gauge. The term over which they disagree, $\mp 2\mu$, is a supertranslation effect of the half-string gauge vector $\xi^\pm_\alpha$; this parity-irregular supertranslation is annulled by mollification. One might think that mollifying in the $xy$ plane at each fixed $z$, rather than performing a three-dimensional convolution, could preserve the supertranslation effect. We have investigated this possibility and found the same negative result as in the three-dimensional approach.

Although it may be possible to find a suitable class of mollified radiation gauges for which the MP and GSF are associated with those in the half-string gauge, the above analysis suggests that these gauges would have to be extremely specialized in order to preserve the supertranslation effect that occurs in the half-string gauge. The same conclusion would generically be reached for a full-string gauge. We conclude that the motion in these gauges, unlike that in the no-string gauge, probably cannot be interpreted in terms of the motion in an associated gauge that is smooth off the particle. This does not preclude a calculation of the motion in the gauges with strings. One may take Eq.~\eqref{Delta z-app} as a definition of the position in those gauges relative to that in the Lorenz gauge and proceed unhindered. But the definition has lost some of the properties we would like it to have, and one would have to carefully ponder the question of whether it can be trusted to provide an accurate approximation of a small, compact object's motion, in the sense of providing accurate predictions of gauge-invariant quantities related to that motion.

There is one exception to this analysis: the equal-weight full-string gauge. When the strings to either side of the particle are equally weighted, the MP is parity-regular, and there is hence no supertranslation effect (see Appendix~\ref{force-with-full-string}). The mollification in this case can be done exactly as it was in the no-string gauge, leading to the result $\lim_{\eps\to0}(\Delta z^a_{1\pm})^\eps = Z^a=\Delta z^a_{1}$. But as we argue at various points in this paper, a full-string gauge is unlikely to be amenable to a numerical mode-by-mode reconstruction.


\section{Self-force in an undeformed radiation gauge: the half- and full-string cases}\label{force-with-string}
In Appendix~\ref{motion-arbitrary-gauge}, we reviewed a method of deriving expressions for the GSF in any sufficiently regular gauge. Here we apply that method to find the GSF in (undeformed) half- and full-string completed radiation gauges. Our results in the half-string gauge come with a caveat attached; we refer the reader back to Appendix~\ref{Colombeau} for a discussion.

\subsection{Half-string gauge}
We first consider the case of a half-string gauge. The gauge vector $\xi_\alpha=\xi^{\rm Lor\to Rad'}_\alpha$ transforming from a Lorenz gauge to a half-string gauge is given by $\xi^\pm_\alpha=-\xi^{0\pm}_\alpha-Z^\pm_\alpha(\tau)+o(1)$, where $\xi^{0\pm}_\alpha$ is found in Eqs.~\eqref{xi0_t-half}--\eqref{xi0_A-half} in Fermi-like coordinates. $Z^\pm_\alpha(\tau)$ has not been determined within our analysis but is assumed to be a $C^2$ function of $\tau$. Substituting this gauge vector into Eq.~\eqref{Delta z} gives 
\beq
\Delta z^a_{1\pm} = \frac{3}{4\pi}\lim_{s\to0}\int n^a n^b\xi_b^{0\pm} d\Omega+Z^a_\pm, 
\eeq
where the integral is over a sphere of radius $s$ and we have used the identity $\int n^a n^bd\Omega=4\pi/3\delta^{ab}$ to evaluate the second term. In terms of angles $(\theta,\phi)$ covering the sphere, the area element is $d\Omega=\sin\theta d\theta d\phi$, the unit vector has components $n^a=(\sin\theta\cos\phi,\sin\theta\sin\phi,\cos\theta)$, and the term in the gauge vector still to be integrated is $\xi_A^{0\pm}=2\mu n^A/(1\pm \cos\theta)$ and $\xi_z^{0\pm}=0$. A simple calculation yields
\beq\label{z1z-half}
\Delta z^z_{1\pm} = Z_z^\pm(\tau) \mp 2\mu
\eeq
and
\beq
\Delta z^A_{1\pm} = Z_A^\pm(\tau).
\eeq
By noting that the tetrad member $e^\alpha_3$ has components $\delta^\alpha_z$ in our local coordinates, we can write the shift in position in the covariant form 
\beq
\Delta z^\alpha_{1\pm} = P^{\alpha\beta}Z_\beta^\pm \mp 2\mu e_3^\alpha. 
\eeq
Here we see that the transformation from Lorenz to half-string induces an intuitive shift in position corresponding to the smooth translation $Z^\pm_a$, but it also induces a finite shift along the direction of the string, due to the discontinuous term $\xi^{0\pm}_A$ in the gauge vector. That second shift can be understood as the effect of a supertranslation. As discussed in Sec.~\ref{gauges}, it does not have as intuitive a physical meaning as one might like.

From the shift in position, the equation of motion can be found simply by taking two derivatives along the worldline. Doing so, we find
\begin{align}
\mu\frac{D^2 \Delta z^\alpha_{1\pm}}{d\tau^2} &= \mu P^{\alpha\beta}\frac{D^2}{d\tau^2}Z_\beta^\pm \mp 2\mu^2 \frac{D^2}{d\tau^2}e_3^\alpha\nonumber\\
	&= -\mu R^\alpha{}_{\mu\beta\nu}u^\mu \Delta z^\beta_{1\pm} u^\nu\nonumber\\
	&\quad +\mu\left(P^{\alpha\beta}\frac{D^2}{d\tau^2}Z_\beta^\pm+R^\alpha{}_\mu{}^\beta{}_\nu u^\mu Z_\beta^\pm u^\nu\right)\nonumber\\ 
	&\quad +\mu\left(\frac{D^2}{d\tau^2}e_\pm^\alpha+R^\alpha{}_{\mu\beta\nu} u^\mu e^\beta_\pm u^\nu\right)\nonumber\\
	&= -\mu R^\alpha{}_{\mu\beta\nu}u^\mu \Delta z^\beta_{1\pm} u^\nu-\delta_{Z^\pm}F^\alpha-\delta_{e_\pm}F^\alpha.\label{acceleration-halfstring}
\end{align}
In going from the first line to the second, we have added zero in the form of Riemann terms, and we have defined
\beq
e^\alpha_\pm \equiv \mp 2\mu e^\alpha_3 = \mp 2\mu \frac{P^\alpha{}_\beta \ell^\beta}{\sqrt{P_{\mu\nu}\ell^\mu \ell^\nu}}.\label{epm}
\eeq
In going from the second line to the third, we have noted that the terms in parentheses are, respectively, equal (up to an overall sign) to $\delta_{Z^\pm}\tilde F^\alpha$ and $\delta_{e_\pm}\tilde F^\alpha$, the changes in a full gravitational force evaluated on $\Gamma$ due to continuous gauge vectors $\xi_\alpha=Z_\alpha^\pm$ and $\xi^\alpha=e^\alpha_\pm$, as given in Eq.~\eqref{dxiF-continuous}.

Equation~\eqref{acceleration-halfstring} contains a geodesic-deviation term plus a GSF term, which we now find to be
\beq
\Delta F^\alpha_\pm = -\delta_{Z^\pm}\tilde F^\alpha-\delta_{e_\pm}\tilde F^\alpha.
\eeq
Therefore, the total GSF in the half-string gauge is
\beq
F^\alpha_\pm = F^\alpha_{\rm Lor} + \Delta F^\alpha_\pm.\label{F-halfstring}
\eeq

\subsubsection{Mode-sum formula}
Equation \eqref{F-halfstring} can be written more usefully in mode-sum form. Simple manipulations yield
\begin{align}
F^\alpha_\pm &= F^\alpha_{\rm Lor} - \delta_{Z^\pm}\tilde F^\alpha - \delta_{e_\pm}\tilde F^\alpha \nonumber\\
	&= \sum_\ell\big[(\tilde F^\alpha_{\rm Lor})^\ell - (\delta_{Z^\pm}\tilde F^\alpha)^\ell - (\delta_{\xi^{0\pm}}\tilde F^\alpha)^\ell - A_\pm^\alpha L \nonumber\\
	&\quad - B^\alpha - C^\alpha/L\big] +\sum_\ell (\delta_{\xi_0^\pm}\tilde F^\alpha)^\ell -\delta_{e_\pm}\tilde F^\alpha,\label{F1-hs}
\end{align}
where we have written $F^\alpha_{\rm Lor}$ in standard mode-sum form, decomposed $\delta_{Z^\pm}\tilde F^\alpha$ into modes, and added and subtracted $\sum_\ell (\delta_{\xi^{0\pm}}\tilde F^\alpha)^\ell$. In all cases, the $\ell$ modes are  evaluated in the limit from the side of the particle containing no string, as discussed in Sec.~\ref{LL-method}.

Notice that in Eq.~\eqref{F1-hs}, the first three terms in square brackets together give $(\tilde F^\alpha)_\pm^\ell$, the modes of the full force in the half-string gauge. Combining them, we arrive at a formula in the traditional mode-sum form
\beq
F^\alpha_\pm = \sum_\ell\big[(\tilde F^\alpha)_\pm^\ell - A_\pm^\alpha L - B^\alpha - C^\alpha/L\big] -D^\alpha_\pm,
\eeq
where $A^\alpha_\pm$, $B^\alpha$, and $C^\alpha$ take their Lorenz-gauge values, and
\beq
D^\alpha_\pm \equiv -\sum_\ell (\delta_{\xi_0^\pm}\tilde F^\alpha)^\ell -\mu\frac{D^2e_\pm^\alpha}{d\tau^2}-\mu R^\alpha{}_{\mu\beta\nu} u^\mu e^\beta_\pm u^\nu.
\eeq
The first term in $D^\alpha_\pm$ is precisely (up to an overall sign) the regularization parameter $\delta D^\alpha_\pm$ defined in Eq.~\eqref{dD} for the GSF in the LL gauge. The second term is easily calculated in any given situation, given the simple form~\eqref{epm} of the vector $e_\pm^\alpha$.

In Sec. \ref{RP-general} we found that when writing the GSF in a certain LL gauge in terms of the modes of the full force in a half-string gauge, we pick up a finite $\delta D^\alpha_\pm$ term. One might hope that here, where we instead write the GSF in the same gauge as the full force, no $D^\alpha$ term would arise. The $e_\pm^\alpha$ term in $D^\alpha_\pm$ could exactly cancel the $\delta D^\alpha_\pm$ term, making $D^\alpha_\pm$ identically zero and leaving the regularization parameters equal to their values in the Lorenz gauge. However, explicit calculations show that this hope is in vain. For example, for the case of circular orbits in Schwarzschild coordinates, we find
\beq
\mu\frac{D^2e_\pm^\alpha}{d\tau^2}+\mu R^\alpha{}_{\mu\beta\nu} u^\mu e^\beta_\pm u^\nu = \frac{\pm 6\mu^2 M f_p}{r_p^{5/2}(r_p-3M)^{1/2}}\delta^\alpha_r.
\eeq
Comparing this to the result for $g^{\alpha\beta}\delta D_\beta^\pm$ from Eq.~\eqref{dDr}, we see that the two terms do not cancel. We conclude that although the motion can be defined directly in a half-string gauge, with no local deformation, the mode-sum formula in such a gauge is altered by a finite $\d D^\alpha$, and that finite $\d D^\alpha$ differs from the one found in the locally deformed half-string gauge.

\subsection{Full-string gauge}\label{force-with-full-string}
We now consider the case of a full-string gauge. Motion in a generic full-string gauge will include the supertranslation effects we observed in the half-string case. We consider instead the special case in which the strings on either side of the particle have equal weight. The gauge vector transforming from a global Lorenz gauge to such a full-string gauge is given by $\xi_\alpha=-\xi^{0}_\alpha-Z_\alpha(\tau)+o(1)$, where $\xi^{0}_\alpha$ is found in Eqs.~\eqref{xi0_t-full}--\eqref{xi0_A-full}. Substituting this into Eq.~\eqref{Delta z}, we find 
\begin{align}
\Delta z^a_1 &= -\frac{3}{4\pi}\lim_{s\to0}\int n^a n^b\xi_b d\Omega\\
 				&= Z^a(\tau).
\end{align}
Because the equal-weight full-string gauge is parity-regular, the supertranslation effects of the strings have cancelled one another, leaving only the effects of an ordinary translation.

Straightforward manipulations, following Sec.~\ref{force-averaging}, show that the motion in the equal-weight full-string gauge obeys the Quinn-Wald-Gralla  equation
\beq
\mu\frac{D^2z_1^\alpha}{d\tau^2} = -\mu R^{\beta}{}_{\mu\gamma\nu}u^{\mu}z_1^\alpha u^{\nu}+\frac{1}{4\pi}\lim_{s\to 0}\int \tilde F^\alpha d\Omega,
\eeq
where $\tilde F^\alpha$ and the integration procedure are as described in Sec.~\ref{force-averaging}.

We argued in Secs.~\ref{Fermi-analysis} and \ref{recon-MP} that the string singularities in the full-string MP are too singular to be expanded in spherical harmonics. Therefore a mode-sum formula in this gauge would be ill-defined.


\section{Transformation from Fermi-like to arbitrary coordinates}\label{trans_to_global}
The local analysis in Sec.~\ref{Fermi-analysis} is performed in Fermi-like coordinates. To make use of local results in devising practical mode-sum schemes, in this section we find the local expansion of the vector $\xi^{\rm Rad'\to Lor}_\alpha$ in an arbitrary coordinate system (and in an arbitrary algebraically special vacuum background). Our strategy begins by expressing a covariant expansion of the vector at a point in terms of its Fermi-like components, found in Eq.~\eqref{xipm} for the half-string case, \eqref{fullstring} for the full-string, and \eqref{nostring} for the no-string. Next, we expand the covariant expression for the vector at a point $x$ in terms of coordinate distances $\delta x^{\alpha'}\equiv x^\alpha-x^{\alpha'}$ relative to a point $x'$ on $\Gamma$. 

\subsection{Covariant expansion}\label{covariant_expansion}
As described in Sec.~\ref{Fermi-analysis}, the Fermi-like coordinates $(\tau,x^a)$ at a point $x$ are defined in terms of a point $\bar x=x_p(\tau)$ on $\Gamma$. We develop a covariant expansion of $\xi_\alpha$ starting from the definition of the three scalar fields
\beq
x^a = -e^a_{\bar\alpha}\sigma^{;\bar\alpha}\label{xa}
\eeq
together with the condition 
\beq
\sigma_{;\bar\alpha}u^{\bar\alpha} = 0,\label{sigmau}
\eeq
which states that the geodesic connecting $x$ and $\bar x$ intersects $\Gamma$ orthogonally. (Recall that $\sigma$ is Synge's world function, equal to one-half the squared geodesic distance from $x$ to $\bar x$.) We will also make use of the fact that the triad $e^a_{\bar\alpha}$ satisfies
\beq
\frac{De^{\bar\alpha}_a}{d\tau} = \omega_a{}^be^{\bar\alpha}_b.\label{Dedtau}
\eeq

Now, since the point $\bar x$ depends on the point $x$, when differentiating a function of the two points, say $f(x,\bar x)$, we have 
\begin{align}
\frac{df(x,\bar x(x))}{dx^\alpha} &= \frac{\partial f}{\partial x^\alpha}+\frac{\partial f}{\partial \bar x^\beta}\frac{d\bar x^\beta}{dx^\alpha}\\
 &= \frac{\partial f}{\partial x^\alpha}+\frac{\partial f}{\partial \bar x^\beta}u^{\bar\beta}\frac{d\tau}{dx^\alpha}.
\end{align}
In terms of one-forms, this reads
\beq
df = \frac{\partial f}{\partial x^\alpha}dx^\alpha +\frac{\partial f}{\partial \bar x^\beta}u^{\bar\beta}d\tau.
\eeq
By applying the same principle, we can differentiate Eq.~\eqref{sigmau} to find
\beq
d\tau = \nu \sigma_{;\bar\alpha\beta}u^{\bar\alpha}dx^{\beta} \label{dtau}
\eeq
where $\nu\equiv -(\sigma_{;\bar\alpha\bar\beta}u^{\bar\alpha}u^{\bar\beta})^{-1}$.
We can differentiate Eq.~\eqref{xa} in the same manner to find
\beq
dx^a = -\frac{De^a_{\bar\alpha}}{d\tau}\sigma^{;\bar\alpha} - e^a_{\bar\alpha}\left(\sigma^{;\bar \alpha}{}_{\beta}dx^\beta+\sigma^{;\bar \alpha}{}_{\bar\beta}u^{\bar\beta}d\tau\right).
\eeq
Substituting Eqs.~\eqref{Dedtau} and \eqref{dtau} into this equation returns
\begin{align}
dx^a & = -e^b_{\bar\alpha}\left[\delta^a_b\sigma^{;\bar\alpha}{}_\alpha+\nu\left(\omega^a{}_b\sigma^{;\bar\alpha}+\delta^a_b\sigma^{;\bar\alpha}{}_{\bar\beta}u^{\bar\beta}\right)\sigma_{;\alpha\bar\gamma}u^{\bar\gamma}\right]dx^\alpha.
\end{align}
We can now write any one-form $\xi_\alpha=(\xi_\tau,\xi_a)$ in covariant form using $\xi_\alpha=\xi_\tau \frac{d\tau}{dx^\alpha}+\xi_a \frac{dx^a}{dx^\alpha}$. 

All of these expressions are exact. Since we require only leading-order behavior in the transformation, we can take advantage of the standard covariant expansions~\cite{Poisson-Pound-Vega:11}
\beq
\sigma_{\bar\alpha\bar\beta} = g_{\bar\alpha\bar\beta}+O(s^2),\quad \sigma_{\alpha\bar\beta} = -g^{\bar\alpha}_\alpha g_{\bar\alpha\bar\beta}+O(s^2),
\eeq
where $g_\alpha^{\bar \alpha}$ is the parallel propagator from $\bar x^\alpha=x^\alpha_p(\tau)$ to $x^\alpha$. At leading order we then find  
\begin{align}
\frac{dx^a}{dx^\alpha} &= g^{\bar\alpha}_{\alpha}e^a_{\bar\alpha}+O(s),\\
\frac{d\tau}{dx^\alpha} &= -g^{\bar\alpha}_{\alpha}u_{\bar\alpha} +O(s^2).
\end{align}
For any $\xi_\alpha$, this allows us to write
\beq\label{xicovariant}
\xi_\alpha = g^{\bar\alpha}_\alpha\left(-\xi_\tau u_{\bar\alpha}+\xi_ae^a_{\bar\alpha}\right)+O(s\xi).
\eeq
Notice that because we work at leading order, the rotation $\omega_a{}^b$ does not come into play.

Equation~\eqref{xicovariant} has not yet utilized any information about $\xi_\alpha$. Recalling Eqs.~\eqref{xipm}, \eqref{fullstring}, and \eqref{nostring}, we see it can be written as 
\begin{equation}
\xi_\alpha = g_\alpha^{\bar \alpha}(-\xi^{0}_{\tau} u_{\bar \alpha}+\xi^{0}_Ae^A_{\bar \alpha}+Z_{\bar\alpha})+o(1),\label{xi_cov}
\end{equation}
where we have made use of the fact that $\xi^{0}_z=0$. We now set about removing the dependence on the choice of triad. First we write Eq.~\eqref{e3} in the simplified form 
\begin{equation}\label{e3v2}
e^\alpha_3 = -\frac{P^\alpha{}_\beta\ell^\beta}{u_\mu\ell^\mu}.
\end{equation} 
From the orthonormality conditions $g_{\alpha\beta}u^\alpha e^\beta_a=0$ and $g_{\alpha\beta}e^\alpha_a e^\beta_b=\delta_{ab}$, it follows that the remaining two legs $e_A^\alpha$ must satisfy
\begin{equation}
u_\alpha e^\alpha_A=0,\qquad \ell_\alpha e^\alpha_A=0, \qquad e_{A\alpha}e^\alpha_B=\delta_{AB}.
\end{equation}
However, we will not require a precise specification of these two legs, because we shall find they appear only in the contracted form $e_{A\alpha}e^A_\beta$. A general expression for this contraction can be found from the completeness relation $-u_{\alpha}u_{\beta}+e_{a\alpha}e^a_{\beta}=g_{\alpha\beta}$. Rearranging, we find $e_{a\alpha}e^a_{\beta}=P_{\alpha\beta}$. Combining this with Eq.~\eqref{e3v2} yields
\beq
e_{A\alpha}e^A_{\beta}=Q_{\alpha\beta},\label{eAeA}
\eeq
where
\beq
Q_{\alpha\beta}\equiv P_{\alpha\beta}-\frac{P_{\alpha\mu}P_{\beta\nu}\ell^{\mu}\ell^{\nu}}{(\ell^{\gamma}u_{\gamma})^2}\label{Q}
\eeq
is a projection operator orthogonal to both $u^{\alpha}$ and $\ell^{\alpha}$.

We now observe that $\xi^0_A$, given in Eq.~\eqref{xi0_A-half}, \eqref{xi0_A-full}, or \eqref{xi0_A-no}, is proportional to $x^A=-e^A_{\bar\alpha}\sigma^{;\bar\alpha}$, allowing us to write
\beq
\xi^{0}_A=-\xi e^{A}_{\bar\alpha}\sigma^{;\bar\alpha},
\eeq
where
\begin{align}
\xi^\pm &\equiv \frac{2\mu}{s\pm z}\quad \text{(half-string case)},\\
\xi &\equiv \frac{2\mu s}{\varrho^2}\quad \text{(full-string)},\\
\xi &\equiv \xi^+\theta^+ + \xi^-\theta^- \quad \text{(no-string)}.
\end{align}
Here we have defined step functions $\theta^\pm$ that are 1 on the side of $\mathcal{S}$ where $\xi_\alpha^\pm$ is regular, and 0 on the other side. 
Substituting this in Eq.~\eqref{xi_cov} and making use of \eqref{eAeA}, we obtain
\beq
\xi_\alpha = -g_\alpha^{\bar \alpha}\left(\xi^0_{\tau} u_{\bar \alpha}+\xi Q_{\bar\alpha\bar\beta}\sigma^{;\bar\beta}-Z_{\bar\alpha}\right)+o(1),\label{xi_cov2}
\eeq
which remains covariant. The scalar fields $s$, $\varrho$, and $z$ could easily be expressed in terms of $u^{\bar\alpha}$, $\ell^{\bar\alpha}$, and $\sigma^{;\bar\alpha}$, but we forgo that step.

\subsection{Coordinate expansion}
We now turn to the task of expanding Eq.\ \eqref{xi_cov2} in terms of coordinate differences $\delta x^{\alpha'}$.  This will require the standard expansions~\cite{Haas-Poisson:06}
\begin{align}
g^{\alpha'}_{\beta}(x,x')&=\delta^{\alpha'}_{\beta'}+O(s),\\
\sigma^{;\alpha'}(x,x')&=-\delta x^{\alpha'}+O(s^2),\\
\sigma_{;\alpha'\beta'}(x,x')&=g_{\alpha'\beta'}+O(s^2).
\end{align}
We wish to relate the Fermi-like coordinates $x^a$ to a coordinate difference $\delta x^{\mu'}$. To do so, we replace the dependence on $\bar x$ with a dependence upon the coordinates $x^{\mu'}=x^\mu_p(\tau')$ at some other convenient location on the worldline. For example, $\tau'$ can be chosen to be the proper time on the worldline at coordinate time $t$, such that $\delta t=t(x)-t(x')=0$. This replacement is effected by defining $x^a(\tau)=-e^a_{\alpha}(x_p(\tau))\nabla^{\alpha}\sigma(x,x_p(\tau))$ and expanding $x^a(\tau)$ about $\tau'=\tau-\delta\tau$. Placing a prime on indices associated with the point $x'$, we have
\begin{align}
x^a(\tau) &= x^a(\tau')+u^{\alpha'}x^a{}_{;\alpha'}(\tau')\delta\tau+O(s^2)\nonumber \\
 &= -e^a_{\alpha'}\left[\sigma^{;\alpha'}(x,x')+u^{\beta'}\sigma^{;\alpha'}_{\ \ \beta'}(x,x')\delta\tau+O(s^2)\right]\nonumber\\
 &= -e^a_{\alpha'}\left[\sigma^{;\alpha'}(x,x')+u^{\alpha'}\delta\tau+O(s^2)\right]\nonumber\\
 &= e^a_{\alpha'}\delta x^{\alpha'}+O(s^2).\label{x_generic}
\end{align} 
In going from the second line to the third we have used $\sigma_{;\alpha'\beta'}=g_{\alpha'\beta'}+O(s^2)$, and in going from the third to the fourth we have used $e^a_{\alpha'}u^{\alpha'}=0$ and $\sigma^{;\alpha'}=-\delta x^{\alpha'}+O(s^2)$. 

Most importantly for us, Eq.~\eqref{x_generic} [with Eq.~\eqref{e3v2}] yields $z=z_0+O(s^2)$, where
\beq 
z_0\equiv-u_{\alpha'}\delta x^{\alpha'}-\frac{\ell_{\alpha'}\delta x^{\alpha'}}{\ell_{\beta'}u^{\beta'}}.\label{z0}
\eeq
We shall not require explicit expressions for $x^A$, which would require specifying $e_A^{\alpha'}$. However, we do require explicit expressions for the distances $s$ and $\varrho^2$. Using $s^2=\delta_{ab}x^ax^b=P_{\bar\alpha\bar\beta}e^{\bar\alpha}_ae^{\bar\beta}_bx^ax^b$, we have $s = s_0+O(s^2)$, where 
\beq \label{s0}
s_0^2\equiv P_{\alpha'\beta'}\delta x^{\alpha'}\delta x^{\beta'}.
\eeq
Next, from $\varrho^2=s^2-z^2$ we obtain $\varrho^2=s^2_0-z^2_0+O(s^3)$.

Returning to the covariant expression \eqref{xi_cov2}, we see that these expansions of $s$, $z$, and $\varrho$ allow us to express $\xi_\tau$ and $\xi$ in terms of the coordinate distance from an arbitrary point $x'$ on $\Gamma$. But the bitensors $g_\alpha^{\bar \alpha}$ and $\sigma^{;\bar\alpha}$ still retain a dependence on $\bar x$. Expanding that dependence about $x'$, we have
\beq
\xi_\beta = -g_\beta^{\alpha'}\left(\xi^0_{\tau} u_{\alpha'}+\xi Q_{\alpha'\beta'}\sigma^{;\beta'}-Z_{\alpha'}\right)+o(1).\label{xi_cov3}
\eeq
We complete our coordinate expansion by using $g^{\alpha'}_{\beta}=\delta^{\alpha'}_{\beta'}+O(s)$ and $\sigma^{;\alpha'}=-\delta x^{\alpha'}+O(s^2)$. 

Our final result is
\begin{equation}
\xi_\alpha = -\xi^{0}_{\tau} u_{\alpha'}+\xi Q_{\alpha'\beta'}\delta x^{\beta'}+Z_{\alpha'}+o(1).\label{xi_final}
\end{equation}
It is reproduced as Eq.~\eqref{xi-coords-final} in Sec.~\ref{Fermi-analysis}, where the form of the right-hand side is written explicitly for each of  the three cases---half-, full-, and no-string.


\section{Local gauge transformation in Eddington-Finkelstein coordinates}\label{Schw-example}
The main thread of this paper begins with local information in Fermi-like coordinates and then transforms to arbitrary coordinates. We complement that approach here by calculating the local gauge transformation $\xi_\alpha^{\rm Rad'\to Lor}$ directly in global coordinates in a particular physical scenario: generic orbits in a Schwarzschild background. The explicit calculations proceed in analogy with Sec.~\ref{Fermi-gauge-transformation}.

\subsection{Setup}
We focus on the case of an ingoing radiation gauge, in which the null vector $\ell^\alpha$ is in the outgoing null direction. Just as in the analysis in Fermi-like coordinates, we simplify the problem by using coordinates adapted to this vector. Here the ideal choice is Eddington-Finkelstein (EF) coordinates $(v,u,\th,\vf)$, in which
\beq
\ell^{\alpha}=(2/f,0,0,0),
\eeq
where $f\equiv 1-2M/r$.

We write the particle's zeroth-order worldline $\Gamma$ in EF coordinates as $x^{\alpha}_p=(v_p,u_p,\th_p,\vf_p)$, and without loss of generality, we set the particle on the equator, where $\th_p\equiv\pi/2$. $\Gamma$ is characterized by the particle's (specific) energy $\mathcal{E}=-(u_u+u_v)$ and angular momentum $\mathcal{L}=u_{\varphi}$. In terms of these constants of motion, the nonvanishing covariant EF components of the 4-velocity are
\beq\label{u}
u_v=-\frac{1}{2}(\mathcal{E}-\rdot_p),\qquad
u_u=-\frac{1}{2}(\mathcal{E}+\rdot_p),\qquad
u_\vf=\mathcal{L}.
\eeq
The 4-velocity in the radial direction is
\beq
\rdot_p \equiv \frac{dr_p(\tau)}{d\tau}= \pm\sqrt{\mathcal{E}^2-(1+\mathcal{L}^2/r_p^2)f_p}, \label{rdot}
\eeq
where $f_p\equiv f(r_p)$.

Our calculations will be performed in terms of local expansions near the worldline, of the sort described in Appendix~\ref{trans_to_global}. For that purpose, we define $\dx^{\alpha'}\equiv x^{\alpha}-x^{\alpha'}(x)$ to be the coordinate difference between a point $x$ off the worldline and a nearby point $x'(x)$ on the worldline, and we expand functions of $x$ in powers of $\dx^{\alpha'}$. Since we begin with functions of $x$, with no dependence on $x'$, we must fix a relationship between $x'$ and $x$. It will prove most convenient to take $x'(x)$ to be the point on $\Gamma$ with the same retarded time as $x$: $x'=x_p(\tau_u(u))\equiv x_p(u)$, where $\tau_u(u)$ is the proper time on $\Gamma$ at retarded time $u$. Explicitly,
\beq
 x^{\alpha'}(x)=(v_p(u),u,\pi/2,\varphi_p(u)), 
\eeq 
and
\begin{align}
\delta x^{\alpha'}&=(v-v_p(u),0,\th-\pi/2,\vf-\vf_p(u))\nonumber\\
				&\equiv (\d v,0,\d\theta,\d\varphi).
\end{align}
In what follows, we will consistently use primed indices to denote components of a field evaluated at $x'$. We will use $v_p$, $\vf_p$, and the derived quantities $r_p$ and $\dot r_p$ to denote $v_p(u)$, $\vf_p(u)$, $r_p(u)$, and $\dot r_p(u)$.

\subsection{Local gauge deformation}
Our goal is to solve the local gauge transformation equations \eqref{gaugetransformation2} and \eqref{gaugetransformation3}. We begin with Eq.~\eqref{gaugetransformation2}. After a few simplifications, and assuming Christoffel terms are subdominant compared to derivatives, Eq.~\eqref{gaugetransformation2} becomes
\begin{eqnarray}
\xi_{v,v}  &=&\frac{a_{vv}}{s_0}+o(s^{-1}), \label{eqvv}\\
\xi_{v,u}  +\xi_{u,v}   & =& \frac{a_{vu}}{s_0}+o(s^{-1}), \label{eqvu}\\
\xi_{v,\th}  +\xi_{\th,v}   & = & o(s^{-1}), \label{eqvth}\\
\xi_{v,\vf}+\xi_{\vf,v} & = & -\frac{a_{v\vf}}{s_0}+o(s^{-1}) \label{eqvfi},
\end{eqnarray}
where $s_0$ is given in Eq.~\eqref{s0}, and the (positive) coefficients are 
\begin{eqnarray}
a_{vv}&=&\frac{\mu}{2}(\mathcal{E}-\rdot_p)^2,
\nonumber\\
a_{vu}&=&\mu f_p\mathcal{L}^2/r_p^2,
\nonumber\\
a_{v\vf}&=&2\mu \mathcal{L}(\mathcal{E}-\rdot_p).
\end{eqnarray}

\subsubsection{General solutions}
We elide the detailed process of solving Eqs~\eqref{eqvv}--\eqref{eqvfi}. One can check by substitution that two general solutions are
\beq\label{xiEF}
\xi^{\pm}_{\alpha}=\mp 2 \mu u_{\alpha'} \ln \Delta_v^{\pm}+\mu\frac{\delta_{\alpha}}{\Delta_v^{\pm}}-\delta v\Xi^\pm_{v,\alpha}+\Xi^\pm_\alpha,
\eeq
where
\begin{align} 
\Delta_v^{\pm}	&\equiv s_0\mp u_{\alpha'}\dx^{\alpha'}, \label{Deltav}\\
\delta_{\alpha}	&\equiv 2\mathcal{L}\left(0,-\frac{\dvf}{u^{u'}},\frac{\dth}{u^{\vf'}},\frac{\dvf}{u^{\vf'}}\right)\!,\label{delta_alpha}
\end{align}
and $\Xi^\pm_\alpha$ are arbitrary functions satisfying
\beq\label{dXi}
\partial_v\Xi^\pm_\alpha=o(s^{-1}).
\eeq
The terms involving $\Delta_v^\pm$ make up two particular solutions, and the terms involving $\Xi^\pm_\alpha$ make up two general solutions to the corresponding homogeneous equation. When checking that they are solutions, one must be mindful of the $u$ dependence in $x'$. Derivatives of $x'$-dependent quantities act as $\frac{dx^{\alpha'}}{du}=\frac{u^{\alpha'}}{u^{u'}}$ and $\frac{d u_{\alpha'}}{du}=O(1)$.

The gauge vector must also satisfy Eq.~\eqref{gaugetransformation3}, which arose from the trace-free condition on $h^{\rm Rad'}_{\alpha\beta}$. One can verify that it imposes no restriction on the two particular solutions, but it requires the homogeneous solutions to satisfy
\begin{align}
\Xi^\pm_{\th,\th}+\Xi^\pm_{\vf,\vf} &= o(s^{-1}),\label{divXi}\\
(\Xi^\pm_{v,\th\th}+\Xi^\pm_{v,\vf\vf}) &= o(s^{-2})\quad \text{for }\d v\neq0.\label{LaplaceXi}
\end{align}
These conditions on $\Xi^\pm_\alpha$, as well as that in Eq.~\eqref{dXi}, are trivially satisfied by any sufficiently smooth piece of $\Xi^\pm_\alpha$, but they do restrict non-differentiable pieces of $\Xi^\pm_\alpha$.

As we found in Sec.~\ref{Fermi-analysis}, the general solutions contain three classes of particular solutions, each with its own distinct type of irregularity away from the particle.

\subsubsection{Half-string solutions}\label{half-string-Schw}
Consider the particular solutions $\xi_\alpha^+$ and $\xi_\alpha^-$ with $\Xi^{\pm}_\alpha=0$. They diverge everywhere where $\Delta_v^\pm\to0$. This includes the particle, where $\delta x^{\alpha'}=0$, but it also includes a half-ray emanating from the particle. To see this, fix $x'$ and consider the behavior along the radial null ray $\du=\dth=\dvf=0$. We find
\beq
\Delta_v^{\pm}(\dth=\dvf=0)=-(|\dv|\pm \dv)u_{v'},
\eeq
which vanishes along $\dv<0$ if the upper sign is chosen, and along $\dv>0$ if the lower sign is chosen. We can gather more precise information by expanding $\Delta_v^{\pm}$ for small $\delta\theta$ and $\delta\vf$ at fixed $\d v\neq0$. The result in the singular half of spacetime is
\beq
\Delta_v^\pm = \pm\frac{r_p^2(\d\th^2+\d\vf^2)}{2u_{v'}\d v}+O(\d\th^3,\d\vf^3),
\eeq
which goes to zero with the square of the distance from the singular half-ray. Hence, the terms proportional to $u_{\alpha'}$ in $\xi_\alpha^+$ blow up logarithmically on the ingoing radial null ray $\dv(u)<0$, and in $\xi_v^{-}$ they blow up logarithmically on the outgoing radial null ray $\dv(u)>0$. The terms involving $\delta_{\alpha'}$ diverge as inverse distance from those same singular half-rays. The solution $\xi_v^{+}$ is regular for $\d v(u)>0$, and $\xi_v^{-}$ is regular for $\d v(u)<0$. 

Using the freedom in $\Xi_\alpha^{\pm}$, we can switch the string from one side of the particle to the other. A straightforward calculation shows that $\Xi_\alpha^\pm=\pm 2\mu u_{\alpha'} \ln (\Delta_v^{+}\Delta_v^{-})$ achieves this goal, and one can confirm that it satisfies Eqs.~\eqref{dXi}, \eqref{divXi}, and \eqref{LaplaceXi}.

We can generate a whole class of half-string solutions by choosing $\Xi_\alpha$ to be arbitrary continuous functions. Since they are continuous, they can be approximated as $\Xi_\alpha(x)=\Xi_\alpha(x')+o(1)$. Furthermore, the terms $\dv\,\Xi^{\pm}_{v,\alpha}$ in $\xi_\alpha^\pm$ become $o(1)$ and may be neglected. Therefore we arrive at the class of half-string solutions 
\beq\label{xi-half-string-Schw}
\xi_\alpha^\pm=\xi^{0\pm}_\alpha+Z^\pm_\alpha(u)+o(1),
\eeq
where $Z^\pm_\alpha(u)\equiv \Xi^\pm_\alpha(x_p(u))$ is arbitrary and
\begin{align}\label{xi0-half-string-Schw}
\xi^{0\pm}_{\alpha}\equiv \mp 2 \mu u_{\alpha'} \ln \Delta_v^{\pm}+\mu\frac{\delta_\alpha}{\Delta_v^{\pm}}.
\end{align}

\subsubsection{Full-string solutions}\label{full-string-Schw}
Since the half-string fields $\xi_\alpha^+$ and $\xi_\alpha^{-}$ given in Eq.~\eqref{xi-half-string-Schw} are each solutions to Eqs.~\eqref{gaugetransformation2} and \eqref{gaugetransformation3}, any linear combination $n\xi_\alpha^++(1-n)\xi_\alpha^-$, $n\in\mathbb{R}$, is also a solution. For $n\neq0,1$, these solutions contain singularities on the entire radial ray $\dth=\dvf=0$ at each retarded time $u$. We can write the gauge vector as $\xi^{(n)}_\alpha=\xi^{0(n)}_\alpha+Z_\alpha(u)+o(1)$, where $Z_\alpha(u)$ is arbitrary and $\xi^{0(n)}_\alpha=n\xi_\alpha^{0+}+(1-n)\xi_\alpha^{0-}$. Generically, the divergences on each side of the particle have differing magnitudes, respectively proportional to $n$ and $1-n$. As a special case, we can consider weighting the divergences identically by choosing $n=1/2$, leading to
\beq\label{xi-full-string-Schw}
\xi_\alpha=\xi^{0}_\alpha+Z_\alpha(u)+o(1),
\eeq
where $Z_\alpha(u)$ is arbitrary and
\beq\label{xi0-full-string-Schw}
\xi_{\alpha}=\mu u_{\alpha'} \ln (\Delta_v^{-}/\Delta_v^{+})+\mu\frac{\eps_0\delta_\alpha}{\Delta_v^{+}\Delta_v^{-}}.
\eeq

\subsubsection{No-string solutions}\label{no-string-Schw}
The full-string solutions were found by summing two half-string solutions. But we can also consider combining two half-string solutions in a different way: by gluing together the regular regions of each. The surface $\mathcal{S}$ along which we glue them can, in principle, be chosen almost arbitrarily, so long as the two half-strings lie on opposite sides of it. Take the simple choice of gluing along the sphere $\dv=0=\d u$. It leads to the gauge vector 
\beq\label{xi-no-string-Schw}
\xi_\alpha=\xi^{0}_\alpha+Z_\alpha(u,v)+o(1),
\eeq 
where 
\begin{align}
\xi^0_\alpha &= \xi^{0+}_\alpha\theta^++\xi^{0-}_\alpha\theta^-,\\
Z_\alpha &= Z^{+}_\alpha(u)\theta^++Z^{-}_\alpha(u)\theta^-.
\end{align}
Here $\theta^\pm\equiv\theta[\pm(r-r_p)]$. These no-string solutions are regular on both sides of the particle, but they contain a jump discontinuity across the sphere of coordinate radius $r=r_p$.

\subsection{Comparison with the results in arbitrary coordinates}
We have now completed the analysis of the gauge vector in EF coordinates. It is instructive to check that these specific results can be recovered from the general results we obtained in arbitrary coordinates, given by Eq.~\eqref{xi_final}. The general results were written in terms of a coordinate distance $\delta x^{\alpha'}=x^\alpha-x^{\alpha'}$, where $x^{\alpha'}=x_p^\alpha(\tau')$. As a first step in reducing them to our specific results, we choose $\tau'=\tau_u$. This implies $\ell_{\alpha'}\delta x^{\alpha'}=\ell_{u'}\delta u=0$, which reduces Eq.~\eqref{z0} to 
\beq
z_0=-u_{\alpha'}\delta x^{\alpha'};\label{z_EF}
\eeq
that is, $z$ at leading order is the proper time from $x'$ to $\bar x$. Although we do not require expressions for the scalars $x^A$, we can easily find them by fixing the orbit to lie in the equatorial plane, as we did at the beginning of this section. Doing so eliminates $u^\theta$ from the orthogonality condition $e^A_{\alpha}u^{\alpha}=0$, allowing us to make the following simple choices for the triad members $e^\alpha_A$: 
\begin{equation}\label{e1_and_e2}
e^\alpha_1 = \frac{1}{r_p}\delta^\alpha_\theta, \qquad e^\alpha_2 = \frac{1}{r_p}\left(\delta^\alpha_\phi-\frac{u_\phi}{u_v}\delta^\alpha_v\right).
\end{equation}
Substituting these expressions into Eq.~\eqref{x_generic}, we find
\begin{eqnarray}\label{coord_transform}
x=r_p\delta\theta+O(s^2),\qquad y=r_p\delta\varphi+O(s^2);
\end{eqnarray}
$x$ and $y$ are simply displacements away from the particle along the sphere described by $r=r_p$. The distance $\varrho$ now reads
\beq\label{dist_transform}
\varrho =r_p\sqrt{\delta\theta^2+\delta\varphi^2}+O(s^2);
\eeq
at leading order it is the distance from the particle on that same sphere. We can also write it in terms of $\Delta^\pm_v$ as $\varrho^2=\Delta^+_v\Delta^-_v+O(s^3)$. Similarly, $s_0\pm z_0$ is simply $\Delta^\pm_v$. Lastly, a short calculation shows that $Q_{\alpha'\beta'}\delta x^{\beta'}=\frac{1}{2}\delta_{\alpha}$. 

With these results in hand, and noting that in the no-string solution the surface of discontinuity, $\mathcal{S}$, is the sphere at $r=r_p$, one finds that the general gauge vector in Eq.~\eqref{xi_final} trivially reduces to Eqs.~\eqref{xi-half-string-Schw} (in the half-string case), \eqref{xi-full-string-Schw} (in the full-string case), or \eqref{xi-no-string-Schw} (in the no-string case).


\section{Gauge transformation of the full force}\label{dF-arbitrary}
In this appendix we prove three important properties of the gauge transformation of the full force, $\delta_\xi \tilde F_\alpha$, generated by a gauge vector $\xi_\alpha$, and we establish local expansions of it necessary for the calculations in Secs.~\ref{LL-method} and \ref{force-no-string} and Appendix~\ref{force-with-string}. We use the most general form of the full force, given in Eq.~\eqref{full_force}, which transforms according to Eq.~\eqref{dxiF} off $\Gamma$. We rewrite that transformation here in the slightly different form
\begin{eqnarray}
\delta_\xi \tilde F_\alpha = -\mu \tilde P_\alpha{}^\beta\left[\tilde u^\mu\tilde\nabla_\mu\left(\tilde u^\nu\tilde\nabla_\nu\xi_\beta\right)-\left(\tilde u^\mu\tilde\nabla_\mu \tilde u^\nu\right)\tilde\nabla_\nu\xi_\beta
\right.
\nonumber\\
\left.+\tilde R_{\beta\mu}{}^\gamma{}_\nu \tilde u^\mu\xi_\gamma \tilde u^\nu\right].\label{dF-arb-extension}
\nonumber\\
\end{eqnarray}
Recall that $\tilde u^\alpha$, $\tilde P_\alpha{}^\beta$, and $\tilde\nabla_\mu$ are any smooth extensions of the four-velocity $u^\alpha$, projection operator $P_\alpha{}^\beta$, and covariant derivative $\nabla_\mu$ off the worldline, and $\tilde R_{\beta\mu}{}^\gamma{}_\nu$ is the Riemann tensor corresponding to $\tilde\nabla_\mu$.

The three properties we will establish are as follows: in an expansion in terms of coordinate distances $\delta x^{\alpha'}= x^\alpha-x^{\alpha'}$ from $\Gamma$ in an arbitrary coordinate system,
\begin{enumerate}
\item the projection of $\xi_\alpha$ along $\tilde u_\alpha$ does not contribute to $\delta_\xi \tilde F_\alpha$ at leading order,
\item the leading-order term in $\delta_\xi \tilde F_\alpha$ is of the same order as the projection $\tilde P_\alpha{}^\beta\xi_\beta$,
\item if the spatial components $\xi_a$ of $\xi_\alpha$ in Fermi-like coordinates have a definite parity under $x^a\to-x^a$ at leading order, then all components of $\delta_\xi \tilde F_\alpha$ have that same parity under $\delta x^{\alpha'}\to -\delta x^{\alpha'}$ at leading order. 
\end{enumerate}
To derive these properties, we assume that all functions satisfy conditions (SR3) and (SR4) of Appendix~\ref{motion-arbitrary-gauge}.

When describing expansions in coordinate differences in Appendix~\ref{trans_to_global}, we allowed the reference point $x'$ on $\Gamma$ to be related in an arbitrary way to $x$. For concreteness, we now define $x'=x'(x)$ to be the position on the worldline at coordinate ``time'' $t$, such that $\delta t=0$, where $t$ is any coordinate that increases monotonically along the worldline. We allow any smooth extensions of $u^\alpha$, $P_\alpha{}^\beta$, and $\Gamma^\alpha_{\beta\gamma}$ off $\Gamma$, implying they have expansions of the form
\begin{align}
\tilde u^\alpha &= u^{\alpha'}+\tilde u^{\alpha'}{}_{,\mu'}\delta x^{\mu'}+O(s^2),\label{u-expansion}\\
\tilde\Gamma^\alpha_{\beta\gamma} &= \Gamma^{\alpha'}_{\beta'\gamma'}+\tilde\Gamma^{\alpha'}_{\beta'\gamma',\mu'}\delta x^{\mu'}+O(s^2),\label{Gamma-expansion}\\
\tilde P_\alpha{}^\beta &= P_{\alpha'}{}^{\beta'}+O(s),\label{P-expansion}\\
\tilde R_{\alpha\beta}{}^\gamma{}_\delta &= R_{\alpha'\beta'}{}^{\gamma'}{}_{\delta'}+O(s).\label{R-expansion}
\end{align}
In these expansions, each of the quantities on the left is a function of the field point $x=x'+\delta x'$, and the expansion coefficients on the right are functions of the worldline point $x'$. The notation $\tilde u^{\alpha'}{}_{\!,\mu'}$ indicates differentiation with respect to $x$ followed by evaluation at $x=x'$.

To evaluate Eq.~\eqref{dF-arb-extension}, we first determine the action of $\tilde\nabla_\mu$ on a dual vector $w_\alpha$, treating it as a function of $x^{\alpha'}$ and $\delta x^{\alpha'}$. Both $x^{\alpha'}$ and $\delta x^{\alpha'}$ are implicitly functions of $x^\alpha$: $x^{\alpha'}=x^{\alpha'}(t)$, and $\delta x^{\alpha'}=x^\alpha-x^{\alpha'}(t)$. When we act with a derivative at $x^\alpha$, we must differentiate these quantities as
\begin{align}
\partial_\alpha x^{\mu'} &= \delta^t_\alpha\frac{u^{\mu'}}{u^{t'}},\label{dx'}\\
\partial_\alpha \delta x^{\mu'} &= \delta^{\mu'}_{\alpha}-\delta^t_\alpha\frac{u^{\mu'}}{u^{t'}}.\label{ddeltax'}
\end{align}
Now define $\hat\partial_{\mu'}$ to be a partial derivative with respect to $x^{\mu'}$, holding $\delta x^{\mu'}$ fixed, and define $\delta_{\mu'}$ to be a partial derivative with respect to $\delta x^{\mu'}$, holding $x^{\mu'}$ fixed. Using Eqs.~\eqref{dx'}--\eqref{ddeltax'}, we find 
\begin{align}
\partial_\mu w_\alpha(x',\delta x') &= \frac{\partial x^{\beta'}}{\partial x^\mu}\hat\partial_{\beta'}w_\alpha+\frac{\partial \delta x^{\beta'}}{\partial x^\mu}\delta_{\beta'}w_\alpha\\
&=\delta_\mu^t\frac{u^{\beta'}}{u^{t'}}\hat\partial_{\beta'}w_\alpha+\left(\delta^{\beta'}_{\mu}-\delta^{t}_\mu\frac{u^{\beta'}}{u^{t'}}\right)\delta_{\beta'}w_\alpha.
\end{align}
Combining this with the expansion of the Christoffel symbols, we arrive at
\begin{eqnarray}
\tilde\nabla_\mu w_\alpha(x',\delta x')=\left[\delta_\alpha^\gamma\delta_\mu^t\frac{u^{\beta'}}{u^{t'}}\hat\partial_{\beta'}+\delta_\alpha^\gamma\left(\delta^{\beta'}_{\mu}-\delta^{t}_\mu\frac{u^{\beta'}}{u^{t'}}\right)\delta_{\beta'}\right.
\nonumber\\
\left.
-\Gamma^{\gamma'}_{\mu'\alpha'}+O(s) \right]w_\gamma.\label{grad-w}
\nonumber\\
\end{eqnarray}
Notice that in this expression, $\hat\partial_{\beta'}$ and $\Gamma^{\alpha'}_{\beta'\gamma'}$ do not affect $w_\alpha$'s parity or its scaling with $s$, while $\delta_{\beta'}$ both reverses the parity and reduces the order by one power of $s$. 

From these results and the expansion of $\tilde u^\mu$ in Eq.~\eqref{u-expansion}, we immediately find
\begin{eqnarray}
\tilde u^\mu\tilde\nabla_{\!\mu} w_\alpha(x',\delta x')=\qquad\qquad\qquad\qquad\qquad\qquad
\nonumber\\
\left[\delta_\alpha^\gamma u^{\beta'}\hat\partial_{\beta'}\right.
+\delta_\alpha^\gamma\left(\tilde u^{\beta'}{}_{,\delta'}-\tilde u^{t'}{}_{,\delta'}\frac{u^{\beta'}}{u^{t'}}\right)\delta x^{\delta'}\delta_{\beta'}
\nonumber\\
\left.
-u^{\mu'}\Gamma^{\gamma'}_{\mu'\alpha'}+O(s)\right]w_\gamma.
\nonumber\\\label{wdot}
\end{eqnarray}
Here we see that for any $w_\alpha$, the operator $\tilde u^\mu\tilde\nabla_\mu$ does not increase the singular behavior of the leading-order term, and it preserves the parity at that order; as we would expect, even though we work off the worldline, there is a sense in which a derivative ``along the worldline" changes neither the parity nor the order. [We assume, in accordance with conditions (SR3)--(SR4) of Appendix~\ref{motion-arbitrary-gauge}, that $\hat\partial_{\beta'}$ does not increase $w_\alpha$'s degree of singularity and that $\delta_{\beta'}$ increases it by no more than one order in $s$.] Therefore, in particular, $\tilde u^\mu\tilde\nabla_\mu \xi_\beta$ and $\tilde u^\nu\tilde\nabla_\nu\left(\tilde u^\mu\tilde\nabla_\mu \xi_\beta\right)$ have the same leading order as $\xi_\beta$ and the same parity at that order. 

Using Eq.~\eqref{wdot}, we can straightforwardly evaluate the first term in the transformation \eqref{dF-arb-extension}. We now move to the second term, $(\tilde u^\mu\tilde\nabla_\mu \tilde u^\nu)\tilde\nabla_\nu\xi_\beta$. An explicit calculation, using the expansions \eqref{u-expansion} and \eqref{Gamma-expansion} and the differentiation rules \eqref{dx'} and \eqref{ddeltax'}, yields 
\begin{align}
\tilde u^\nu\tilde\nabla_{\nu}\tilde u^\mu &=& \bigg[u^{\alpha'}\tilde u^{\mu'}{}_{\!,\beta'\alpha'}
			+2u^{\alpha'}\Gamma^{\mu'}_{\alpha'\gamma'}\tilde u^{\gamma'}{}_{\!,\beta'}
			+\tilde u^{\alpha'}{}_{\!,\beta'}\tilde u^{\mu'}{}_{\!,\alpha'} \nonumber\\
			&& +u^{\alpha'}\tilde \Gamma^{\mu'}_{\alpha'\gamma',\beta'}u^{\gamma'}\bigg]\delta x^{\beta'}+O(s^2).\label{a-expansion}
\end{align}
We note that this expression is the only place in which the choice of extension $\tilde\Gamma^\alpha_{\beta\gamma}$ enters into our calculation. Defining  $\tilde a^\mu \equiv \tilde u^\nu\tilde\nabla_{\nu}\tilde u^\mu$, the above result can be written compactly as $\tilde a^\mu=\tilde a^{\mu'}{}_{\!,\nu'}\delta x^{\nu'}+O(s^2)$. Combining this with Eq.~\eqref{grad-w}, we find
\beq
\left(\tilde u^\mu\tilde\nabla_{\!\mu} \tilde u^\nu\right)\!\tilde\nabla_\nu\xi_\beta = \left(\!\tilde a^{\gamma'}{}_{\!,\mu'}-\tilde a^{t'}{}_{\!,\mu'}\frac{u^{\gamma'}}{u^{t'}}\!\right)\!\delta x^{\mu'}\delta_{\gamma'}\xi_{\beta}+O(s\xi).\label{a-term}
\eeq 
As with the first term in Eq.~\eqref{dF-arb-extension}, we find that this second term preserves the parity and order of $\xi_\beta$.

The final expression for the change in full force can be found by substituting the expansions \eqref{wdot} and \eqref{a-term}, together with \eqref{P-expansion}, \eqref{R-expansion}, and \eqref{u-expansion}, into Eq.~\eqref{dF-arb-extension}. Regardless of the choice of extension, the resulting expression for $\delta_\xi \tilde F_\alpha$ at leading order receives no contribution from the component of $\xi_{\alpha}$ parallel to $\tilde u_\alpha$. To see this, replace the parallel projection $\tilde u_\alpha \tilde u^\beta\xi_{\beta}$ with its leading term $u_{\alpha'}u^{\beta'}\xi_{\beta}$, and do likewise for $\tilde P_\alpha{}^\beta$ in Eq.~\eqref{dF-arb-extension}. The desired conclusion then follows from (i) $u^{\mu'}\nabla_{\mu'} P_{\alpha'}{}^{\beta'}=0$, which together with Eq.~\eqref{wdot} implies that the projection operator commutes with the derivatives $\tilde u^\alpha\tilde\nabla_{\!\alpha}$ at leading order, (ii) $P_{\alpha'}{}^{\beta'}u_{\beta'}=0$, and (iii) $R_{\beta'\mu'}{}^{\gamma'}{}_{\nu'} u^{\mu'}u_{\gamma'} u^{\nu'}=0$, by virtue of the symmetries of the Riemann tensor.

This establishes the first property enumerated at the beginning of this appendix. The remaining two properties follow immediately from the first, given that (i) for any extension, each term in $\delta_\xi \tilde F_\alpha$ has the same parity and scaling with $s$ as does $\xi_\alpha$ itself, as we have noted over the course of the calculation, (ii) the projection $P_{\alpha'}{}^{\beta'}\xi_{\beta}$ inherits the parity of $\xi_a$, as shown in Appendix~\ref{trans_to_global}.

In the following two subsections, we write $\delta_\xi \tilde F_\alpha^{\pm}$ explicitly for two choices of extension.

\subsection{Example 1: rigid extension}\label{rigid extension}
In the simplest extension, which we call ``rigid'', the coordinate components of both $\tilde u^\alpha$ and $\tilde\Gamma^\alpha_{\beta\gamma}$ are extended as constant fields, i.e., they are taken to have the same coordinate values at $x$ as at $x'$. If we adopt this extension, then the partial derivatives of these quantities in the $\delta x'$ direction (i.e., $\tilde u^{\alpha'}{}_{,\beta'}\delta x^{\beta'}$ and $\tilde\Gamma^{\alpha'}_{\gamma'\delta,\beta'}\delta x^{\beta'}$) all vanish. We immediately find 
\begin{align}
\delta_\xi \tilde F_\alpha &= -\mu P_{\alpha'}{}^{\beta'}u^{\mu'}\hat\nabla_{\!\mu'}(u^{\nu'}\hat\nabla_{\!\nu'}\xi_\beta)
					-\mu R_{\alpha'\mu'}{}^{\gamma'}{}_{\nu'}u^{\mu'}\xi_{\gamma}u^{\nu'}\nonumber\\
					&\quad + O(s\xi),\label{dF-rigid-extension}
\end{align}
where $\hat\nabla_{\!\mu'}$ is the covariant derivative that acts on the $x'$ dependence of its argument while holding the $\delta x'$ dependence fixed, meaning it acts as $\nabla_{\!\mu'}\omega_\nu=\hat\partial_{\mu'}\omega_\nu-\Gamma^{\rho'}_{\mu'\nu'}\omega_{\rho}$.

We make use of this extension when calculating explicit corrections to regularization parameters in Sec.~\ref{LL-method}. While it might not be the most useful in practice, since it is not an extension for which the Lorenz-gauge parameters $A_\alpha$, $B_\alpha$, $C_\alpha$ are available~\cite{Barack-Ori:03b,Barack:09}, it serves to illustrate our main conclusions. It is also pertinent when comparing with the extant literature, because it is implicitly the one used by Shah \emph{et al.} in their calculation of the radiation-gauge GSF~\cite{Keidl-etal:10,Shah-etal:11}.

\subsection{Example 2: rigid extension of $u^\alpha$, natural extension of $\Gamma^\alpha_{\beta\gamma}$}
Another obvious option is to use a rigid extension of the four-velocity while allowing the Christoffel symbols to take their natural values (i.e., $\tilde \Gamma^\alpha_{\beta\gamma}=\Gamma^\alpha_{\beta\gamma}$). With this choice, we find 
\begin{align}
\delta_\xi \tilde F_\alpha &= -\mu P_{\alpha'}{}^{\beta'}u^{\mu'}\hat\nabla_{\!\mu'}(u^{\nu'}\hat\nabla_{\!\nu'}\xi_\beta)
					-\mu R_{\alpha'\mu'}{}^{\gamma'}{}_{\nu'}u^{\mu'}\xi_{\gamma}u^{\nu'}\nonumber\\
					&\quad +\mu P_{\alpha'}{}^{\beta'} \left[\Gamma^{\mu'}_{\gamma'\delta',\nu'}-\frac{u^{\mu'}}{u^{t'}}\Gamma^{t'}_{\gamma'\delta',\nu'}\right]u^{\gamma'}u^{\delta'}\delta x^{\nu'}\delta_{\mu'}\xi_{\beta}
					\nonumber\\
					&\quad +O(s\xi).\label{dF-rigidu-extension}
\end{align}
Since the extension of $P_{\alpha}{}^{\beta}$ does not come into play at leading order, this result holds for any choice.

We make use of this extension in Sec.~\ref{force-no-string} and Appendix~\ref{force-with-string}. It is an example of the types used most often in the literature, where $P^{\alpha\beta}$ and $u^\alpha$ are extended in some way or another but the Christoffel symbols are left with their natural values; see, e.g., Refs.~\cite{Mino-Nakano-Sasaki:03,Barack-Ori:03b,Barack:09}. For example, Eq.~\eqref{dF-rigidu-extension} is valid (after raising the indices $\alpha$ and $\alpha'$) in the extension used throughout Ref.~\cite{Barack:09}, where $\tilde F^\alpha$
(as opposed to $\tilde F_\alpha$) is written with a rigid extension of both $P^{\alpha\beta}$ and $u^\alpha$.

\bibliography{modrad}

\end{document}